\pdfoutput=1
\documentclass[usenatbib]{mn2e}
\usepackage{apjfonts}
\usepackage{amssymb}
\usepackage{amsmath}
\usepackage{ctable}
\usepackage{fixltx2e} 

\newcommand{\be}{\begin{equation}}
\newcommand{\ee}{\end{equation}}

\newcommand{\paperone}{Paper {\small I}} 
\newcommand{\papertwo}{Paper {\small II}} 
\newcommand{\paperthree}{Paper {\small III}} 

\newcommand{\ffirst}{f_{f}}
\newcommand{\flast}{f_{\ell}}
\newcommand{\Sprime}{S^{\prime}}

\newcommand{\Mach}{\mathcal{M}}
\newcommand{\cloudsub}{_{\rm cl}}
\newcommand{\cloudsubzero}{_{{\rm cl},\,0}}
\newcommand{\Machcompressive}{\Mach_{\rm compressive}}
\newcommand{\Machdisk}{\Mach_{h}}
\newcommand{\Machcloud}{\Mach\cloudsub}
\newcommand{\Machcloudzero}{\Mach\cloudsubzero}
\newcommand{\rhocrit}{\rho_{\rm crit}}
\newcommand{\rhocloud}{\rho\cloudsub}
\newcommand{\vtcloud}{v_{t,\,{\rm cl}}}
\newcommand{\cscloud}{c\cloudsub}
\newcommand{\cscloudzero}{c_{0}}
\newcommand{\xcloud}{\tilde{r}}

\newcommand\plotonesize[2]
 {\centering \leavevmode \includegraphics[width={#2\columnwidth}]{#1}}
\newcommand{\plotsidesize}[2]
 {\centering \leavevmode \includegraphics[width={#2\textwidth}]{#1}}
\newcommand{\acknowledgments}{\begin{small}\section*{Acknowledgments}\end{small}}
\newcommand\altaffilmark[1]{$^{#1}$}
\newcommand\altaffiltext[1]{$^{#1}$}
\title[Gravo-Turbulent Fragmentation]{A General Theory of Turbulent Fragmentation
\vspace{-0.5cm}}

\author[Hopkins]{
\parbox[t]{\textwidth}{ 
Philip F. Hopkins\altaffilmark{1}\thanks{E-mail:phopkins@astro.berkeley.edu}} 
\vspace*{6pt} \\
\altaffiltext{1}{Department of Astronomy, University of California
  Berkeley, Berkeley, CA 94720\vspace{-1.1cm}} \\
}

\date{Submitted to MNRAS, September, 2012\vspace{-0.6cm}}
\begin{document}
\maketitle
\label{firstpage}

\begin{abstract}

We develop an analytic framework to understand fragmentation in turbulent, self-gravitating media. In previous work, we showed how some properties of turbulence can be predicted by application of the excursion-set formalism. Here, we generalize this to understand fully time-dependent gravo-turbulent fragmentation and collapse. We show that turbulent systems are always gravitationally unstable in a probabilistic sense. The fragmentation mass spectrum, size/mass/density/linewidth relations of collapsing objects, their correlation functions and clustering, the range of spatial scales over which fragmentation occurs, and the time-dependent rate of collapse/fragmentation (as a function of size/mass) are  analytically predictable. We show how these depend on bulk properties of turbulence; fragmentation is promoted at higher Mach numbers and shallower power spectra. We also generalize the model to properly include rotation, complicated gas equations of state, collapsing/expanding backgrounds, magnetic fields, intermittency, and non-normal statistics (with inherently correlated fluctuations). This allows us to formally derive how fragmentation is suppressed with ``stiffer'' equations of state (e.g.\ higher polytropic index $\gamma$) or differently driven turbulence (solenoidal vs.\ compressive). The suppression appears at an ``effective sonic scale'' where $b\Mach(R_{\rm s},\,\rhocrit [R_{\rm s}])\approx1$, where $\rhocrit$ is the (scale-dependent) critical density for fragmentation. Gas becomes stable against collapse below this scale for $\gamma>4/3$; however fragmentation still occurs on larger scales. We show that the scale-free nature of turbulence and gravity generically drives mass spectra and correlation functions towards universal shapes (observed in a wide variety of astrophysical phenomena), with weak residual dependence on many properties of the media. We find that correlated fluctuations on different scales, non-Gaussian density distributions, and intermittency have surprisingly small effects on the fragmentation process. We demonstrate that this is because fragmentation cascades on small scales are generically ``frozen in'' when large-scale fluctuations push the ``parent'' region above the collapse threshold; though they collapse, their statistics are only weakly modified by the collapse process. Finally, with thermal or turbulent support, structure develops ``top-down'' in time via a fragmentation cascade, but we show that significant rotational/angular momentum support reverses the sense of structure formation to ``bottom-up'' growth via mergers of bound clumps, and introduces a characteristic ``maximal instability scale'' distinct from the Toomre scale.

\end{abstract}

\begin{keywords}
hydrodynamics --- instabilities --- turbulence --- star formation: general --- galaxies: formation --- galaxies: evolution --- protoplanetary discs --- cosmology: theory
\vspace{-1.0cm}
\end{keywords}

\vspace{-1.1cm}
\section{Introduction}
\label{sec:intro}

Fragmentation and the collapse of gas under the influence of self-gravity in turbulent media is a process central to a wide range of astrophysics. Galaxy formation within dark matter halos, the formation of giant molecular clouds (GMCs) and structure within the interstellar medium (ISM), the formation of protostellar cores within GMCs, formation of binary and multiple stellar systems in proto-stellar disks, and planet formation within proto-planetary disks may all be fundamentally related to these basic physics \citep[for reviews, see e.g.][]{scalo:2004.turb.fx.review,elmegreen:2004.obs.ism.turb.review,mac-low:2004.turb.sf.review}. As well, many processes as diverse as fusion within convective stars, grain growth in the ISM, and magnetic reconnection may be dramatically influenced by turbulent density and velocity fluctuations. 

Empirically, many independent lines of evidence suggest these phenomena are driven by some fundamental shared underlying physics and scale-free processes. For example, the mass functions (MFs) of stars at formation (IMF; for a review see \citealt{chabrier:2005.review2}), and of GMCs (\citealt{blitz:2004.gmc.mf.universal}), as well as protostellar cores \citep[e.g.][]{enoch:2008.core.mf.clustering,sadavoy:2010.cmfs}, star clusters \citep{portegies-zwart:2010.starcluster.review}, and HI ``holes'' or underdense bubbles in the ISM \citep[][and references therein]{oey:1997.HI.hole.models,walter:1999.ic2574.hI.holes} are nearly universal in shape. More remarkably, these mass functions are actually close to self-similar {\em to one another}, and even similar to the MF of (seemingly) completely unrelated systems such as dark matter halos! These all feature a power law-like range, with slopes close to ${\rm d}n/{\rm d}M\propto M^{-2}$, i.e.\ the slope at which there is equal mass per logarithmic interval in mass (a generic expectation of scale-free processes), with an exponential-like cutoff at low/high masses. Within these systems, quantities such as the mass-size relation (of GMCs, protostellar cores, ISM voids, massive star clusters, certain structures within turbulence regulating e.g.\ stellar temperature fluctuations, and dark matter halos) follow simple power-laws with near-universal slopes (see references above). The clustering (auto-correlation function) of young stars, proto-stellar cores, GMCs or dense (molecular) gas in the ISM, star clusters, and even galaxies also follows approximate power-laws (to first order), with -- surprisingly again -- apparently self-similar (near-universal) slopes \citep[compare e.g.][]{zhang:2001.antennae.starcluster.clustering,zehavi:local.clustering,stanke:2006.core.mf.clustering,enoch:2008.core.mf.clustering,hennekemper:2008.smc.stellar.clustering,kraus:2008.stellar.clustering,scheepmaker:2009.m51.cluster.clustering}. Extending this to the distribution of radial separations in binary and multiple stellar systems (where there is again a quasi-universal power-law distribution of separations) suggests that the same statistics may extend self-similarly to fragmentation within proto-stellar disks \citep[see][and references therein]{simon:1997.stellar.clustering,kraus:2008.stellar.clustering}. And higher-order statistics (such as column density distribution shapes and velocity structure functions) of the turbulent ISM appear consistent with simple scalings based on self-similar models that also apply to a wide range of laboratory MHD turbulence \citep[e.g.][]{boldyrev:structfn.tests,ridge:2006.lognormal.pdf,wong:2008.gmc.column.dist,schmidt:2009.isothermal.turb,lombardi:2010.larsonlaws.extinction.lognormal}. Indeed, there is a long history of phenomenological models treating these phenomena as different aspects of fractal-like (i.e.\ strictly self-similar) systems \citep[see e.g.][and references therein]{elmegreen:2002.fractal.cloud.mf}. 

Unfortunately, our theoretical understanding of the formation of self-gravitating structures within turbulent systems remains limited. Early work on stability and fragmentation focused on smooth media dominated by thermal pressure and/or rotation \citep[e.g.][]{jeans:1902.stability,toomre:Q,goldreich:1965.spiral.stability,lin.shu:spiral.wave.dispersion}. Subsequent analytic work extended this to include the effect of turbulent ``support'' (energy, momentum, or ram pressure) in resisting collapse, in the dispersion relation for linear density perturbations in turbulent, rotating, and possibly magnetized disks \citep[][]{chandrasekhar:1951.turb.jeans.condition,vandervoort:1970.dispersion.relation,elmegreen:1987.cloud.instabilities,bonazzola:1987.turb.jeans.instab}. But these derivations still assumed that the media of interest were homogenous and steady-state, despite the fact that perhaps the most important inherent property of turbulent systems is their inhomogeneity. These analytic considerations provide no means to calculate the statistical properties of fluctuations in these turbulent media, let alone the statistics of complicated objects forming within those media or their time-dependence.

This is not surprising: the systems of interest (fully developed turbulence) are chaotic, non-linear, inhomogenous (with large stochastic fluctuations), span an enormous dynamic range (Reynolds numbers, or ratio of driving to dissipations scales as large as $\sim10^{5}-10^{8}$), intermittent (with shocks in super-sonic turbulence producing very large local perturbations), time-dependent, and include complicated thermal, magnetic, and radiative processes as well as differential rotation. As a result, most progress in the last couple decades has come from numerical simulations. 

Such simulations have led to a number of important breakthroughs. Our understanding of the basic properties of astrophysical turbulence is rapidly improving, and it appears to obey at least some surprisingly simple scalings in the velocity fields, albeit with significant intermittency \citep{ossenkopf:2002.obs.gmc.turb.struct,federrath:2010.obs.vs.sim.turb.compare,
block:2010.lmc.vel.powerspectrum,bournaud:2010.grav.turbulence.lmc}. In the ideal case of isothermal, non self-gravitating turbulence, at least, it appears that density distributions can be described (approximately) as log-normal, with a dispersion that scales in a simple predictive manner with the large-scale compressive Mach number \citep{vazquez-semadeni:1994.turb.density.pdf,padoan:1997.density.pdf,scalo:1998.turb.density.pdf,ostriker:1999.density.pdf}. Turbulence and the induced density fluctuations evolve (or decay, depending on the driving) on a crossing time \citep[][and references therein]{pan:2010.turbulent.mixing.times}. For the problems of interest in star formation, the relation of turbulent driving, (super-sonic) Mach number, and density fluctuations in ideal ``driven boxes'' to the collapse rate of small self-gravitating regions is increasingly well-mapped \citep{vazquez-semadeni:2003.turb.reg.sfr,li:2004.turb.reg.sfr,padoan:2011.new.turb.collapse.sims,federrath:2012.sfr.vs.model.turb.boxes}. And in differentially rotating disks, some improvements to criteria for ``effective stability'' against fragmentation (beyond the original Toomre criterion) have been developed \citep[][]{gammie:2001.cooling.in.keplerian.disks,cai:2008.protoplanet.disk.w.rad,hopkins:inflow.analytics,elmegreen:2011.two.component.disk.instab}. Larger-scale simulations have attempted to follow the full ``fragmentation cascade'' in the ISM from GMCs to protostellar cores, with self-consistently driven turbulence from stellar feedback \citep[e.g.][]{kim:2002.mhd.disk.instabilities,tasker:2011.photoion.heating.gmc.evol,hopkins:fb.ism.prop,hopkins:stellar.fb.mergers}. These simulations and others like them have, individually, been able to reproduce many of the specific observations described above \citep[e.g.\ mass functions, size-mass relations, or correlation functions of some of the quantities of interest; see references above and][]{klessen:2000.cluster.formation,jappsen:2005.imf.scale.thermalphysics,krumholz:2011.rhd.starcluster.sim,hansen:2012.lowmass.sf.radsims}. 

But there are many caveats to the simulations. Numerical resolution is limited, often well below the tremendous dynamic range in spatial scale, mass, and Reynolds numbers over which these processes operate. The space of interesting parameters describing the turbulence, fluid properties, and backgrounds is enormous and only a very small fraction can be surveyed. Sampling extremely rare events requires running simulations for durations which are infeasible. And even when simulations are possible, the chaotic and non-linear nature of the systems means that interpreting the results, let alone extracting the important physics and understanding how to extrapolate it to systems not simulated -- critical to understand the links between diverse astrophysical phenomena -- is immensely challenging. 


Therefore considerable analytic theory has also been developed, much of which has used the basic results of these simulations to develop predictions for a wide range of scales and turbulent properties. Simple analytic derivations underpin our understanding of the approximately lognormal character of turbulent density PDFs \citep[e.g.][]{passot:1998.density.pdf,nordlund:1999.density.pdf.supersonic}, and \citet{passot:1998.density.pdf} extended this further to predict the non-lognormal character in non-isothermal flows. A range of ``cascade models'' have been developed that appear to successfully describe the scale-by-scale character of hierarchical velocity fluctuations in fully non-linear turbulence \citep[for a review, see][]{shezhang:2009.sheleveque.structfn.review}, and these have been increasingly applied to compressible density fluctuations \citep{boldyrev:structfn.tests,kowal:2007.log.density.turb.spectra,liufang:2008.logpoisson.cosmic.baryons}. Analytic models for star formation have been developed, which use the above scalings to calculate the mass or volume fraction exceeding some threshold density where self-gravity becomes important, and in turn use this to estimate the stellar initial mass function \citep{padoan:2002.density.pdf,hennebelle:2008.imf.presschechter,veltchev:2011.frag}, mass distribution of clusters \citep{klessen:2001.sf.cloud.pdf} and/or the integrated star formation rate \citep{krumholz.schmidt,hennebelle:2011.time.dept.imf.eps}. 

Building on these results, \citet{hopkins:excursion.ism} (hereafter \paperone) recently showed that the excursion-set formalism could be applied to calculate the statistics of bound objects in the density field of the turbulent ISM. This is a general mathematical formulation for random-field statistics, well known from cosmological applications as a means to calculate halo mass functions and clustering in the ``extended Press-Schechter'' approach from \citet{bond:1991.eps}. This is one of the most powerful theoretical tools in the study of large scale structure and galaxy formation, and the foundation for our analytic understanding of quantities such as halo mass functions, clustering, mergers, and accretion histories \citep[for a review, see][]{zentner:eps.methodology.review}. The  application to the ISM therefore represents a potential major breakthrough, providing a means to calculate many quantities analytically that normally would require numerical simulations. 

In \paperone, we focused on the specific question of GMCs in the ISM, and considered the case of isothermal gas with an exactly lognormal density distribution. We used this to construct certain statistics of the ``first-crossing distribution'': the statistics of bound objects defined on the largest scales on which they are self-gravitating. And we found that the predicted mass function and correlation functions/clustering properties agree remarkably well with observations of GMCs on galactic scales \citep[formalizing many of the earlier calculations in e.g.][]{klessen:2001.sf.cloud.pdf}. In \citet{hopkins:excursion.imf} (\papertwo), we extended the formalism (with identical assumptions) to the ``last crossing distribution'' -- specifically, the mass function of bound objects defined on the smallest scales on which they remain self-gravitating but do not have self-gravitating sub-regions (i.e.\ are not fragmenting). We argued that these should be associated with proto-stellar cores, and in \papertwo\ showed that the resulting core MF agrees well with canonical Milky Way core MF and (by extrapolation) stellar IMFs. This connects earlier theoretical work in \citet{padoan:2002.density.pdf} and \citet{hennebelle:2008.imf.presschechter} to a fully self-consistent galactic-scale framework. And in \citet{hopkins:excursion.clustering} (\paperthree) and \citet{hopkins:excursion.imf.variation} we showed how to calculate other properties of these ``last-crossings'' such as their correlation functions and dependence on the turbulent power spectrum. In addition to providing critical analytic insights into these quantities, this formulation allows us to simultaneously treat an enormous dynamic range in scales and consider extremely rare fluctuations, which are impossible to follow statistically in current simulations. However, the general applicability of these first models is limited by some of the assumptions made; in \paperone-\paperthree, we consider only isothermal, non-magnetized, isotropic gas, obeying strictly lognormal statistics with uncorrelated fluctuations on different scales, and evaluate properties at a fixed instant in stationary backgrounds (i.e.\ do not follow time-dependent collapse). We also considered only a narrow range of disk stability parameters, compressive-to-solenoidal ratios in the turbulence, and highly super-sonic Mach numbers. 

In this paper, we develop these initial models further, and generalize them in many critical ways to enable the development of a theory of turbulent fragmentation applicable to a wide range of interesting astrophysical systems. Specifically, we generalize the models to include arbitrary turbulent power spectra, different degrees of rotational support, complicated (multivariable) gas equations of state, collapsing (or, in principle, expanding) backgrounds, magnetic fields and anisotropic media, intermittency, and non-Gaussian statistics (with or without inherently correlated fluctuations on different scales). We also develop a time-dependent version of the theory, both for statistically stationary and for globally evolving backgrounds and collapsing self-gravitating ``fragments.'' We show how, for all of these cases, quantities such as the mass spectrum of self-gravitating objects, their size-mass-density-linewidth relations, correlation functions/clustering, spatial/mass scales of the fragmentation cascade, and rate of formation, collapse, and fragmentation can be predicted. We present these as general models, applicable to a wide range of turbulent systems, but we also show that we can already reach a number of quite general conclusions regarding e.g.\ the ``near-universality'' of quantities like mass functions, size-mass relations, and correlation functions, as well as the conditions and characteristic scales (and range of scales) at which fragmentation occurs.

\vspace{-0.5cm}
\subsection{Paper Overview}

\begin{figure*}
    \centering
    \plotsidesize{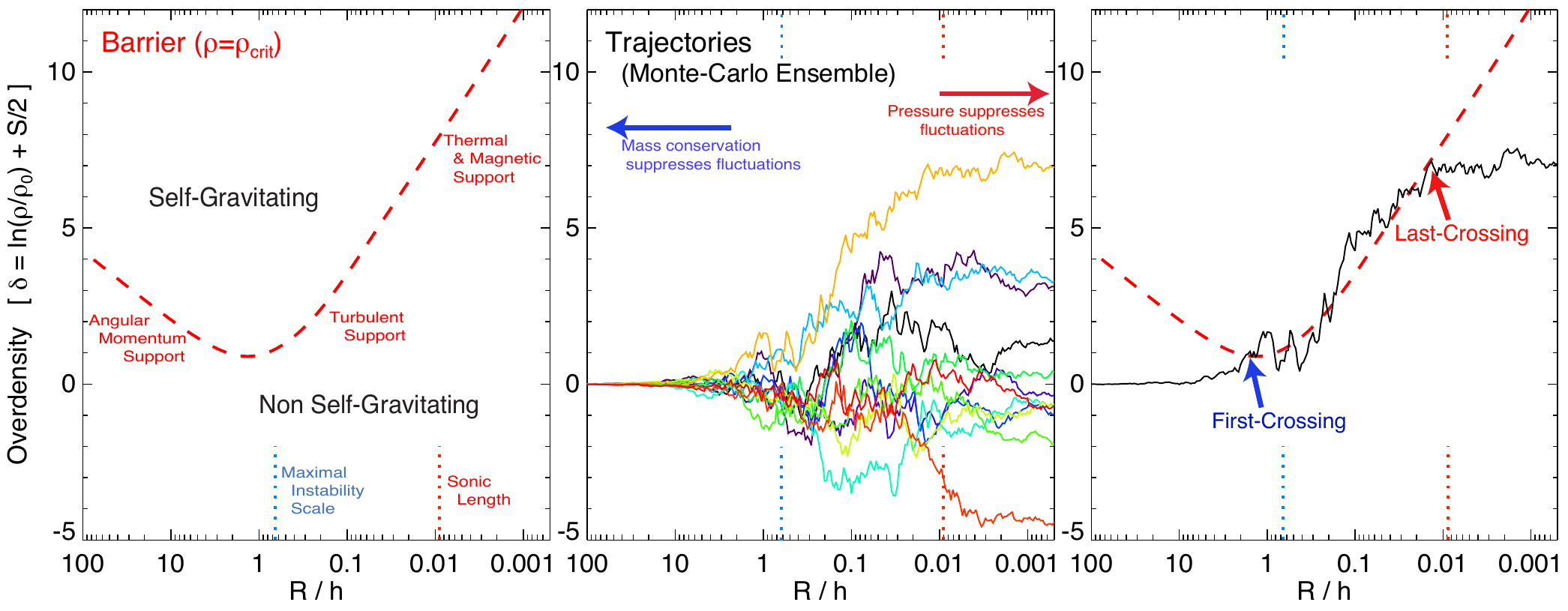}{1.0}
    \caption{Illustration of the model and key concepts presented here. 
    {\em Left:} The ``barrier,'' or critical density above which a given region (with average density $\rho(R)$ in a spherical radius $R$) inside a gaseous disk will collapse under self-gravity (Eqs.~\ref{eqn:rhocrit.1}-\ref{eqn:rhocrit}). We label the dominant term resisting collapse at each scale: at large scales ($\gtrsim h$) this is angular momentum, at intermediate scales $h\gtrsim R \gtrsim R_{\rm sonic}$ this is turbulence, and at small scales $R\lesssim R_{\rm sonic}$ this is thermal and magnetic pressure. We label the sonic length $R_{\rm sonic}$ (Eq.~\ref{eqn:r.sonic}) below which the Mach number $\Mach\lesssim1$, hence thermal support dominates, and the ``maximal instability scale'' (Eq.~\ref{eqn:r.maximal}) near $\sim h$ where fragmentation is most efficient. 
    {\em Center:} A Monte-Carlo ensemble of several ``trajectories.'' Each is an independent random realization of the density power spectrum, averaged in a window of variable size $R$ about a random point in space ${\bf x}$ within the disk. Comparing to the barrier, most points in space (i.e.\ most of the volume) are not self-gravitating at any scale, but rare regions do cross the threshold. At large scales, density fluctuations must be suppressed by mass conservation, so $\rho\rightarrow\rho_{0}$. At scales $\lesssim R_{\rm sonic}$, thermal pressure suppresses turbulent density fluctuations, so the smaller-scale fluctuations in each trajectory are damped and the curves become ``frozen'' at the values set by fluctuations over the scales where turbulence dominates. 
    {\em Right:} Comparison of one high-density trajectory to the barrier. ``First-crossing'' is the largest scale $R$ on which the region about the point ${\bf x}$ is self-gravitating, with enclosed mass given by Eq.~\ref{eqn:mass.radius.exact}. Multiple crossings below this scale represent independent sub-regions each collapsing within the parent, i.e.\ a fragmentation cascade. ``Last-crossing'' is the smallest scale on which the region remains self-gravitating -- below this region no sub-scales are independently self-gravitating so there is no successive fragmentation. 
    \label{fig:demo}}
\end{figure*}

The paper is organized as follows. A general illustration of the methodology is given in Fig.~\ref{fig:demo}, and a number of important terms and variables that will be used throughout the paper are defined in Table~\ref{tbl:defns}, along with relevant equations for certain variables.

In \S~\ref{sec:methods}-\S~\ref{sec:intermittency} \&\ \S~\ref{sec:frag:time} we develop our methodology and generalize the theory in a number of important ways. \S~\ref{sec:methods} reviews the basic methodology developed in \paperone-\paperthree\ (for isothermal gas), shows how the model generalizes to rotating disks or collapsing subregions, and derives analytic solutions for mass functions and correlation functions. \S~\ref{sec:frag:poly}-\ref{sec:frag:baro} generalize this to polytropic and multi-variate equations of state. \S~\ref{sec:frag:driving} discusses how different driving (compressive vs.\ solenoidal) enters the theory. \S~\ref{sec:mhd} generalizes to include magnetic fields and anisotropy (in the velocity and/or density fields as well as collapsing structures). \S~\ref{sec:intermittency} extends to intermittent turbulence and highly non-Gaussian density statistics. In \S~\ref{sec:frag:time}, we generalize all of these to be time-dependent, both following time-dependent collapse and evolution in statistically stationary background (\S~\ref{sec:frag:time:general}) as well as time-dependent collapse and development of fragmentation within objects that are themselves collapsing (\S~\ref{sec:collapse:poly}). 

In \S~\ref{sec:results}, we show and discuss the results, for fragmentation in fully-developed turbulence at fixed time. \S~\ref{sec:basic} derives some basic key quantities and scaling relations, such as the ``maximal instability'' and sonic scales, and mass-radius-density-velocity dispersion relationships. \S~\ref{sec:mf} considers the mass function of objects that form both on the largest self-gravitating scales and on the smallest scales in the fragmentation cascade, and how it depends on all of the parameters describing the medium (above). \S~\ref{sec:frag.dynrange} derives the distribution of the ``dynamic range'' of fragmentation, i.e.\ the extent of fragmentation cascades. \S~\ref{sec:results.corr.fn} presents the correlation functions and how they depend on properties of the medium. 

In \S~\ref{sec:timedep.results}, we consider the time-dependence of these results. \S~\ref{sec:global}-\ref{sec:top.vs.bottom} show how the global mass functions develop in time, how structures grow hierarchically and form fragments, and calculate the global rate of collapse of bound structures. \S~\ref{sec:cloud.collapse} shows how sub-fragmentation develops in collapsing clouds with time, and how this depends on the initial properties of the cloud. \S~\ref{sec:frag.trees} outlines the methodology needed to link these into a statistical ensemble of time-dependent ``fragmentation trees.'' 

Finally, in \S~\ref{sec:discussion} we summarize our results and discuss future work. Several more technical details are presented in the Appendices (\ref{sec:nongaussian}-\ref{sec:appendix:window}).  

\begin{footnotesize}
\ctable[
  caption={{\normalsize Definitions of Important Terms \&\ Variables Used Throughout This Paper}\label{tbl:defns}},center,star
  ]{lll}{
}{
\hline\hline
\multicolumn{1}{l}{Term} &
\multicolumn{1}{l}{Definition} & 
\multicolumn{1}{l}{Eq.} \\ 
\hline
barrier & critical density above which a region becomes self-gravitating (or any other criterion ``of interest'') \\ 
trajectory & random realization of the density field (density smoothed in successive scales $R$ about a point ${\bf x}$) \\ 
sonic scale & scale $R$ below which the rms compressive turbulent velocities are sub-sonic ($b\,v_{t}(R)<c_{s}$) \\ 
maximal instability scale & scale $R$ at which fragmentation occurs most rapidly (near the disk scale height/turbulent driving scale) \\ 
first-crossing & largest scale on which a region is self-gravitating (trajectory crosses the barrier) \\ 
last-crossing & smallest scale on which a region is self-gravitating (has no self-gravitating/fragmenting sub-regions) \\ 
crossing distribution & the distribution of spatial scales (or corresponding masses) on which first or last-crossings occur \\
clustering & strength (amplitude) of the two-point correlation function (Eq.~\ref{eqn:xi.defn}) \\ 
\hline
\\
\hline
$R$ & spatial scale of a region (over which quantities are smoothed) & -- \\
$k$ & Fourier mode $k\sim1/R$ & -- \\
$\rho$, $\rho(R)$ & density ($\rho(R)$ is volume-averaged over a region of size $R$) & -- \\ 
\hline
$S$, $S(R)$ & variance in $\ln{(\rho)}$ (volume-averaged within $R$) & \ref{eqn:P0},\,\ref{eqn:S.R} \\ 
$B$, $B(R)$ & barrier (see above), for a region of size $R$ & \ref{eqn:flast},\,\ref{eqn:rhocrit.1} \\ 
$\nu$ & number of standard deviations corresponding to $B$ ($\nu\equiv |B|/S^{1/2}$) & -- \\
\hline
$P$ & probability distribution function (PDF) & -- \\
$P_{0}$ & Gaussian (normal) PDF & \ref{eqn:P0} \\ 
\hline
$\ffirst$ & first-crossing distribution (defined above) & \ref{eqn:ffirst} \\ 
$\flast$ & last-crossing distribution (defined above) & \ref{eqn:flast} \\
\hline
$\rhocrit(R)$ & critical density for collapse (density associated with $B(R)$) & \ref{eqn:rhocrit.1} \\ 
$\rho_{0}$ & volume-averaged mean density of the system (mid-plane density in a disk) & \ref{eqn:P0} \\ 
\hline
$h$ & (exponential) disk scale height & \ref{eqn:rhocrit.1} \\ 
$R\cloudsub$ & radius of an isolated cloud or turbulent box & -- \\ 
\hline
$\kappa$ & epicyclic frequency ($\kappa\equiv 2\Omega\,r_{g}^{-1}\,{\rm d}(r_{g}^{2}\Omega)/{\rm d}r_{g}$ where $r_{g}$ is the disk-centric radius, and $\tilde{\kappa}\equiv \kappa/\Omega$) & \ref{eqn:rhocrit} \\ 
$\Omega$ & orbital frequency ($\Omega\equiv V_{c}/r_{g}$ where $V_{c}(r_{g})$ is the circular velocity) & -- \\ 
\hline
$Q$ & Toomre $Q$ parameter (defined on scale $h$, $Q\equiv (\sigma_{g}[h]\,\kappa)/(\pi\,G\,\Sigma_{\rm gas}$)) & \ref{eqn:rhocrit} \\ 
$Q^{\prime}$ & virial parameter for an isolated cloud/box ($Q^{\prime}\equiv \sigma_{\rm g}[R\cloudsub]^{2}\,R\cloudsub/(G\,M\cloudsub)$) & \ref{eqn:rhocrit} \\ 
\hline
$M$, $M(R)$ & mass ($M(R)$ is the critical mass for collapse within $R$, i.e.\ mass within $R$ at $\rho=\rhocrit$) & \ref{eqn:mass.radius.exact} \\ 
$W(x,\,R)$ & window function for smoothing the density field on scale $R$ ($\tilde{W}(k,\,R)$ is the Fourier transform) & \ref{eqn:S.R} \\ 
\hline
$c_{s}$, $c_{s}(\rho,\,R)$ & gas sound speed & -- \\ 
$v_{t}$, $v_{t}(R)$ & rms turbulent velocity dispersion (averaged on scale $R$) & -- \\
$v_{\rm A}(\rho,\,R)$ & Alfv{\'e}n speed ($v_{\rm A}=B/\sqrt{4\pi\,\rho}$) & -- \\
$\sigma_{\rm g}(\rho,\,R)$ & total effective gas dispersion ($\sigma_{\rm g}^{2}=c_{s}^{2}+v_{t}^{2}+v_{\rm A}^{2}$) & \ref{eqn:sigmagas} \\
\hline
$\Mach$ & Mach number $\Mach(R) \equiv v_{t}(R)/c_{s}$ & -- \\ 
$\Machdisk$ & Mach number at the disk scale height $\Machdisk\equiv v_{t}[h]/c_{s}$ & -- \\ 
\hline
$b$ & mean fraction of turbulent velocity in compressive (longitudinal) modes & \ref{eqn:S.R} \\
$p$ & turbulent spectral index over the inertial range ($E(k)\propto k^{-p}$, $v_{t}^{2}\propto R^{p-1}$) & -- \\
$\gamma$ & gas polytropic index ($c_{s}^{2}\propto \rho^{\gamma-1}$) & \ref{eqn:cs.polytrope} \\ 
\hline
$\xi(r\,|\,M)$ & correlation function for objects of mass $M$, as a function of separation $r$ & \ref{eqn:xi.defn} \\ 
\hline
$\beta$ & intermittency parameter of the velocity field 
($\beta=1$ is non-intermittent, $\beta=0$ infinitely strong) & \ref{eqn:sheleveque.structurefn} \\ 
$\beta_{\rho}$ & effective intermittency parameter for the density field ($\beta_{\rho}=\beta^{1/3}$) & -- \\
\hline
$t_{\rm cross}(R)$ & turbulent crossing time on a scale $R$ ($t_{\rm cross}(R)\equiv R/v_{t}(R)$) & -- \\ 
\hline
$R_{\rm maximal}$, $M_{\rm maximal}$ & maximal instability scale in radius or mass ($M_{\rm maximal} = M(R_{\rm maximal})$) & \ref{eqn:r.maximal} \\ 
$R_{\rm sonic}$, $M_{\rm sonic}$ & sonic scale in radius or mass ($M_{\rm sonic}=M(R_{\rm sonic})$) & \ref{eqn:r.sonic} \\ 
\hline\hline\\
}
\end{footnotesize}

\vspace{-0.5cm}
\section{The First \&\ Last-Crossing Distributions: Outline}
\label{sec:methods}

\vspace{-0.1cm}
\subsection{General Methodology}
\label{sec:methods.general}

In \paperone\ and \papertwo\ we outline some of the methodology for applying the excursion-set formalism to study the properties of self-gravitating structures in a turbulent medium (for simple cases). We briefly review this before generalizing this model to more complicated systems. 

If gas is isothermal, density fluctuations in {both} sub and super-sonic turbulence are (approximately) lognormal, so (by definition) the variable $\delta({\bf x})\equiv \ln{[\rho({\bf x})/\rho_{0}]}+S/2$, 
where $\rho({\bf x})$ is the density at a point ${\bf x}$, 
$\rho_{0}$ is the global mean density and $S$ is the variance in $\ln{\rho}$, 
is normally distributed according to the PDF:\footnote{The $+S/2$ term is just the subtracted mean 
in $\delta$, required so that the integral of $\rho\,P_{0}(\rho)$ correctly 
gives $\rho_{0}$ with $\langle \delta\rangle = 0$. If we consider the mass-weighted density distribution instead, then this becomes $-S/2$.} 
\be
\label{eqn:P0}
P_{0}(\delta\,|\,S) = \frac{1}{\sqrt{2\pi\,S}}\,\exp{\left(-\frac{\delta^{2}}{2\,S} \right)}
\ee
More generally, we can evaluate the field $\delta({\bf x}\,|\,R)$, 
which is the $\delta({\bf x})$ field averaged around the point ${\bf x}$ with 
some window function of characteristic radius $R$. As shown in \paperone, this is also normally distributed, with a variance $S(R)$ as a function of scale given by the integral in Fourier space over the variance in $k$-modes (i.e.\ the logarithmic density power spectrum) on scales $0<k<R^{-1}$ (discussed below).\footnote{Some subtleties related to the convolution of linear products of lognormally-distributed variables in real-space over a varying window (and important tests of the scale-by-scale assumption in numerical simulations) are discussed in Appendices~\ref{sec:appendix:lognormal.convolve}-\ref{sec:appendix:window}, but these do not alter our results.}

We define $\delta({\bf x}\,|\,R)$ as the ``trajectory'' about a given random point in Eulerian space. If this is above some critical value $B(R)$, it defines an ``object of interest'' on the relevant scale. In principle, this can be almost anything; for this paper, we focus on self-gravitating collapse and fragmentation, so the obvious value of $B(R)$ corresponds to the critical density above which a region on the scale $R$ is self-gravitating. Of course, a given trajectory may cross $B(R)$ many times in an arbitrarily narrow interval, corresponding to the region being self-gravitating on some scales, but not on others. We must therefore be careful what we mean by ``self-gravitating'' and how we select the relevant scales.

In \paperone, we focus on the ``first-crossing distribution'': the distribution of self-gravitating regions defined on the {\em largest} scales on which they are self-gravitating. Physically, this corresponds to GMCs in the ISM (the largest self-gravitating gas structures). In \papertwo, we define the ``last-crossing'' distribution instead, the distribution of self-gravitating regions defined on the {\em smallest} self-gravitating scales.\footnote{Of course, once a region becomes self-gravitating and collapses, its density can increase with time, departing from the lognormal density distribution. We treat this in \S~\ref{sec:cloud.collapse}. This does not, however, change our derivations, since what we are interested in is the statistics of regions {\em becoming} self-gravitating via turbulent density fluctuations in the first place.}

The properties of these regions can be derived in a Monte-Carlo manner. Recall, $\delta({\bf x}\,|\,R_{w})$ is smoothed over some window -- i.e.\ it is the convolution $\delta{({\bf x}\,|\,R_{w})} \equiv \int{{\rm d}^{3}x^{\prime}\,W(|{\bf x}^{\prime}-{\bf x}|,\,R_{w})\,\delta{({\bf x}^{\prime})}}$. So if we Fourier transform, we obtain $\delta{({\bf k}\,|\,R_{w})} \equiv W({\bf k}\,|\,R_{w})\,\delta({\bf k})$; the amplitude $\delta{({\bf x}\,|\,R_{w})}$ is simply the (window-weighted) integral of the contribution from all Fourier modes $\delta({\bf k})$. Now begin at some sufficiently large scale $R_{1}$, where mass conservation implies $S_{1}(R_{1})\rightarrow 0$ and $\delta_{1}({\bf x}\,|\,R_{1})\rightarrow 0$. Then consider a ``step'' from this scale to $R_{2}<R_{1}$ ($S_{2}>S_{1}$). We can then draw a new $\delta_{2}(R_{2}\,|\,\delta_{1}[R_{1}]) \equiv \delta_{1} + \Delta\delta$ from the conditional PDF given $\Delta S\equiv S_{2}(R_{2})-S_{1}(R_{1})$. This follows from Fourier transforming the distribution of $\delta_{1}$, taking a ``step'' in Fourier space where we integrate the contribution from all contributing Fourier modes in the intermediate scales, and Fourier transforming back to real space \citep[for a detailed discussion, see][]{bond:1991.eps,zentner:eps.methodology.review}. For the simplifying case where our window function is a Fourier-space top-hat,\footnote{Other window function choices are discussed in Appendix~\ref{sec:appendix:window}.} this gives the simple result 
\begin{align}
p(\delta_{1}+\Delta\,\delta)\,{\rm d}\,\Delta\,\delta &= 
\frac{1}{\sqrt{2\pi\,\Delta\,S}}\,\exp{{\Bigl(}-\frac{(\Delta\,\delta)^{2}}{2\,\Delta\,S}{\Bigr)}}
{\rm d}(\Delta\,\delta) 
\end{align}
Taking sufficiently small steps in $\Delta R$ to ensure convergence, the trajectory $\delta(R)$ is then 
\be
\label{eqn:trajectory.sum}
\delta(R_{i}) \equiv \sum_{j}^{R_{j}>R_{i}} \Delta\,\delta_{j}\ .
\ee
We can then construct an arbitrarily large Monte Carlo ensemble of trajectories, to evaluate various statistical properties of the medium.\footnote{Note that we have implicitly made an important assumption (in addition to assuming the PDFs are log-normal in the first place): namely, that the phases of different Fourier modes are un-correlated. This is unlikely to be true in detail in turbulent flows, where fluid parcels may have complicated correlation structures. In Appendix~\ref{sec:appendix:lognormal.convolve} we note that this cannot, in fact, be true if the PDF is lognormal on all scales; although we show that (based on simulations) it can be a good approximation for the quantities of interest if we focus on a portion (say, the high-density tail) of the PDF. We also show in \S~\ref{sec:frag:poly} that considering a non-isothermal gas forces us to explicitly include strong mode correlations (and discuss how to treat this). And in \S~\ref{sec:intermittency} we consider some models for intermittency, which implicitly include an adjustable degree of correlation structure, and therefore give some idea of how large the consequences may be. Even if the Fourier modes are intrinsically un-correlated, averaging in real-space itself introduces correlations between modes; this can be treated through the use of different window functions, discussed in Appendix~\ref{sec:appendix:window}. In any case, improving our understanding of the hierarchical structure in the density field is an important goal of ongoing and future numerical work, one which directly informs the class of analytic models we develop here.}

\vspace{-0.5cm}
\subsection{Analytic Solutions}
\label{sec:analytic.mfs}

\citet{zhang:2006.general.moving.barrier.solution} show that the first-crossing distribution\footnote{We define this, $\ffirst(S)$, as the fraction of trajectories which first cross above the critical density (i.e.\ become self-gravitating at $S$ without being so on any larger scales), per differential unit ${\rm d}S$ (i.e.\ per unit scale, with the scale variable being the variance).} $\ffirst$ for a normally distributed variable, as defined above, can be determined purely analytically; and in \papertwo\ we derive an analogous expression for the last-crossing distribution $\flast$. This is given by the numerical solution to the Volterra integral equation:
\begin{align}
\label{eqn:flast}
\flast(S) = g_{1}(S) + \int_{S}^{S_{i}}\,{\rm d}S^{\prime}\,\flast(S^{\prime})\,g_{2}(S,\,S^{\prime})
\end{align}
where 
\begin{align}
g_{1}(S) &= {\Bigl [}2\,{\Bigl|}\frac{dB}{dS}{\Bigr|} -\frac{B(S)}{S}{\Bigr]}\,P_{0}(B(S)\,|\,S)\\
g_{2}(S,\,\Sprime) &= {\Bigl[}\frac{B(S)-B(\Sprime)}{S-\Sprime} 
+\frac{B(S)}{S}-2\,{\Bigl|}\frac{dB}{dS}{\Bigr|}{\Bigr]}\times\\
\nonumber& P_{0}[B(S)-B(\Sprime)\,(S/\Sprime)\,|\,(\Sprime-S)\,(S/\Sprime)]
\end{align}
and $B(S)$ is the minimum value of the overdensity $\delta({\bf x}\,|\,R)$ which defines objects of 
interest (here, self-gravitating regions). The expression for first-crossing is qualitatively similar but with some modifications (see \papertwo):
\begin{align}
\label{eqn:ffirst}
\ffirst(S) = \tilde{g}_{1}(S) + \int_{0}^{S}\,{\rm d}S^{\prime}\,\ffirst(S^{\prime})\,\tilde{g}_{2}(S,\,S^{\prime})
\end{align}
where 
\begin{align}
\tilde{g}_{1}(S) &= {\Bigl [}-2\,{\Bigl|}\frac{dB}{dS}{\Bigr|} +\frac{B(S)}{S}{\Bigr]}\,P_{0}(B(S)\,|\,S)\\
\tilde{g}_{2}(S,\,\Sprime) &= {\Bigl[}-\frac{B(S)-B(\Sprime)}{S-\Sprime} 
+2\,{\Bigl|}\frac{dB}{dS}{\Bigr|}{\Bigr]}\times\\
\nonumber& P_{0}[B(S)-B(\Sprime)\,|\,\Sprime-S]
\end{align}

These solutions are valid for {\em any} $B$ and $S$, provided that $P_{0}$ is Gaussian. We will use these expressions where possible, since they removes the need for Monte Carlo evaluation of trajectories, but emphasize that simply counting the fraction of trajectories which first cross $B(S)$ (or fall below $B(S)$ for the last time) gives an identical result (provided the number of trajectories is sufficiently large to converge). 

For the particularly simple case of a linear barrier ($B = B_{0}+\mu\,S$), these yield the closed-form solutions: 
\begin{align}
\label{eqn:flast.linear.barrier}
\flast(S\,|\,B=B_{0}+\mu\,S) &= \frac{\mu}{\sqrt{2\pi\,S}}\,\exp{{\Bigl(}-\frac{B^{2}}{2\,S}{\Bigr)}} \\
\label{eqn:ffirst.linear.barrier}
\ffirst(S\,|\,B=B_{0}+\mu\,S) &= \frac{B_{0}}{S\,\sqrt{2\pi\,S}}\,\exp{{\Bigl(}-\frac{B^{2}}{2\,S}{\Bigr)}} 
\end{align}
Although the barrier is never exactly linear for any realistic turbulent regime, this is not a bad approximation in many regimes; for example, when ${\rm d}B/{\rm d}S\gg B/S \gg 1$ we recover the same limiting expression for $\flast$, and in the intermediate regime where ${\rm d}B/{\rm d}S\ll B/S$ we recover the same limiting expression for $\ffirst$.

\vspace{-0.5cm}
\subsection{Dependence of the Variance and Critical Density on Turbulent Properties}

In \paperone\ we derive $S(R)$ and $B(S)$ from simple theoretical considerations for all scales in an isothermal, turbulent galactic disk. It is well-established that the contribution to density variance from the velocity variance goes as $S\approx\,\ln{(1+\Machcompressive^{2})}$, where $\Machcompressive$ is the (compressive) Mach number \citep[][and references therein]{federrath:2010.obs.vs.sim.turb.compare,konstantin:mach.compressive.relation}. For a given turbulent power spectrum, this suggests that we can approximate $S(R)$ by summing the contribution from the velocity variance on all scales $R^{\prime}>R$
\begin{align}
\label{eqn:S.R}
S(R) &= \int_{0}^{\infty} 
|\tilde{W}(k,\,R)|^{2}
\ln{{\Bigl [}1 + 
\frac{b^{2}\,v_{t}^{2}(k)}{c_{s}^{2} + \kappa^{2}\,k^{-2}}
{\Bigr]}} 
{\rm d}\ln{k} 
\end{align}
where $W$ is the window function for the smoothing\footnote{For convenience 
we take this to be a $k$-space tophat inside $k<1/R$, which is implicit in our 
previous derivation, but we consider alternative window functions and their consequences in Appendix~\ref{sec:appendix:window}.}, $v_{t}(k)$ is the turbulent velocity dispersion averaged on a scale $k$ (trivially related to the turbulent power spectrum), $c_{s}$ is the thermal sound speed, and $b\sim1$ is the fraction of the turbulent velocity in compressive (longitudinal) motions. Here $\kappa$ is the epicyclic frequency; this must enter in the same way relative to $c_{s}$ as it appears in the dispersion relation for density perturbations \citep[e.g.][]{toomre:spiral.structure.review,lau:spiral.wave.dispersion.relations}; physically, this represents angular momentum suppressing large-scale fluctuations (and such a term must be present to ensure mass conservation, since the fluctuation amplitude must vanish sufficiently quickly on large scales). The constant $b$ depends on certain properties of the turbulence, but we vary this below. For a derivation of $S(R)$, see \paperone; we stress that the form we adopt is only an approximation, calibrated from numerical simulations, and discuss some alternatives in Appendix~\ref{sec:appendix:S.of.R}.

Given $\delta(R) \equiv \ln{[\rho(R)/\rho_{0}]}+S/2$ defined above, then $B(R)$ follows from the dispersion relation for a density perturbation in a disk with self-gravity, turbulence, thermal pressure, and angular momentum/shear \citep{vandervoort:1970.dispersion.relation,aoki:1979.gas.disk.instab.crit,elmegreen:1987.cloud.instabilities,romeo:1992.two.component.dispersion}: 
\be
\label{eqn:rhocrit.1}
B(R) = \ln{\left(\frac{\rhocrit}{\rho_{0}} \right)} + \frac{S(R)}{2}
\ee
where $\rhocrit$ is the critical density above which a region is self-gravitating. This is\footnote{\label{foot:exponential.disk}Eqs.~\ref{eqn:rhocrit}-\ref{eqn:mass.radius.exact} are derived exactly for a disk with an exponential vertical profile. They are generically asymptotically exact at small and large $|k|$ and (comparing with numerical calculations) tend to be within $\sim10\%$ of the exact solution at all $|k|$ for a wide range of observed vertical profiles 
\citep[see][]{kim:2002.mhd.disk.instabilities}.}
\begin{align}
\label{eqn:rhocrit}
\frac{\rhocrit}{\rho_{0}} \equiv \frac{Q}{2\,\tilde{\kappa}}\,\left(1+\frac{h}{R} \right)
{\Bigl[} \frac{\sigma_{g}^{2}(R)}{\sigma_{g}^{2}(h)}\,\frac{h}{R}  + 
\tilde{\kappa}^{2}\,\frac{R}{h}{\Bigr]} 
\end{align}
where $\rho_{0}$ is the mean midplane density of the disk, 
$h$ is the disk scale height, 
$\tilde{\kappa}\equiv\kappa/\Omega=\sqrt{2}$ ($\Omega\equiv V_{c}/R$) for a constant-circular velocity ($V_{c}$) disk, and 
$Q\equiv (\sigma_{g}[h]\,\kappa)/(\pi\,G\,\Sigma_{\rm gas})$ is the Toomre $Q$ parameter. The dispersion $\sigma_{g}$ is 
\be
\label{eqn:sigmagas}
\sigma_{g}^{2}(R) = c_{s}^{2} + \langle v_{t}^{2}(R) \rangle  + v_{\rm A}^{2}
\ee 
($v_{\rm A}$ is the Alfv{\'e}n speed). Again, the full derivation of these relations is presented in \paperone; for now, note that this ensures not only that a region is locally self-gravitating, but also that it can resist destruction by tidal shear (the $\kappa$ term) and/or energy input from shocks or the turbulent cascade (the $v_{t}$ term). 

The mapping between radius and mass$^{\ref{foot:exponential.disk}}$ is 
\be
\label{eqn:mass.radius.exact}
M(R) \equiv 4\,\pi\,\rhocrit\,h^{3}\,
{\Bigl[}\frac{R^{2}}{2\,h^{2}} + {\Bigl(}1+\frac{R}{h}{\Bigr)}\,\exp{{\Bigl(}-\frac{R}{h}{\Bigr)}}-1 {\Bigr]}
\ee
It is easy to see that on small scales, these scalings reduce to the Jeans criterion 
for a combination of thermal ($c_{s}$) and turbulent ($v_{t}$) support, with $M=(4\pi/3)\,\rhocrit\,R^{3}$; on large scales it becomes the Toomre criterion 
with $M=\pi\Sigma_{\rm crit}\,R^{2}$. 

Recall $\ffirst(S)\,{{\rm d}S}$ (or $\flast(S)\,{{\rm d}S}$) gives the differential fraction trajectories that have a first (last) crossing in a narrow range ${\rm d}S$ about the scale $S[R]$ (corresponding to mass $M=M[R]$). Each trajectory randomly samples the Eulerian volume, so the differential number of first (last)-crossing regions is related by $V_{\rm cl}(M)\,{\rm d}N(M) = V_{\rm tot}\,\flast(S[M])\,{{\rm d}S}$ (where $V_{\rm cl}(M) = M/\rhocrit(M)$ is the cloud volume at the time of last-crossing and $V_{\rm tot}$ is the total volume sampled). Hence, the mass function -- the number density ${\rm d}n = {\rm d}N/V_{\rm tot}$ in a differential interval -- is given by
\be
\frac{{\rm d}n}{{\rm d}M} = 
\frac{\rhocrit(M)}{M}\,f(M)\,{\Bigl |}\frac{{\rm d}S}{{\rm d}M} {\Bigr |}
\ee
(for both $\ffirst$ and $\flast$). 

The absolute units of the problem completely factor out; so the results generalize to broad classes of systems. However we need to specify the global stability parameter $Q$, as well as the basic properties of the turbulence (e.g.\ the parameter $b$ and the power spectrum). For convenience, we will generally focus on inertial-range turbulence $E(k)\propto k^{-p}$, where $p$ is the turbulent spectral index (usually $p\approx5/3-2$).\footnote{As in \paperone, we note that this must turn over above $R\gg h$ in a disk (to $E(k)\propto k^{-1}$) to avoid an energy divergence, so we impose the flattening $E(k)\rightarrow E(k)\,(1+|k\,h|^{-2})^{(1-p)/2}$, which gives a good fit to simulation results around the inertial scale \citep{bowman:inertial.range.turbulent.spectra}. But as shown in \paperone, because the $\kappa$ terms dominate Eq.~\ref{eqn:S.R}-\ref{eqn:rhocrit} on these scales, the form of this turnover has weak effects on our results.} The normalization of this is defined by the Mach number at the scale $h$, $\Machdisk^{2} \equiv \langle v_{t}^{2}(h)\rangle/c_{s}^{2}$.

\vspace{-0.3cm}
\subsection{Specification Within Collapsing ``Regions''}
\label{sec:frag:subcmf}

We will sometimes focus here on spherical, collapsing regions with radius $R\cloudsub$ (defined generally, but analogous to GMCs or protostellar cores). In \paperthree\ we derive the solution for the ``two-barrier'' problem for collapsing sub-regions within a parent distribution; for our purposes here, the important result is that if we consider the internal properties of initial clouds on some scale we can factor out the contribution from super-cloud scale fluctuations (effectively transforming to $S\rightarrow S-S_{0}$ where $S_{0}=S(R\cloudsub)$ without loss of generality). We can therefore treat each region independently, with the variance ``beginning'' on the parent scale $R\cloudsub$ (i.e.\ treating this as a maximum radius as opposed to $R\rightarrow\infty$); we do not need to know the behavior of the parent scales to model the continued evolution of the sub-region. 

If we consider small sub-regions, $R\cloudsub\ll h$ for inertial-range turbulence, we note that the variance and critical density simplify to 
\begin{align}
\label{eqn:S.R.new}
 S(R) \rightarrow & \int_{R/R\cloudsub}^{1} 
\ln{{\Bigl [}1 + b^{2}\,\Machcloud^{2}\,{\Bigl(} \frac{R}{R\cloudsub}{\Bigr)}^{p-1}
{\Bigr]}} \,
{\rm d}\ln{{\Bigl(}\frac{R}{R\cloudsub} {\Bigr)}} \\ 
\frac{\rhocrit}{\rho_{0}(R\cloudsub)} &\rightarrow Q^{\prime}\frac{1+\Machcloud^{2}\,(R/R\cloudsub)^{p-1}}{1+\Machcloud^{2}}\,{\Bigl(}\frac{R}{R\cloudsub} {\Bigr)}^{-2} \label{eqn:rhocrit.box}
\end{align}
where $Q^{\prime}\approx1$ is trivially related to $Q$, but can for our purposes be taken to be an arbitrary virial parameter of the region, and $\rho_{0}(R\cloudsub) \equiv M\cloudsub/(4\pi\,R\cloudsub^{3}/3)$. The density ratio, then depends only on $R/R\cloudsub$ at fixed $\Machcloud\equiv\Mach(R\cloudsub)$. Thus, isothermal clouds form a one-parameter family in their behavior in $\Machcloud$.

\vspace{-0.5cm}
\subsection{Correlation Functions}
\label{sec:corr.fn}

The auto-correlation function $\xi$ of a given population is defined as the excess probability 
of finding another member of the population within a differential volume at a radial separation $r$ from one such member.
\be
\label{eqn:xi.defn}
1 + \xi_{{\rm MM}}(r\,|\,M) \equiv \frac{\langle N(r\,|\,M) \rangle}{\langle n(M) \rangle {\rm d}V} 
\ee
where $N(r\,|\,M)$ is the number with mass $M$ in the separation $r$ and $n(M)$ is the average number density of objects of mass $M$. This is directly related to the conditional density PDF and first/last crossing distributions; specifically, to the probability that, given a first/last crossing on a given scale, the density field on a larger scale $r$ has some value $\delta$, and then the probability that the field with some $\delta$ on a larger scale contains multiple first/last crossings of the same mass. But these are both derivable from the existing field information -- for example, by sampling the space of all possible ``random walks'' as described above. 

In \paperone-\paperthree\ we derive the correlation function for both first and last-crossing distributions. For example, for last-crossings:
\begin{align}
\nonumber 1 + \xi_{{\rm MM}}^{\ell}(r\,|\,M) &= \int_{\delta_{0}}\,{\Bigl (}\frac{\flast(M\,|\,\delta_{0})}{\flast(M)} {\Bigr)}\, P_{\ast}(\delta_{0}\,|\,S_{0}[r])\,{\rm d}\delta_{0} \\ 
&= \int_{-\infty}^{\infty}\,{\Bigl (}\frac{\flast(M\,|\,\delta_{0})}{\flast(M)} {\Bigr)}^{2}\,P_{0}(\delta_{0}\,|\,S_{0}[r])\,{\rm d}\delta_{0}
\label{eqn:xi.last.defn}
\end{align}
In the above, $\flast(M\,|\,\delta_{0})$ is the last-crossing distribution (at mass $M$) determined for trajectories that begin with the initial condition $\delta_{0}$ at scale $S[r]=S_{0}$. $P_{\ast}(\delta_{0}\,|\,S_{0}[r])$ is the probability of $\delta(S_{0})$ having the value $\delta_{0}$ on the scale $S_{0}$, {\em given} that $\delta(S[M])=B(S[M])$ -- i.e.\ that there is a barrier crossing (a collapsing object) at the smaller scale, which is simply related to the conditional probability of $\delta(S[M])$ given the initial condition $\delta_{0}(S_{0})$ by Bayes's theorem. Here $\flast(M\,|\,\delta_{0})$ is the conditional last-crossing distribution, i.e.\ the last-crossing distribution given an initial density $\delta_{0}$ on a scale $S_{0}[r]$; we show in \paperthree\ that this is equivalent to $\flast(M)$ solved with the new initial condition for trajectories ``beginning'' at $r$ with $S\rightarrow S-S_{0}$ and $B\rightarrow B-\delta_{0}$. 

For first-crossings, 
\begin{align}
\label{eqn:xi.cm.defn}
1+\xi_{\rm MM}^{f}(r\,|\,M)  
&= \int_{-\infty}^{\delta_{\rm crit}(r)}\,{\Bigl (}\frac{\ffirst(M\,|\,\delta_{0})}{\ffirst(M)} {\Bigr)}^{2}\,q(\delta_{0}\,|\,S_{0}[r])\,{\rm d}\delta_{0}
\end{align}
where the integral is over all $\delta_{0}<\delta_{\rm crit}(r)$, 
and $q(\delta_{0}\,|\,S_{0})$ is a weighting factor defined 
in \citet{bond:1991.eps} as the probability that the overdensity at a random 
point, smoothed on a scale $r$, is $\delta_{0}$ and does not exceed 
$\delta_{\rm crit}(r)$ on any larger smoothing scale.

It is similarly straightforward to calculate the cross-correlation between objects of different masses;\footnote{If fluctuations on different scales are uncorrelated; the correlation must be explicitly calculated from the Monte Carlo correlated random walk if there is explicit correlation structure in the Fourier modes of the density field. Wherever required, we use this method of calculation.} this amounts to replacing $f(M\,|\,\delta_{0})^{2}$ with $f(M_{1}\,|\,\delta_{0})\,f(M_{2}\,|\,\delta_{0})$. And the projected correlation function is just given by the line-of-sight integral (weighted by the appropriate number density). For details we refer to \paperone\ \&\ \paperthree.

\vspace{-0.5cm}
\section{Generalization for Polytropic Gas}
\label{sec:frag:poly}

The above derivation considers only isothermal gas. We now generalize this to locally polytropic gases: systems with $c_{s}^{2}\propto \rho^{\gamma-1}$. 

\vspace{-0.5cm}
\subsection{The Collapse Threshold}
\label{sec:frag:poly:thold}

Following \citet{hennebelle:2009.imf.variation}, it is straightforward to generalize our expression for the critical physical density above which a region will become self-gravitating, by replacing 
\be
\label{eqn:cs.polytrope}
c_{s}^{2} \rightarrow \cscloudzero^{2}\,{\Bigl (}\frac{\rho}{\rho_{0}} {\Bigr)}^{\gamma-1}
\ee
using the fact that $c_{s}(\rho_{0}) \equiv \cscloudzero$. This replacement in Eq.~\ref{eqn:sigmagas} makes Eq.~\ref{eqn:rhocrit} an implicit equation for $\rhocrit$:
\begin{align}
\label{eqn:rhocrit.gamma}
\frac{\rhocrit}{\rho_{0}} \equiv \frac{Q}{2\,\tilde{\kappa}}\,\left(1+\frac{h}{R} \right){\Bigl[} \frac{\cscloudzero^{2}\,(\rhocrit/\rho_{0})^{\gamma-1}+v_{t}^{2}(R)}{\cscloudzero^{2}+v_{t}^{2}(h)}\,\frac{h}{R}  + 
\tilde{\kappa}^{2}\,\frac{R}{h}{\Bigr]} 
\end{align}
which must in general be solved numerically.

For collapsing spherical sub-regions ($R\cloudsub\ll h$) where we recover the Jeans criterion (i.e.\ drop the $\kappa$ terms), $\rhocrit \ge k^{2}\,[v_{t}(k)^{2}+c_{s}^{2}]/(4\pi\,G)$, so Eq.~\ref{eqn:rhocrit} simplifies to 
\begin{align}
\frac{\rhocrit}{\rho_{0}} 
&= \frac{Q^{\prime}}{1+\Machcloud^{2}}\,{\Bigl(}\frac{R}{R\cloudsub} {\Bigr)}^{-2}\,{\Bigl[} 
{\Bigl(} \frac{\rhocrit}{\rho_{0}}  {\Bigr)}^{\gamma-1} + 
\Machcloud^{2}\,{\Bigl(}\frac{R}{R\cloudsub} {\Bigr)}^{p-1}
{\Bigr]}
\end{align}
where we define $\Machcloud \equiv v_{t}(R\cloudsub)/ c_{s}(\rho_{0})$.
It is straightforward to solve this numerically for $\rhocrit(R)$. Note that so long as $\gamma<2$ there is always a unique solution. 

For $\Machcloud^{2}\,(R/R\cloudsub)^{p-1}\gg1$ turbulence dominates the support and we obtain $\rhocrit \approx \rho_{0}\,(R/R\cloudsub)^{p-3}$, identical to the isothermal case (independent of $\gamma$). On the other hand for $\Mach^{2}=\Machcloud^{2}\,(R/R\cloudsub)^{p-1}\ll1$ we obtain $\rhocrit\approx\rho_{0}\,(1+\Machcloud^{2})^{-1/(2-\gamma)}\,(R/R\cloudsub)^{-2/(2-\gamma)}$; the sub-sonic collapse density becomes a steeper function of $R$ as $\gamma$ increases.

\vspace{-0.5cm}
\subsection{The Density PDF}
\label{sec:frag:poly:pdf}

The generalization of the density PDF in the non-isothermal case is more delicate. Numerical studies have shown that it is no longer exactly lognormal in this case \citep[see][]{passot:1998.density.pdf,scalo:1998.turb.density.pdf,nordlund:1999.density.pdf.supersonic,ballesteros-paredes:2011.dens.pdf.vs.selfgrav}. These authors point out, however, that the inviscid, unforced Navier-Stokes equations are invariant under the substitution $\Mach^{2} \rightarrow \Mach^{2}_{\rm eff}(\rho)\equiv \Mach_{0}^{2}\,(\rho/\rho_{0})^{-(\gamma-1)}$ (just the replacement of $c_s(\rho_{0})$ by $c_{s}(\rho)$). The same arguments driving the isothermal PDF to lognormality can be generalized, then, to predict the PDF of the form 
\begin{align}
\label{eqn:pdf.gamma}
P{\Bigl[}s\equiv \ln{{\Bigl(}\frac{\rho}{\rho_{0}} {\Bigr)}}{\Bigr]} &= \exp{{\Bigl(}-\frac{s^{2}}{2\,S(R,\,s)} + \psi(R)\,s + \phi(R) {\Bigr)}}
\end{align}
where we replace the isothermal, constant variance $S(R) = S(R,\,\Mach[R])$ with the local ``effective variance'' $S(R,\,s) = S(R,\,\Mach_{\rm eff}[R,\,s])$. The $\psi$ and $\phi$ are simply determined for any $S(R,\,s)$ by the normalization conditions $\int P(s)\,{\rm d}s = 1$ and $\int P(s)\,\exp{(s)}\,{\rm d}s = 1$ (mass conservation). The replacement in $S$ simply amounts to 
\begin{align}
\label{eqn:S.R.gamma}
S(R,\,\rho) &= \int_{0}^{\infty} 
|\tilde{W}(k,\,R)|^{2}
\ln{{\Bigl [}1 + 
\frac{b^{2}\,v_{t}^{2}(k)}{c_{s}^{2}(\rho[k]) + \kappa^{2}\,k^{-2}}
{\Bigr]}} 
{\rm d}\ln{k} 
\end{align}
This should be valid so long as the turbulence obeys locality. Indeed, a variety of numerical experiments have shown that this is a good approximation, at least over the range $0.3\le\gamma\le1.7$ \citep[see e.g.][]{passot:1998.density.pdf,scalo:1998.turb.density.pdf,li:2003.turb.mc.vs.eos}

\vspace{-0.5cm}
\subsection{Constructing a Random Walk}
\label{sec:frag:poly:walk}

We now need to consider how we evaluate trajectories $\delta$, if the PDF is non-Gaussian. There is a considerable literature on this topic for cosmological non-Gaussianity \citep[see][and references therein]{matarrese:2000.nongaussian.numdens,
afshordi:2008.nongaussian.collapsestatistics,
maggiore:2010.nongaussian.eps}; however, most of the methods discussed therein are appropriate only for very small deviations from Gaussianity (the cosmological case of interest), whereas for $\gamma$ significantly different from unity we are confronted with entirely non-normal PDFs. However, we can take advantage of the fact that in a sufficiently small range of $s$, Eq.~\ref{eqn:pdf.gamma} is {\em locally} Gaussian-like. 

As in the isothermal case, when we consider a ``step'' from $R_{1}\rightarrow R_{2}=R_{1}-\Delta R$ and correspondingly $s_{2}(R_{2}) = s_{1}+\Delta s$, we transform to Fourier space, integrate the contribution from all modes (according to our window function) and transform back. We simplify this considerably first by the use of a top-hat Fourier-space window function. In the limit of infinitesimal steps $\Delta s$, this reduces to the coupled stochastic differential equation 
\begin{align}
\nonumber \Delta s_{i} &= \langle s_{i}(R+\Delta R,\,s+\Delta s)\rangle - \langle s_{i}(R,\,s)\rangle \\
& + \sqrt{S[R+\Delta R,\,s_{i}+\Delta s_{i}] - S[R,\,s_{i}]}\,\mathcal{R}_{i}
\end{align}
where $\mathcal{R}_{i}$ is an independent Gaussian random variable, and $s_{i}$ is the value for a single trajectory $i$. This is, to linear order 
\begin{align}
{\Bigl(}{1-\frac{\partial \langle s \rangle}{\partial s}}{\Bigr)}
\,\Delta s & \approx 
\frac{\partial \langle s \rangle}{ \partial R}\,\Delta R + 
\mathcal{R}\,
\sqrt{\frac{\partial S}{\partial R}\,\Delta R + \frac{\partial S}{\partial s}\,\Delta s} \\ 
\Delta s &\approx \frac{\Delta \langle s \rangle {\bigr|}_{s} + \mathcal{R}\,\sqrt{\Delta S {\bigr|}_{s}}}
{1 - {\bigl(} \frac{\partial\langle s\rangle}{\partial s} + \frac{1}{2\sqrt{\Delta S {|}_{s}}}\,\frac{\partial S}{\partial s}\,\mathcal{R} {\bigr)}} \\ 
\langle s \rangle &= \langle s(R,\,s) \rangle = \psi(R)\,S(R,\,s)
\end{align}
And the expansions are straightforward (albeit tedious) to higher orders. 
This can be solved numerically for each step of the trajectories in a Monte Carlo ensemble, populating $P[s,\,R]$, if we restrict our ``step size'' in $\Delta R$ to sufficiently small increments such that the expansions are valid to the chosen order (we caution that this requires some care). We solve the full equation with an iterative method to ensure the appropriate constraint is satisfied for all trajectories at all radii. 
Essentially, this amounts to locally near-Gaussian behavior, with an ``effective'' variance $S_{e}=S(R,\,s)$, and a shift/bias in the mean $\langle s \rangle$ because Eq.~\ref{eqn:pdf.gamma} is equivalent to a Gaussian with mean $\psi(R)\,S(R,\,s)$ {if} $S(R,\,s)$ were constant; the $\partial/\partial s$ terms in the denominator account to leading order for the locally-introduced skewness and kurtosis from the dependence of $S(R,\,s)$ on $s$. 


We have explicitly tested the above (for both first and second-order expansions), with arbitrary $c_{s}(R_{0}\,|\,\rho_{0})$ and $v_{t}(R)$ to generate $S(R,\,\rho)$ in Eq.~\ref{eqn:S.R.gamma}, to confirm that the Monte Carlo PDF that results does indeed agree well with the analytic Eq.~\ref{eqn:pdf.gamma} at each $R$ (even far out in the tails of the distribution, where $P\lesssim 10^{-7}$). We find that for the range of $\gamma$ of interest, a linear expansion is sufficiently accurate to recover all of the interesting behaviors provided the dependence of $S(R,\,s)$ on $s$ is sufficiently weak that ${\bigl|} \frac{\partial\langle s\rangle}{\partial s} + \frac{1}{2\sqrt{\Delta S {|}_{s}}}\,\frac{\partial S}{\partial s}\,\mathcal{R} {\bigr|} \ll 1$; otherwise a more exact solution is required.

We stress that here ($\gamma\ne1$) the steps are {\em always} correlated at some level, because the  density $s$ (itself determined by fluctuations on other scales) enters the equation to determine $\Delta s$. If the steps were not correlated, the central limit theorem would drive the PDF back to a Gaussian. An important question for study in numerical simulations in future work, is whether or not this correlation structure approximated above is truly universal, and whether or not it implies different intermittency structures (since intermittency also implies correlated fluctuations) from the isothermal case.

\vspace{-0.5cm}
\section{Barotropic Gases \&\ Bivariate Gas Equations of State}
\label{sec:frag:baro}

It is straightforward to see that our derivation for polytropic gases trivially generalizes to any barotrope $c_{s} = c_{s}(\rho)$. The polytropic index $\gamma$ has no unique role in the above other than defining the barotrope, so is easily replaced by a more complex function \citep[for examples of this, see e.g.][]{jappsen:2005.imf.scale.thermalphysics,hennebelle:2009.imf.variation,veltchev:2011.frag}.

Moreover, since we separate the various scales $R$ in the above, it is also trivial to replace a barotrope with any bivariate function $c_{s} \rightarrow c_{s}(\rho,\,R)$ of the density and radius. Or any quantity which is a function of the density and/or radius, such as the total mass inside a region (a product of $\rho$ and $R^{3}$), the average turbulent velocity dispersion within the region scale (using $v_{t}(R)$), or the surface density of the region. 

In this paper, we simplify our analysis by restricting to polytropes. However, these bivariate distributions may be extremely relevant for a number of astrophysical cases: for example if the equation of state is influenced by a uniform photo-ionizing background, or turbulent shocks. By simply replacing $c_{s}\rightarrow c_{s}(\rho,\,R)$ in Eqs.~\ref{eqn:rhocrit} \&\ \ref{eqn:S.R.gamma}, the appropriate PDF, barrier, and method of following the Monte Carlo tree trivially follow. 

\vspace{-0.5cm}
\section{Dependence on Driving Mechanisms}
\label{sec:frag:driving}

In this model \citep[as in previous analytic and numerical work;][]{hennebelle:2009.imf.variation,padoan:2011.new.turb.collapse.sims,federrath:2012.sfr.vs.model.turb.boxes}, the dependence of our results on the turbulent driving mechanisms, at otherwise fixed parameters, is entirely encapsulated in two parameters. First, the turbulent power spectrum (freely varied below). Second, the parameter $b$, which represents the mean velocity component in compressive (longitudinal) as opposed to solenoidal (transverse) motions. This enters the equations for $S(R)$ since longitudinal motion is (by definition) what actually drives variation in $\rho$ \citep[see][]{konstantin:mach.compressive.relation}. If the turbulent driving is purely compressive, $b=1$. For pure solenoidal driving, $b=1/3$ (a simple geometric consequence in isotropic, three-dimensional turbulence). For random driving with no ``preferred'' mode, $b=1/2$ \citep{padoan:2002.density.pdf}; for more detailed behavior in intermediate cases, see \citet{federrath:2010.obs.vs.sim.turb.compare,price:2011.density.mach.vs.forcing,konstantin:mach.compressive.relation}.

\vspace{-0.5cm}
\section{Magnetic Fields \&\ Anisotropic Collapse}
\label{sec:mhd}

It is, in principle, also straightforward to generalize our results for magnetized media, with a few important caveats. A wide range of numerical experiments have shown that both sub and super-sonic turbulence in magnetized media {also} develop PDFs that obey many of the above scalings; for example, for $\gamma=1$ the PDF is also lognormal \citep{ostriker:1999.density.pdf,
klessen:2000.pdf.supersonic.turb,kowal:2007.log.density.turb.spectra,lemaster:2009.density.pdf.turb.review,padoan:2011.new.turb.collapse.sims}.\footnote{In fact, as shown in \citet{hopkins:2012.intermittent.turb.density.pdfs}, from these simulations and those in \citet{molina:2012.mhd.mach.dispersion.relation}, the log-normal approximation typically becomes {\em more} accurate with increasing magnetic field strength.} Provided the shape of the PDF remains consistent with the class of barotropic solutions above, then there are a few additional ways magnetic fields may modify our conclusions. It is possible that MHD effects modify the turbulent velocity power spectra; however, for nearly all conditions of astrophysical interest which have been studied in numerical simulations, the shape of the velocity power spectrum remains close to $p\sim5/3$ (within the range $p\sim4/3-2$ that we survey in this paper; see references above and \citealt{kritsuk:2011.mhd.turb.comparison}).

The consequences of magnetic fields ``resisting collapse'' depend on the field geometry, which must be assumed. At one extreme, we can follow previous analytic approximations \citep[e.g.][]{kim:2002.mhd.disk.instabilities,hennebelle:2009.imf.variation} and consider fields which are locally tangled on very small scales. In this limit, the field acts as an isotropic pressure term. In the criterion for collapse $\rhocrit$ (Eq.~\ref{eqn:rhocrit}), we simply take
\be
\sigma_{g}^{2}(\rho,\,R) \rightarrow c_{s}^{2}(\rho) + \langle v_{t}^{2}(R)\rangle + v_{\rm A}^{2}(\rho,\,R)
\ee
where $c_{s}^{2}(\rho)$ allows for a non-isothermal equation-of-state and $v_{\rm A}$ is the Alfven speed $v_{\rm A}^{2}\equiv |{\bf B}|^{2}/4\pi\rho \equiv \beta_{A}^{-1}\,c_{s}^{2}$. Likewise, in calculating the variance $S(R)$ (Eqs.~\ref{eqn:S.R} \&\ \ref{eqn:S.R.gamma}) we add the magnetic pressure term resisting collapse, $c_{s}^{2}(\rho)\rightarrow c_{s}^{2}(\rho) + v_{\rm A}^{2}(\rho,\,R)$. We allow the Afven speed to vary with density and radius -- the variance with density is an effective ``equation of state,'' while that with radius simply follows the magnetic power spectrum. Having specified this, the magnetic fields are mathematically degenerate with a bivariate equation of state ($c_{s}^{2}\rightarrow c_{s}^{2}(\rho,\,R)\,[1+\beta_{A}^{-1}(\rho,\,R)]$). \citet{kim:2002.mhd.disk.instabilities} and \citet{kunz:2010.mhd.sf.solutions} compare this simple approximation with much more detailed numerical simulations and discuss where it may (and may not) apply; they suggest that where applicable, a near-constant $v_{\rm A}$ is also a reasonable approximation, further simplifying the relations above. 

At the opposite extreme, consider locally ordered magnetic fields ${\bf B}({\bf x}\,|\,R) = B_{\|}\,\hat{B}$. Now, there is a preferred field axis, which has some important consequences. We can decompose the field and turbulence into components along each axis:
\begin{align}
v_{t}^{2} &= v_{t,\,x}^{2} + v_{t,\,y}^{2} + v_{t,\,z}^{2} = v_{t,\|}^{2} + v_{t,\,\bot}^{2} 
\end{align}
where $v_{t,\,x}$ is the rms $x$-component of the turbulent field, and $v_{t,\,\bot}$, $v_{t,\,\|}$ are the rms perpendicular/transverse (two-dimensional) and parallel/longitudinal turbulent components with respect to the $B$ field. Compressions and expansions in this regime feel magnetic pressure perpendicular to the field lines, but not along the field lines. Therefore the differential term in the variance $S(R)$ becomes approximately 
\begin{align}
\nonumber \frac{v_{t}^{2}(k)}{c_{s}^{2}(\rho[k])+\kappa^{2}\,k^{-2}} 
&\rightarrow
\frac{v_{t,\,\bot}^{2}(k)}{c_{s}^{2}(\rho[k])+v_{A}^{2}(\rho,\,k) + \kappa^{2}\,k^{-2}} \\ & + 
\label{eqn:vt.corr.bfield}
\frac{v_{t,\,\|}^{2}(k)}{c_{s}^{2}(\rho[k])+ \kappa^{2}\,k^{-2}} 
\end{align}
If we do not {\em a priori} know the magnetic field orientation, or it is changing in time (as is typical in turbulent fields), then we can further approximate the time-average by assuming isotropic turbulence, i.e.\ $v_{t,\,\bot}^{2}\approx(2/3)\,v_{t}^{2}$ and $v_{t,\,\|}\approx(1/3)\,v_{t}^{2}$. In a strongly, magnetized medium, Eq.~\ref{eqn:vt.corr.bfield} then simply becomes $\approx (1/3)\,v_{t}^{2}/(c_{s}^{2}+\kappa^{2}k^{-2})$; this is a pure geometric correction, as we assume the parallel turbulent component is able to introduce compressions but the perpendicular components do not (similar to the scaling of $b$ with solenoidal vs.\ compressive driving). While obviously approximate, this appears to capture the most important result (for our purposes) of MHD turbulence simulations, namely the dispersion-Mach number relation.\footnote{Eq.~\ref{eqn:vt.corr.bfield}, with the assumption of isotropic turbulence, in an ideal driven turbulent box, predicts a density-Mach number relation 
\begin{align}
\nonumber S &\approx \ln{[ 1 + b^{2}\,\Mach^{2}\,(1/3 + 2/3\,c_{s}^{2}/(c_{s}^{2}+v_{\rm A}^{2})) ]} \\ 
\nonumber &=\ln{[1 + b^{2}\,\Mach^{2}\,(2/3+\beta_{\rm A})\,(2+\beta_{\rm A})]}
\end{align}
so the $v_{\rm A}$-dependence can be absorbed into a ``renormalized'' ($\beta$-dependent) $b$ value. If we compare this to simulations in \citet{lemaster:2009.density.pdf.turb.review,padoan:2011.new.turb.collapse.sims,molina:2012.mhd.mach.dispersion.relation}, we find it gives a good approximation to the dispersion-Mach number relation and density power spectra (comparing as described in Appendix~\ref{sec:appendix:S.of.R}), from field-strengths $\beta_{\rm A}\sim20-10^{-3}$ surveyed therein (equivalent to an ``renormalized'' $b\rightarrow 0.58\,b - 0.97\,b$). \citet{molina:2012.mhd.mach.dispersion.relation} propose a $\beta_{\rm A}$-dependent $b$ correction slightly different from that here, which works comparably well. For our purposes, both approximations are functionally identical in terms of how they enter our formalism.}

It is possible in principle (though outside the scope of this paper) to remove the isotropy assumption, and allow for the random walk in the density distribution to be inherently anisotropic. This amounts to considering the turbulent modes in Fourier space not as isotropic (i.e.\ functions of $k$ alone) but as vectors ${\bf k}$. The procedure for treating this is straightforward in the Monte Carlo method. Consider evaluating the random field that is the compressive component of the local Mach number (i.e.\ just the local velocity field). This decomposes into $v_{t,\,x}$, etc., as above. So instead of evaluating one Gaussian random field variable as a function of scale for each trajectory, we would simply evaluate three (independent) variables (the three components $\Mach_{x}$, $\Mach_{y}$, $\Mach_{z}$). We associate with each a variance $S_{M,\,x}(R)$, $S_{M,\,y}(R)$, $S_{M,\,z}(R)$; these are determined from the power spectrum just as in the isotropic case, but measured separately for each component \citep[see an application of this approach for the angular momentum content of dark matter halos in][]{sheth:2002.linear.barrier}. For example, on large scales in a disk, it may be more accurate to similarly decompose $v_{t}^{2}$ into azimuthal, radial, and vertical modes. These can obey different power spectra \citep[as seen in some simulations;][]{block:2010.lmc.vel.powerspectrum,bournaud:2010.grav.turbulence.lmc}, and similar to Eq.~\ref{eqn:vt.corr.bfield} have different support: the $\kappa$ term applies to radial modes, a similar term in $\Omega$ (and its derivatives) applies to azimuthal modes, and no angular momentum term resists vertical modes. In what follows, for simplicity and clarity, we always evaluate quantities in spherical annuli, but in principle one could allow for triaxial collapse with this method.

With that in mind, we see that the lowest-order corrections from magnetic fields -- {\em within the context of our model assumptions} (which are admittedly highly simplified) -- are mathematically identical to changes in the freely-varied parameters. In other words, for magnetic field strength $\beta_{\rm A}$ in either limit described above, the formal approach and results are identical to those for a ``pure hydrodynamical'' case with appropriately re-mapped values of $p,\,\gamma,\,\sigma_{g}$, and $b$. For reasonable field strengths which have been explored numerically, the re-mapped ``effective'' values fall well within the range we treat as freely varied. Therefore, we will not {\em explicitly} denote magnetic vs.\ non-magnetic cases below, but note that the simple relations above can be used to determine the approximate effects for any specific case of interest (the general sense of increasing magnetic field strength being identical to lowering the Mach number and/or making the forcing ``more solenoidal''). In future work, we hope to explore a more detailed approach that includes explicit treatment of anisotropy as well as higher-order effects such as ambipolar diffusion; however, this necessitates extensions to our formalism beyond the scope of this paper.

\vspace{-0.5cm}
\section{Intermittency \&\ Correlated Turbulent Fluctuations}
\label{sec:intermittency}

We have thus far assumed that fluctuations on different scales are (in the isothermal case) un-correlated and continuous, which leads (via the central limit theorem) to locally Gaussian statistics (see the discussion in \S~\ref{sec:methods}). However, intermittency implies violations of this assumption (manifest in e.g.\ the deviations of the structure functions from self-similar, normal statistics); physically, this corresponds to discrete, coherent structures, such as shocks, sound waves, and vortices in the turbulent flow. 

However, a wide range of studies have shown that many statistical properties of intermittent turbulence can be at least phenomenologically described by cascade models such as that in \citet{sheleveque:structure.functions}. We briefly review these here. In self-similar Kolmogorov turbulence (obeying purely lognormal statistics), the velocity structure functions $S_{n}(R)=\langle \delta v(R)^{n} \rangle \equiv \langle |{\bf v}({\bf x}) - {\bf v}({\bf x}+{\bf R})|^{n} \rangle$ should scale as a power-law $\propto R^{\zeta_{n}}$ with $\zeta_{n} = n/3$. The \citet{sheleveque:structure.functions} model was originally proposed as an alternative scaling law for the structure functions, predicting 
\be
\label{eqn:sheleveque.structurefn}
\zeta_{n} = (1-\gamma^{\prime})\,\frac{n}{3} + \frac{\gamma^{\prime}}{1-\beta}\,{\Bigl(}1-\beta^{n/3}{\Bigr)}
\ee
with the original choices $\beta=\gamma^{\prime}=2/3$ (giving $\zeta_{n} = n/9 + 2\,[1-(2/3)^{n/3}]$).\footnote{The label $\gamma^{\prime}$ owes to historical notation, it is completely unrelated to the polytropic index $\gamma$.} The meaning of these variables is discussed below but this gives a means to parameterize the degree of ``non-self-similarity'' and non-Gaussianity in turbulence. 

\citet{shewaymire:logpoisson} and \citet{dubrulle:logpoisson} showed that the scaling in Eq.~\ref{eqn:sheleveque.structurefn} was the exact result of a general class of log-Poisson statistics. They assume extended self-similarity, i.e.\ $\delta v(R)^{n} \propto R^{n/3}\,\epsilon_{R}^{n/3}$ (where $\epsilon_{R}$ is the dissipation term between scales), and a general hierarchical symmetry between scales -- such that the statistics of $\epsilon_{R}$ can be encapsulated in a term $\pi_{R}$ such that $\epsilon_{R} \sim \pi_{R}\,\epsilon_{R}^{\infty}$ (where $\epsilon_{R}^{\infty}$ describes the scaling of the average properties of the most extreme/singular objects, e.g.\ one-dimensional shocks). Under these conditions, the \citet{sheleveque:structure.functions} scaling is equivalent to the statement that $\pi_{R}$ obeys log-Poisson statistics of the form
\begin{align}
\label{eqn:logpoisson}
P(\pi_{R})\,{\rm d}\pi_{R} &= {\rm d}Y\,\sum_{m}\,P_{\lambda(R)}(m)\,G_{R}(Y,\,m),\ \ \ Y \equiv \frac{\ln{\pi_{R}}}{\ln{\beta}} 
\end{align}
with
\begin{align}
P_{\lambda}(m) &= \frac{\lambda^{m}}{m!}\,\exp{(-\lambda)}
\end{align}
and $G_{R}$ is {any} well-defined, infinitely divisible probability distribution function (physically depending on the driving and character of ``structures'' in the turbulence). Since $\delta v(R) \propto \epsilon_{R}^{1/3} \propto \pi_{R}^{1/3}$, $\ln{(\delta v/\langle \delta v\rangle)}=\ln{\pi^{1/3}}=(1/3)\ln{\pi}$ is a linear transformation and should obey the same statistics as $(1/3)\,\ln{\pi_{R}}$ \citep[see e.g.][]{shewaymire:logpoisson,dubrulle:logpoisson}. This describes a general class of random multiplicative processes that obey certain basic symmetry properties.

Now recall, the basic assumption behind our approximation (and essentially all previous analytic work) for the dispersion-Mach number relation and $S(R)$ is that the density field is the product of compressive modes in the velocity field, i.e.\ $\delta\rho\equiv \rho/\rho_{0}-1$ obeys the same statistics as $\delta v(R)$ (for small ``steps'' in scale). So, under these assumptions, the statistics of $\ln{\rho}$ should have the same form as $(1/3)\,\ln{\pi_{R}}$; and from Eqn.~\ref{eqn:logpoisson}, the statistics of $(1/3)\,\ln{\pi_{R}}$ are identical to the statistics of $\ln{\pi_{R}}$ for a value of $\beta\rightarrow \beta^{1/3}$. Thus, 
\be
\label{eqn:general.intermittent.pdf}
P(\ln{(\rho/\rho_{0})}) \approx \frac{1}{\ln{\beta_{\rho}}}\sum_{m=0}^{\infty}P_{\lambda(R)}(m)\,G_{R}{\Bigl(}\frac{\ln{(\rho/\rho_{0})}}{\ln\beta_{\rho}},\, m {\Bigr)}
\ee
with $\beta_{\rho} = \beta^{1/3}$. We emphasize that this is an assumption: after deriving some basic consequences (\S~\ref{sec:intermittency:steps}) we discuss whether it is justified, and how the above formula compares to density PDFs in both experiment and numerical simulations (\S~\ref{sec:intermittency:motivation}).


\vspace{-0.5cm}
\subsection{Monte-Carlo ``Steps'' In an Intermittent Hierarchy \&\ Their Meaning}
\label{sec:intermittency:steps}

Now consider the application of the log-Poisson statistics above on the density ``walk'' as a function of scale. Following \citet{liufang:2008.logpoisson.cosmic.baryons}, first consider the simplest possible form of the driving function $G_{R}$, a Dirac $\delta$ function (the simplest case considered in \citet{shewaymire:logpoisson,dubrulle:logpoisson}). This corresponds to all driving ``events,'' shocks, and other structures which force velocity/density changes being strictly quantized and having identical fractional magnitudes. In this case, $P(\ln{\rho}) = P_{\lambda}(m = \ln{\rho}/\ln{\beta_{\rho}})$. 

In such a log-Poisson random multiplicative process, the (linear) variables, in this case $\rho/\rho_{0}$, are related by a statistical hierarchy between a larger scale $R_{1}$ and smaller scale $R_{2}$:
\be
\label{eqn:logpoisson.step}
\rho_{R_{2}} = W_{R_{1}R_{2}}\,\rho_{R_{1}}
\ee
with 
\be
W_{R_{1}R_{2}} = \beta_{\rho}^{m}\,(R_{1}/R_{2})^{\gamma^{\prime}}
\ee
Here $m$ is a Poisson random variable with $P(m)=P_{\lambda}(m)$ and mean 
\begin{align}
\label{eqn:logpoisson:lambda.def}
\lambda=\lambda_{R_{1}R_{2}} &= \frac{\gamma^{\prime}}{1-\beta_{\rho}}\,\ln{(R_{1}/R_{2})}
\end{align}
from mass conservation. Because the log-Poisson statistics of Eq.~\ref{eqn:logpoisson} are infinitely divisible, we can break up the statistics at any scale $R$ or $\lambda(R)$ into a sum of ``steps'' ($\lambda_{R_{1}R_{2}}$), and will always recover the same statistics independent of the step structure.

Note that this process does not inherently increase or decrease the variance in the distribution relative to Gaussian steps. In fact, the variable $\gamma^{\prime}$ is related {\em by definition} to the differential change in the variance of the distribution over a ``step,'' by\footnote{This is the same relation (replacing the variance in $\ln{\rho}$ with that in $\ln{v}$) used to define $\gamma^{\prime}$ in the traditional \citet{sheleveque:structure.functions} model for the velocity moments. There, $\gamma^{\prime}$ is approximately a constant, because of the approximately power-law behavior of $v_{t}(R)$ over the inertial range. Likewise here, $\gamma^{\prime}$ is nearly constant over the same range. We could, purely mathematically speaking, hold $\gamma^{\prime}$ fixed to some value and either vary $\beta$ to match $S$ or allow $S$ to diverge with scale, but this leads to unphysical results.} 
\be
\gamma^{\prime} = {\Bigl |}\frac{\Delta S}{\ln{(R_{1}/R_{2})}} {\Bigr|}\,\frac{1-\beta_{\rho}}{(\ln{\beta_{\rho}})^{2}}
\rightarrow {\Bigl |}\frac{{\rm d}S}{{\rm d}\ln{R}} {\Bigr|}\,\frac{1-\beta_{\rho}}{(\ln{\beta_{\rho}})^{2}}
\ee
For clarity, and because the variance-Mach number relation is well-established even in systems with large intermittency \citep[see e.g.\ the cases in][]{schmidt:2008.turb.structure.fns,schmidt:2009.isothermal.turb,federrath:2008.density.pdf.vs.forcingtype,federrath:2010.obs.vs.sim.turb.compare,price:2010.grid.sph.compare.turbulence} we compare the effects of intermittency at fixed variance as a function of scale (i.e.\ fixed $\Delta S$). We stress that this represents no loss of generality. 

This scaling is sufficient, then, for us to re-construct all of our previous predictions, for any choice of $\beta_{\rho}$ (and $\gamma^{\prime}$), via the Monte-Carlo trajectory sampling method. We simply replace the Gaussian random step in an interval $\ln{(R_{1}/R_{2})}$ with the log-Poisson step here, using the same calculation of ${\rm d}S/{\rm d}\ln{R}$ and critical density $\rhocrit$ for collapse.\footnote{Strictly speaking, we should also account for intermittency in the turbulent velocity field, since $v_{t}(R)$ enters the barrier and variance. However, at the level of our approximation only the second moment of $v_{t}(R)$ enters, so for a given power spectrum the results are fixed. Accounting for higher-order effects would require a multi-dimensional model which can account for explicit correlations between the local velocity and density fluctuations. Numerical simulations suggest these correlations are small \citep[e.g.\ local $v_{t} \sim \rho^{-0.04}$; see][]{passot:1998.density.pdf,kritsuk:2007.isothermal.turb.stats,federrath:2010.obs.vs.sim.turb.compare}, but they may be important for the most extreme structures.} 

The parameter $\beta_{\rho}=0-1$ represents the degree of intermittency (and non-Gaussianity) in the distribution: as $\beta_{\rho}$ decreases from $1$ to $0$, the strength of correlations between fluctuations on different scales increases and the PDF becomes more non-Gaussian. When $\beta_{\rho}\rightarrow1$, $\lambda\rightarrow \infty$, and the Poisson distribution becomes approximately Gaussian and continuous, hence the log-Poisson statistics become lognormal (this is true for any $G_{R}$ in Eq.~\ref{eqn:logpoisson}). When $\beta_{\rho}\rightarrow0$, steps are perfectly correlated and the density PDF if infinitely skew. In the structure function derivation of \citet{sheleveque:structure.functions}, the parameter $\gamma^{\prime}$ represents the ``degree of singularity'' of the most intermittent structures on the appropriate scale; as ${\rm d}S/{\rm d}R$ increases at fixed $\beta_{\rho}$ these become more singular, raising the variance but in concentrated (and locally correlated) structures. 

We should note that the extreme limit where $G_{R}$ is a $\delta$-function and the effects of ``events'' are strictly quantized (as in \citealt{sheleveque:structure.functions}; see also \citealt{liufang:2008.logpoisson.cosmic.baryons}) is almost certainly too extreme -- this predicts a quantized (non-continuous) density distribution. But it provides a useful upper limit for how strong the effects of intermittency might be. In Appendix~\ref{sec:appendix:alt.intermittency}, we consider alternative descriptions with continuous $G_{R}$, which appear to better describe the density statistics in simulations. For our purposes, however, the results are similar.

\vspace{-0.5cm}
\subsection{Comparison with Simulations and Observations}
\label{sec:intermittency:motivation}

We caution that intermittency in the density field of compressible turbulence remains poorly understood, and is an important topic for future, more detailed study. 
That said, our assumption mapping the intermittent compressive (longitudinal) velocity statistics to the density field 
appears to apply in numerical simulations of super-sonic turbulence
\citep{kowal:2007.log.density.turb.spectra,liufang:2008.logpoisson.cosmic.baryons,schmidt:2009.isothermal.turb}.\footnote{\citet{schmidt:2009.isothermal.turb} noted large non-lognormal features in such calculations, and argued that they followed from velocity intermittency, with log-density structure functions following directly from the velocity structure functions. \citet{kowal:2007.log.density.turb.spectra} systematically measured the log-density PDFs and structure functions in both strongly and weakly-magnetized turbulence spanning $\Mach\sim0.1-7$, and found in all cases that the structure functions could be well-described by the multi-fractal cascade models above. And \citet{he:2006.logpoisson.cosmic.baryons,liufang:2008.logpoisson.cosmic.baryons} showed specifically that a log-Poisson hierarchy accurately describes the density structure functions and scale-scale correlations in simulated cosmological baryonic fluids.}

To test this, in \citet{hopkins:2012.intermittent.turb.density.pdfs}, we compile a large suite of such numerical simulations (including those above, with and without magnetic fields, and $\Mach\sim0.1-15$), and confirm that our ``intermittent'' density PDF model appears to be a (surprisingly) good approximation. We show that in every case considered, a density PDF of the form in Eq.~\ref{eqn:general.intermittent.pdf} (see Appendix~\ref{sec:appendix:alt.intermittency}; Eq.~\ref{eqn:PV}) describes the simulations accurately (moreso than a pure log-normal), and moreover that the ``degree of intermittency'' (e.g.\ $\beta$ and $\gamma$, or $T$ in Appendix~\ref{sec:appendix:alt.intermittency}) estimated by fitting to the density PDFs is consistent with that determined directly from the velocity structure functions. 

This follows from the fact that the density PDF is identical to the volume density distribution of a passive scalar in the limit of zero diffusion (pure advection) -- i.e.\ mass is a non-diffusive passive scalar. The statistics of passive scalars are well-studied, albeit primarily in weakly-compressible turbulence \citep[for a review, see][]{warhaft:2000.passive.scalar.turb.review}. Both analytic arguments \citep[from multi-fractal models as well as directly from the Navier-Stokes equations; see][]{yakhot:1997.scalar.field.turb.pdfs,he:1998.bivariate.logpoisson.advected.scalars} and numerical simulations \citep[e.g.][]{chen:1997.scalar.turb.bivariate.logpoisson} have argued for the same conclusion that if the velocity field obeys a log-Poisson hierarchy, the PDF of advected scalars should as well. This scaling has been measured in laboratory fluid turbulence for the scalar difference PDF shape vs.\ scale (the most important property here) as well as the scalar structure functions \citep[see e.g.][]{ruiz-chavarria:1996.passive.scalar.turb.poisson.tests,zhou:2010.scalar.field.logpoisson.pdfs}. 

Empirically, \citet{burlaga:1992.multifractal.solar.wind.density.velocity} and \citet{marsch:1994.multifractal.solar.wind.fluct.scalar.vel} showed that the observed PDF, Lagrangian structure functions, and scale-correlations of the density fluctuations in the solar wind follow the same multi-fractal scaling as the velocity fluctuations (exactly what we have assumed here). More recently, \citet{leubner:2005.solar.wind.pdf.castaing,leubner:2005.solar.wind.castaing.fits.density} showed specifically that the density statistics agree well with the class of log-Poisson models described in Appendix~\ref{sec:appendix:alt.intermittency} and \citet{hopkins:2012.intermittent.turb.density.pdfs}. For a review, see \citet{chang:2009.space.plasma.complexity.review}. 
And in a series of papers, \citet{he:2006.logpoisson.cosmic.baryons,liufang:2008.logpoisson.cosmic.baryons,lu:2009.logpoisson.neutral.hydrogen,yuan:2009.logpoisson.cluster.gas.sz.lum,lu:2010.lyalpha.logpoisson.cosmo} have specifically argued that both numerical cosmological simulations and observations of the density field in the IGM, cluster gas, and Ly$\alpha$ forest appear to exhibit density PDF shapes, structure functions, and scale-scale correlations following the log-Poisson model. The same model for the density fluctuation PDF and structure functions also appears to successfully describe laboratory MHD plasma turbulence \citep[see][and references therein]{budaev:2008.tokamak.plasma.turb.pdfs.intermittency}. 

\vspace{-0.5cm}
\subsection{Analytic Solutions}
\label{sec:intermittency:analytic}

Remarkably, the generalized formulation here still admits analytic solutions for the mass function of the nature described in \S~\ref{sec:analytic.mfs}, for a given assumption about $G_{R}$. The full derivation of solutions are presented in Appendices~\ref{sec:appendix:analytic.continuous}-\ref{sec:appendix:analytic.beta}.

\begin{figure*}
    \centering
    \plotsidesize{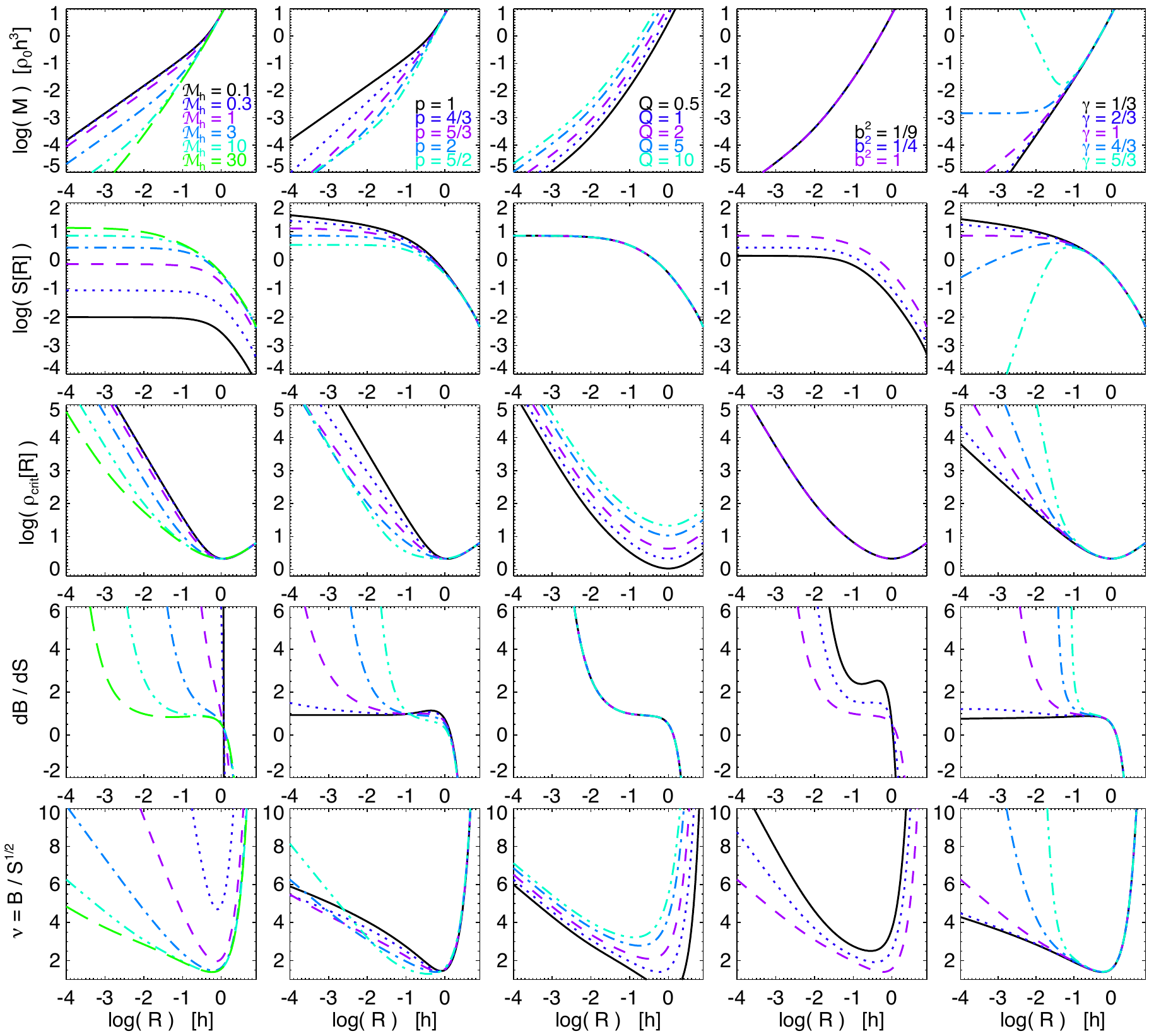}{0.95}
    \caption{Some basic properties of self-gravitating regions in turbulent media. 
    Each assumes a disk with (arbitrary) mean midplane density $\rho_{0}$, scale-height $h$, and the following ``default'' properties, each varied in turn. 
    {\bf (1)} Large-scale Mach number $\Machdisk=\Mach_{\rm rms}(h)=10$, varied from $0.1-30$. 
    {\bf (2)} Turbulent spectral index ($E(k)\propto k^{-p}$), $p=2$ for Burgers (highly super-sonic) turbulence is our default and $p=5/3$ is Kolmogorov, values outside this range are rare. 
    {\bf (3)} Global Toomre parameter (default $Q=1$, for marginal stability). 
    {\bf (4)} Parameter $b$ in Eq.~\ref{eqn:S.R} (the Mach number-density dispersion relation), $b^{2}=1$ (default) for $\Machdisk$ defined as only the compressive part of the turbulence; more generally $b^{2}=1/3$ in strongly magnetized media, $b^{2}=1/4$ in randomly forced supersonic turbulence, $b^{2}=1/9$ for purely solenoidal forced turbulence. 
    {\bf (5)} Equation of state polytropic index $\gamma$ (isothermal gas with $\gamma=1$ by default, which produces a lognormal density PDF; other choices lead to non-lognormal distributions). 
    Rows show different properties. 
    {\em Top:} Mass-radius relation; this follows Eq.~\ref{eqn:mass.radius.approx} and generically produces weakly-varying surface densities over a wide dynamic range. 
    {\em Second:} Variance $S(R)$ in the density field; for non-isothermal cases ($\gamma\ne1$) we plot $S(R,\,\rhocrit)$, the variance at $R$ for a narrow density range about the critical density. Larger $\Machdisk$ (and $b$) systematically increase $S$, stiffer $\gamma$ suppresses it on small scales. 
    {\em Third:} Critical density (in units of $\rho_{0}$) for self-gravitating collapse. This increases with $Q$ and $\gamma$ (on small scales); but at fixed $Q$ it is lower at higher $\Machdisk$. 
    {\em Fourth:} ${\rm d}B_{\rm crit}/{\rm d}S$ ($B_{\rm crit}$ is the collapse threshold or ``barrier'' in Fig.~\ref{fig:demo}); when $\gg1$, the collapse threshold rises ``too fast'' relative to density fluctuations and the field becomes non-self gravitating (the ``last-crossing regime''), when $>0$ but $\ll1$ regions are self-gravitating up to large scales (``first crossing'' regime); when $\ll 0$ on large-scales, direct collapse is suppressed and structure growth switches from ``top-down'' to ``bottom-up'' via mergers. 
    {\em Bottom:} $\nu\equiv B_{\rm crit}/S^{1/2}$; the number of standard deviations of a collapsing fluctuation on that scale. Minima here correspond to the most unstable scale ($\sim h$). 
    \label{fig:mr.scalings}}
\end{figure*}

\begin{figure*}
    \centering
    \plotsidesize{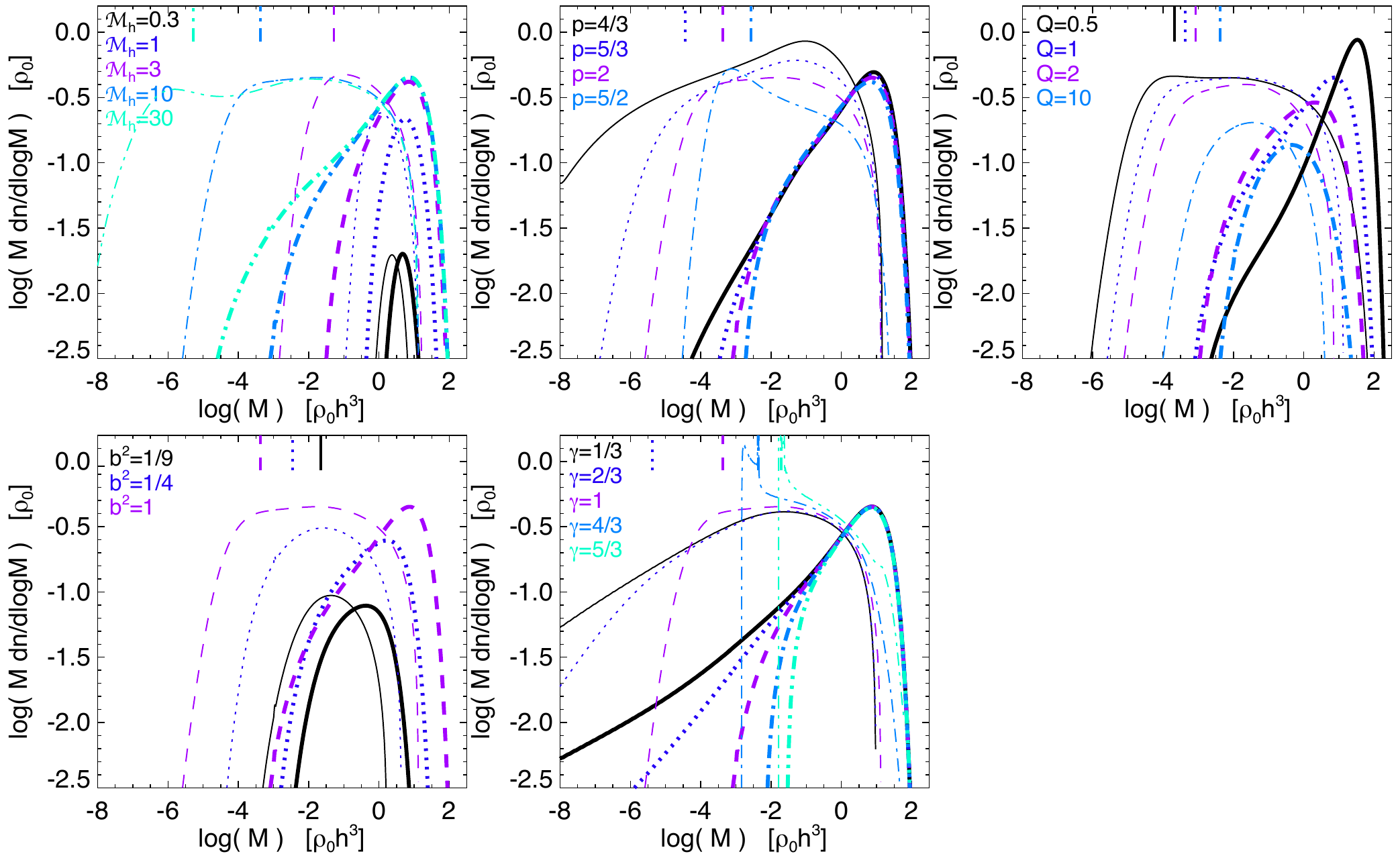}{0.95}
    \caption{Mass functions (MFs) of self-gravitating objects, for the same parameter variations as Fig.~\ref{fig:mr.scalings}. We plot $M\,{\rm d}n/{\rm d}\log{M}$, the mass per logarithmic interval in object mass (a MF with ${\rm d}n/{\rm d}M\propto M^{-\alpha}$ with $\alpha=2$ is a horizontal line). 
    {\em Thick} lines show the first-crossing distribution (mass spectra of self-gravitating regions defined on the {\em largest} self-gravitating scale). These characteristically have faint-end slopes $\alpha<2$ and the mass is concentrated near the ``maximal instability'' scale (collapse on scale $\sim h$), with weak dependence on other properties except at low masses. 
    {\em Thin} lines show the last-crossing distribution (mass spectra on the {\em smallest} self-gravitating, i.e.\ non-fragmenting, scale). These have slopes slightly steeper than $\alpha>2$ over a broad range, then turn over below the sonic scale, defined by Eq.~\ref{eqn:m.sonic} and labeled in each panel by the short vertical line (when $\Machdisk<1$, this is not defined, but the relevant scale is now again the maximal instability scale). 
    The normalization of the MF decreases for $\Machdisk\ll1$ or $b\ll1$ but is otherwise similar (saturated when an order-unity mass fraction is self-gravitating). The dynamic range of first \&\ last crossings increases with $M_{\rm maximal}/M_{\rm sonic}$. This is a strong function of $\Machdisk$, modest function of $Q$ and $p$ (though over the expected range $p=5/3-2$ the effect is small) and weak function of $b$ (except $b\ll 1$). Non isothermal $\gamma$ have no effect in the turbulence-dominated scales $>M_{\rm sonic}$, but increasing $\gamma$ sharply suppresses first and last crossings below $M_{\rm sonic}$, with the cutoff becoming steeper with higher $\gamma$ (infinitely steep for $\gamma \ge 4/3$). Intermittency has small effects on the MFs (see Appendix~\ref{sec:nongaussian}-\ref{sec:appendix:alt.intermittency}).
    \label{fig:mfs}}
\end{figure*}

\begin{figure*}
    \centering
    \plotsidesize{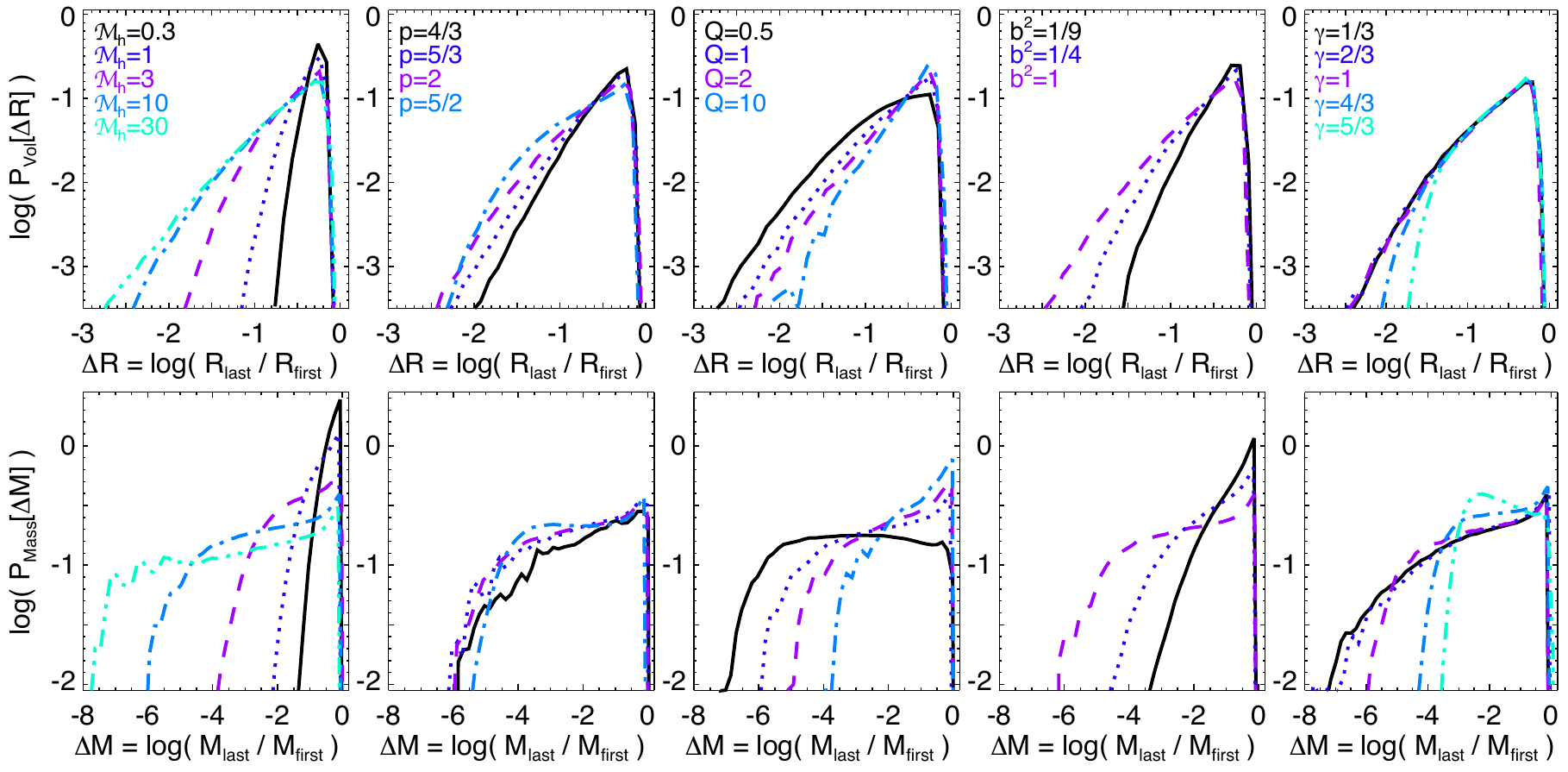}{0.95}
    \caption{Dynamic range of the ``fragmentation cascades'' that make up the MFs in Fig.~\ref{fig:mfs}. 
    {\em Top:} Probability distribution, for a given random point in space (i.e.\ random volume element) which is self-gravitating on some scale, of the logarithmic dynamic range between the scale of first crossing $R_{\rm first}$ and scale of last-crossing $R_{\rm last}$. Most volume within a self-gravitating region is only self-gravitating over a narrow dynamic range, i.e.\ is not part of a much denser fragmenting substructure. The ``tail'' extends in dynamic range with the ratio $M_{\rm sonic}/M_{\rm maximal}$, as the gap between first/last crossing in Fig.~\ref{fig:mfs}. 
    {\em Bottom:} Same, but for a random self-gravitating {\em mass element}, as a function of the gap in first/last crossing mass scales. This is distributed over a wider dynamic range (note the difference in figure scales). A given Lagrangian element is often self-gravitating over several dex in mass scale (i.e.\ locally self-gravitating structures prefer much larger ``parent'' structures). 
    The predicted dynamic range generally scales with $M_{\rm sonic}/M_{\rm maximal}$. However varying $p$ produces weaker (or even opposite) effects than what might be expected: when $p$ is ``flatter'' (more turbulent power on all scales), the last-crossing MF and $M_{\rm sonic}$ do extend to smaller masses, but these often correspond to smaller-scale first-crossings as well (rather than being entirely seeded inside much larger structures). 
    \label{fig:fragrange}}
\end{figure*}

\vspace{-0.5cm}
\section{Results at a Fixed ``Instant''}
\label{sec:results}

Having developed these models, we now examine various ``instantaneous'' properties of self-gravitating objects in fully-developed turbulence (before considering below how these evolve in time in \S~\ref{sec:frag:time}). For the sake of clarity, we will vary several parameters in turn, but otherwise fix our ``reference model'' parameters. This is a rotating disk with Toomre $Q=1$ (marginally stable), large-scale Mach number $\Machdisk=10$ (super-sonic turbulence), $p=2$ (Burgers turbulence, appropriate in the supersonic case), $b^{2}=1$ (i.e.\ $\Machdisk$ defined in terms of the compressive Mach number) and $\gamma=1$, with no intermittency. The dependence on all of these choices is discussed below.

\vspace{-0.5cm}
\subsection{Basic Properties and Scaling Relations}
\label{sec:basic}

First, we consider some basic scaling relations which arise even before we evaluate the full density field statistics. 

Fig.~\ref{fig:mr.scalings} plots several such scalings that follow immediately from the definition of the critical density and variance $S(R)$. This includes the critical density $\rhocrit(R)$ and variance $S(R)$ themselves, the mass-size relation for self-gravitating objects, the ``barrier'' $B$ that defines self-gravitating objects at each radius in units of the variance ($\nu\equiv B/S^{1/2}$), and the derivative of the barrier with respect to the variance ${\rm d}B/{\rm d}S$. 

In all cases, the variance $S(R)$ on very large scales ($\gg h$) falls rapidly; this must be true simply because of mass conservation (finite total mass). The variance grows with decreasing $R$ over intermediate scales, where turbulence produces density fluctuations. As expected, $S(R)$ increases systematically with $\Machdisk$ and $b$ (larger compressive velocity fluctuations). Lower $p$ (shallower turbulent power spectra) contribute more power at small scales, and stiffer $\gamma$ suppress the power on small scales more strongly. Eventually the variance converges on small scales below the sonic length (discussed below). For non-isothermal cases, the ``effective'' variance $S(R,\,\rho)$ is also a function of the density (since the sound-speed, hence $\Mach$ is density-dependent) -- we therefore plot the variance $S(R,\,\rhocrit)$ near the collapse density. For otherwise fixed parameters, the variance at the mean density $S(R,\,\rho_{0})$ is just the plotted value for the isothermal case $\gamma=1$. The ``turnover'' at high-$\gamma$ corresponds to strong suppression of high-density fluctuations on small scales. 

The critical density rises on large scales with support from rotation ($\kappa$), and on small scales with support from thermal pressure and turbulence. Increasing $Q$ linearly increases $\rhocrit$ in turn; $\rhocrit$ also increases with a stiffer equation of state (higher $\gamma$) owing to thermal pressure on small scales. Interestingly, at otherwise fixed parameters, $\rhocrit$ is actually lower when the Mach number is increased. This is because $Q$ on large scales is held constant, so lower $\Machdisk$ means lower $c_{s}/\sigma_{g}(h)$, i.e.\ a lower threshold where thermal pressure becomes important. 

These scalings highlight two critical scales in this problem.

\vspace{-0.5cm}
\subsubsection{The Maximal Instability (Quasi-Toomre) Scale}
\label{sec:toomre}

On large scales, there is a clear ``most unstable scale'' $R_{\rm maximal}$ in Fig.~\ref{fig:mr.scalings}. This is closely related to the minimum in $\rhocrit$ at $R\approx h$, with rotation making the system more stable on larger scales, and thermal support ``stabilizing'' smaller scales (as in the traditional Toomre stability analysis). However the ``most unstable scale'' is not simply where $\rhocrit$ is minimized, nor is it the Toomre wavelength (most unstable wavelength in a constant-$\rho$ disk). 

To estimate the scale which will contain most of the collapsing mass in a {\em non}-uniform density disk, the more relevant parameter is the ratio $\nu=B/S^{1/2}$. If the density PDF is lognormal, then 
\be
F_{\rm coll} = \frac{1}{2}{\rm Erfc}{{\Bigl [} \frac{\nu}{\sqrt{2}} {\Bigr ]}}
\ee
is the fraction of mass which is self-gravitating on that scale (independent of the exact size/mass spectrum), so $F_{\rm coll}$ is maximized where $\nu$ is minimized. We see in Fig.~\ref{fig:mr.scalings} that the rise in $\rhocrit$ on small scales is sufficiently rapid that the scale where $\nu$ is minimized is always close to $R\sim h$ (with weak dependence on any parameters). 

Although there is no trivial analytic solution, it is easy (if desired) to numerically solve for $R_{\rm maximal}\equiv R[\nu_{\rm min}]$. And if we assume turbulence dominates over thermal pressure at the scale of interest\footnote{If thermal support dominates near $\sim h$ (e.g.\ $\Machdisk\ll1$), we derive an expression for $R_{\rm maximal}$ which is identical to Eq.~\ref{eqn:r.maximal} with $p=1$.} and expand about $R[\nu_{\rm min}]\sim h$, we obtain 
\begin{align}
\label{eqn:r.maximal}
\ln{{\Bigl(}\frac{R_{\rm maximal}}{h} {\Bigr)}} &\approx \frac{A_{1}-\sqrt{A_{1}^{2}+2\,A_{2}\,\ln{(2\,Q\,[1+\tilde{\kappa}^{2}])}}}{2\,A_{2}} \\ 
\nonumber
A_{1} &= \frac{2\,(3-p)+(1+\tilde{\kappa}^{2})\,(|{\rm d}S/{\rm d}\ln{R}|_{h}-1)}{4\,(1+\tilde{\kappa}^{2})} \\
\nonumber
A_{2} &= \frac{3}{16} + \frac{3\,\tilde{\kappa}^{2}\,(3-p)^{2}}{4\,(1+\tilde{\kappa}^{2})}
\end{align}
This scaling almost always gives $R[\nu_{\rm min}]\sim h$, but includes the weak dependence on other parameters. Higher $Q$ values and steeper power spectra systematically shift $R_{\rm maximal}$ to smaller scales (they raise $\rhocrit(R=h)$, so require going to smaller $R$, where $S(R)$ is larger, to reach the minimum in $\nu$).

These effects, while weak, are {\em not} captured in the traditional Toomre analysis. In fact a Toomre-style linear stability analysis of a thin disk with constant density gives the most unstable wavelength $R/h\sim |k_{\rm min}\,h|^{-1} = Q\,\tilde{\kappa}^{-1}$ (using $h=\sigma/\Omega$). The Toomre wavelength at the mean density {\em increases} linearly with $Q$; however, the ``most unstable'' wavelength -- in terms of where most of the mass will fragment under self-gravity --  actually {\em decreases} with increasing $Q$ in a turbulent medium.

\vspace{-0.5cm}
\subsubsection{The Sonic Scale}
\label{sec:sonic}

Another critical scale is the sonic length/mass scale. For isothermal gas this is typically defined as the scale below which the rms $\Mach<1$, $R_{\rm sonic} = R(\Mach=1)$. The corresponding ``critical mass'' is $M_{\rm sonic} = M(R_{\rm sonic})$. But recall, what matters for the density distribution is really the compressive component of the Mach number $\Machcompressive$, which enters the variance as $\Machcompressive=b\,\Mach$, hence the effective sonic scale of interest for the mass function is $R(b\,\Mach=1)$.

For the polytropic case, there is no one spatial scale where $b\,\Mach=1$, but since we care about objects crossing the critical density, we should consider $\Mach$ at this density, giving $R_{\rm sonic}=R(b\,\Mach[\rhocrit(R)]=1)$. Although the general solution for $\rhocrit$ and $R_{\rm sonic}$ for arbitrary $\gamma$ must be numerical, it is possible to analytically solve for $R_{\rm sonic}$ and $M_{\rm sonic}$ if we assume either $R_{\rm sonic}\ll h$ or non-rotating turbulence (i.e.\ neglect the large-scale/$\kappa$ terms, generally a good approximation if $\Machdisk\gtrsim1$). This gives 
\begin{align}
\label{eqn:r.sonic}
\frac{R_{\rm sonic}}{R\cloudsub} &= {\Bigl [}\frac{2\,Q^{\prime}}{1+\Machcloud^{2}}\,(b\,\Machcloud)^{\frac{2\,(\gamma-2)}{\gamma-1}} {\Bigr]}^{n_{1}} \\
\label{eqn:m.sonic}
\frac{M_{\rm sonic}}{M\cloudsub} &\approx (b\,\Machcloud)^{\frac{2}{\gamma-1}}\,{\Bigl(}\frac{R_{\rm sonic}}{R\cloudsub}{\Bigr)}^{3 + \frac{p-1}{\gamma-1}} \\
\nonumber n_{1} &= \frac{\gamma-1}{2\,(\gamma-1)-(p-1)\,(\gamma-2)} 
\end{align}
For $\Machcloud\gg1$ and $p=2$, the case of greatest interest in the ISM, this reduces to $(R_{\rm sonic}/R\cloudsub) \approx (2Q^{\prime})^{(\gamma-1)/\gamma}\,b^{(2\,\gamma-4)/\gamma}\,\Machcloud^{-2/\gamma}$ and $(M_{\rm sonic}/M\cloudsub) \approx (2Q^{\prime})^{(3\gamma-2)/\gamma}\,b^{(6\,\gamma-8)/\gamma}\,\Machcloud^{-4/\gamma}$. 

Stronger turbulence (higher $\Machdisk$), softer equations of state $\gamma$, and shallower (more bottom-heavy) turbulent power spectra (lower $p$) decrease $R_{\rm sonic}$ and $M_{\rm sonic}$ as expected. The sonic scale is independent of $Q$ for isothermal gas but for ``stiff'' equations of state, higher $Q$ implies more support and higher $R_{\rm sonic}$ while for ``soft'' equations of state it implies more turbulence on small scales, and lower $R_{\rm sonic}$. 

Below $R_{\rm sonic}$, the contributions to the density variance from successive turbulent fluctuations are small. This scale is obvious in Fig.~\ref{fig:mr.scalings}, as the scale below which $S(R)$ saturates and $\rhocrit$ begins to steeply rise, producing a very sharp increase in ${\rm d}B/{\rm d}S$.

\vspace{-0.5cm}
\subsubsection{Mass-Radius Relation, Densities, \&\ Surface Densities}
\label{sec:mass.radius}

From $\rhocrit(R)$, it is trivial to calculate the corresponding mass-size relation of self-gravitating objects at their formation (Eq.~\ref{eqn:mass.radius.exact}). We see in Fig.~\ref{fig:mr.scalings} that these obey approximate power-laws over a very wide dynamic range, albeit with some gradual shifts in slope. 

The critical densities $\rhocrit$ shown in Fig.~\ref{fig:mr.scalings} rise at large scales because of angular momentum support ($\kappa$), then rise again below $h$ from a combination of turbulent support (above $R_{\rm sonic}$) and thermal support (below) which must be overcome by gravity. Over each regime where thermal, turbulent, or rotation support dominates, we can determine the approximate power-law: 
\begin{align}
\label{eqn:mass.radius.approx}
\frac{M(R)}{\rho_{0}\,h^{3}} \approx 
\begin{cases}
      {\displaystyle {\Bigl(}\frac{4\pi}{3}{\Bigr)}\,{\Bigl(}\frac{Q}{2\tilde{\kappa}[1+\Machdisk^{2}]}{\Bigr)}^{\frac{1}{2-\gamma}}\,{\Bigl(}\frac{R}{h}{\Bigr)}^{\frac{1-3\,(\gamma-1)}{1-(\gamma-1)}}} \hfill {\tiny (R\ll R_{\rm sonic})} \\ 
      \\
      {\displaystyle {\Bigl(}\frac{2\pi Q}{3\tilde{\kappa}}{\Bigr)}\,{\Bigl(}\frac{R}{h}{\Bigr)}^{p}} \hfill {\tiny (R_{\rm sonic}\ll R \ll h)} \\
      \\
      {\displaystyle (\pi\tilde{\kappa}Q)\,{\Bigl (}\frac{R}{h}{\Bigr )}^{3}} \hfill {\tiny (R\gg h)} \
\end{cases}
\end{align}

In the turbulence-dominated regime ($R_{\rm sonic}\ll R \ll h$), the size-mass relation is simply set by the turbulent power spectrum. For typical turbulent properties, $p\approx5/3-2$, this generically gives nearly-constant surface densities $\Sigma\sim M/R^{2}$ in collapsing objects. In the thermal-dominated regime ($R\ll R_{\rm sonic}$), the fact that $M(R)$ is multivalued for $\gamma>4/3$ is intimately related to the suppression of fragmentation: if an object attempted to fragment or contract to smaller spatial scales, it would have to acquire {more} mass to overwhelm the rapidly increasing thermal pressure barrier.

\begin{figure}
    \centering
    \plotonesize{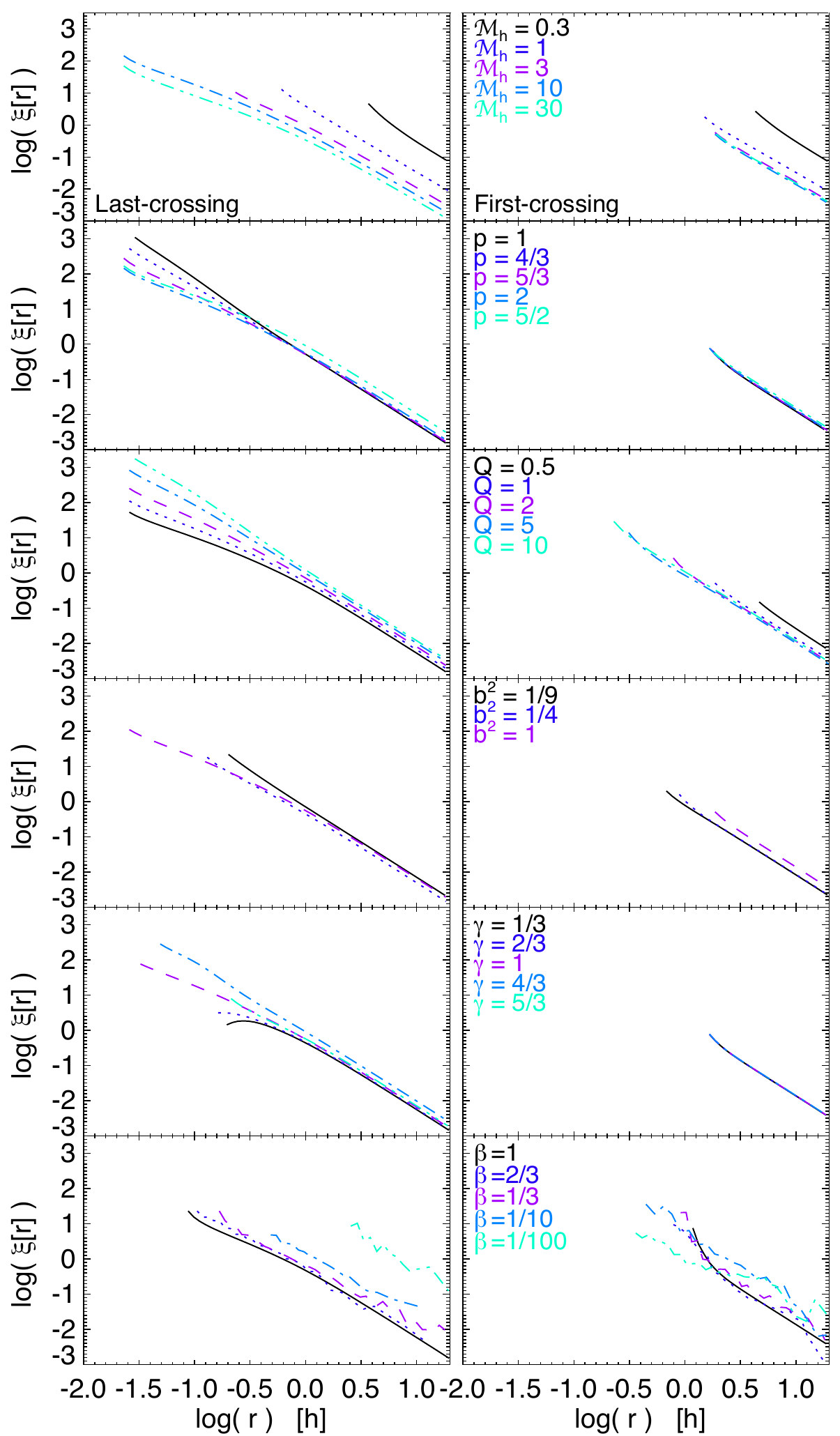}{0.95}
    \caption{Correlation function $\xi(r)$ (excess probability of a neighboring object within a distance $r$ of a given object, Eqs.~\ref{eqn:xi.last.defn}-\ref{eqn:xi.cm.defn}) 
     for last-crossing ({\em left}) and first-crossing ({\em right}) objects of a given mass. 
     {\em Left:} Last-crossings with mass $M_{\rm last}$ (defined as the mass at the maximum of the MF; the peak of $M\,{\rm d}n_{\rm last}/{\rm d}\ln{M}$ in Fig.~\ref{fig:mfs}). In general $M_{\rm last}\sim M_{\rm sonic}$. Note $\xi(r)$ is undefined below $r\approx R(M_{\rm last})$. 
     {\em Right:} Same for first-crossings at $M_{\rm first}$ (the peak of $M\,{\rm d}n_{\rm first}/{\rm d}\ln{M}$), roughly $M_{\rm first}\sim M_{\rm maximal}$. For each, we consider a range of parameters. 
     Measured at these ``characteristic'' masses of the MF, $\xi(r)$ is nearly universal. There is a dependence on $\Machdisk$ when $\Machdisk\lesssim1$; here $\xi(r\,|\,M)$ increases as the number density $n(M)$ decreases. The shape is generically power law-like, with $\xi(r)\propto r^{-2}$ at large separations transitioning to shallower $\propto r^{-1}$ at smaller separations.
    \label{fig:corr.fn.varmass}}
\end{figure}

\begin{figure}
    \centering
    \plotonesize{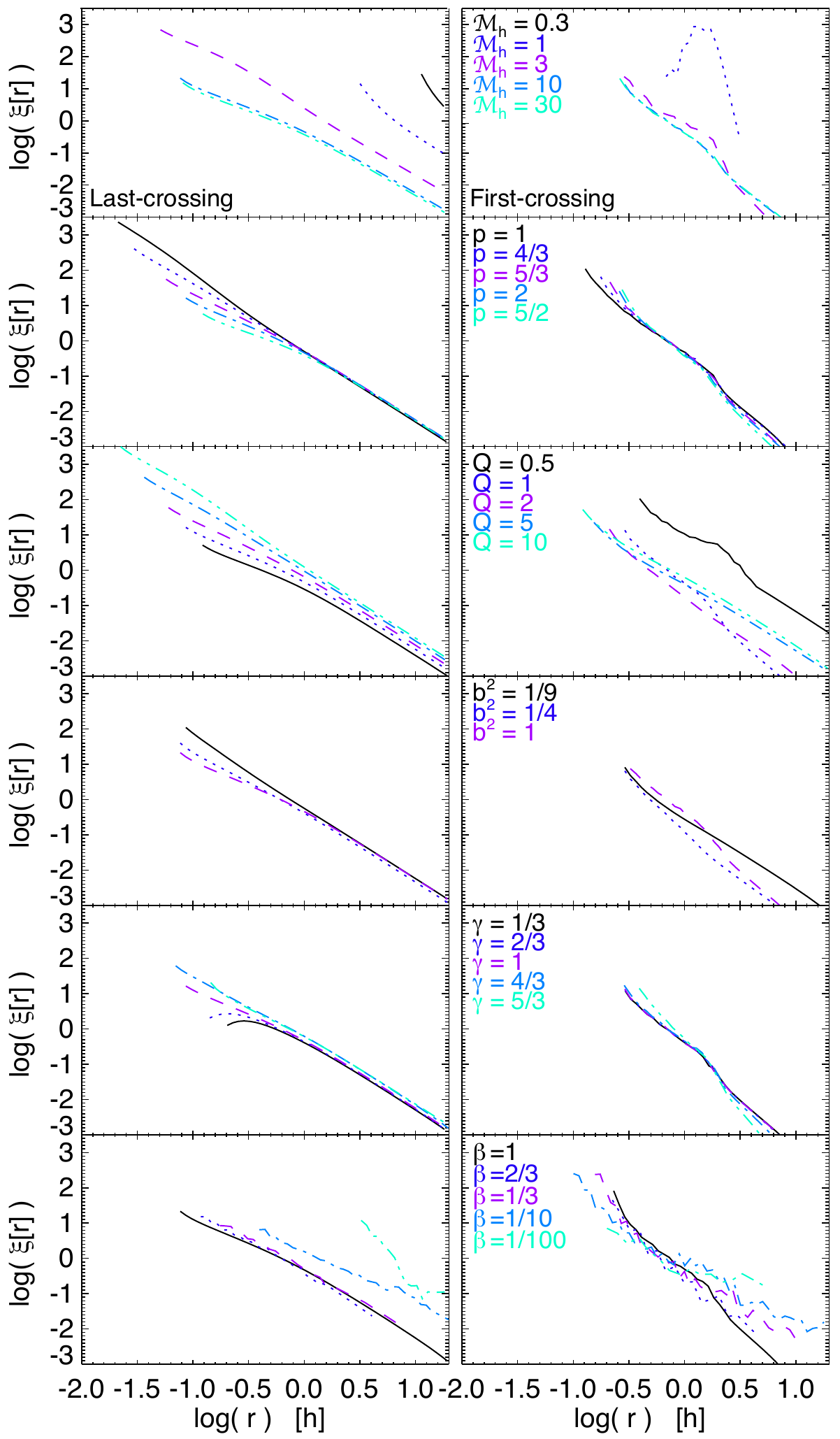}{0.95}
    \caption{The correlation function $\xi(r)$ as Fig.~\ref{fig:corr.fn.varmass}, but for objects with fixed mass $M$. {\em Left:} $\xi(r)$ for last-crossings with mass $M=0.01\,\rho_{0}\,h^{3}$. {\em Right:} $\xi(r)$ for first-crossings with mass $M=\rho_{0}\,h^{3}$. The dependence on other parameters is still weak, but stronger than in Fig.~\ref{fig:corr.fn.varmass}. This dependence comes from sampling different parts of the MF relative to the peak (sonic or maximal instability) mass scale. The amplitude of $\xi(r)$ systematically increases with decreasing number density at fixed $M$ (Eq.~\ref{eqn:numden.bias}). The last-crossing $\xi(r)$ converges to universal shape at large $r$, but becomes more shallow at small radius when the variance is larger (explaining the trends in $Q$, $b$, $p$, and $\gamma$). 
    \label{fig:corr.fn.fixmass}}
\end{figure}

\vspace{-0.5cm}
\subsection{The Mass Function: Gravitating Structures and Sub-Structures}
\label{sec:mf}

Fig.~\ref{fig:mfs} shows the mass functions of both ``first-crossing'' objects (self-gravitating objects with masses/sizes defined on the largest regions on which they are self-gravitating) and ``last-crossing'' objects (defined on the smallest self-gravitating scale). 

\vspace{-0.5cm}
\subsubsection{First Crossings -- The Largest Collapsing Objects}

The first-crossing MF has most of the mass concentrated near the maximal instability (quasi-Toomre) scale defined above (the scale most vulnerable to fragmentation). This is close to $\sim h$ in all cases, but with a weak dependence on other parameters (decreasing weakly with $Q$, the {\em opposite} of the behavior of the traditional Toomre length/mass; see \S~\ref{sec:toomre}). 

On larger scales, angular momentum produces an exponential cutoff of the MF (as the variance falls of rapidly and $\nu$ grows with $R$). On smaller scales, there is a power-law like run over $\sim1-4$ dex in mass, the range where turbulence dominates. As discussed in \paperone, if we use the analytic first-crossing solution for the linear barrier to approximate this, we can indeed predict the generic slope ${\rm d}n/{\rm d}M\propto M^{-\alpha}$ with $\alpha$ slightly smaller than $2$:
\begin{align}
\alpha_{\rm first\ cross} \sim -2 - \frac{(3-p)^{2}}{2\,p^{2}\,S}\,\ln{{\Bigl(}\frac{M}{M_{\rm maximal}}{\Bigr)}}
\end{align}

The value $\alpha\sim2$ is generic because gravity is scale-free, so if the critical density and variance per unit scale were constant we would recover exactly equal mass per logarithmic interval in mass ($\alpha=2$). These quantities are not, however, constant; but they are close to it: $B$ and $S$ are {\em logarithmic} functions of $R$ and $M$, because the density distribution follows a random multiplicative process. As a result, there is only a weak correction: because $\nu$ increases towards smaller $R$ (see Fig.~\ref{fig:mr.scalings}) the MF has less mass at small-$M$ than $\alpha=2$ (i.e.\ we predict $\alpha<2$). 

At sufficiently small scales, the MF again cuts off. We expect this to occur as we approach the sonic mass, since density fluctuations are strongly suppressed (Fig.~\ref{fig:demo}); plotting this in each case in Fig.~\ref{fig:mfs}, we see this is indeed a good predictor of the low-mass MF cutoff.  In all cases, masses near $M_{\rm sonic}$ contain only a small fraction of the total mass in first-crossing objects. 

Interestingly, the first-crossing MF slope (in the power-law range) is remarkably independent of the turbulent spectral index $p$, despite the fact that this sets the run in $M(R)$ and $\rhocrit(R)$. It appears that the steeper run in $S$ (steeper at higher $p$) nearly completely cancels the steeper run in $p$, giving the quite similar behavior in $\nu(R)$ in Fig.~\ref{fig:mr.scalings}, as well as a nearly identical ${\rm d}B/{\rm d}S$ from $\sim h$ down to $R_{\rm sonic}$. This is not trivially expected from the first-crossing solution for a linear barrier, and depends on the full (non-linear) solution of the MF. The index $p$ does make some difference at low masses, where it defines how rapidly the transition from turbulent to thermal support occurs.

The MF does depend significantly on $\Machdisk$ and $b$. Increasing these increases the variance at all scales and so broadens the range where collapse is predicted. The effect is weak near the peak of the MF in the supersonic case and/or when $b^{2}\gtrsim1/4$. In these cases the peak of the MF is ``saturated'' (i.e.\ reaches near order-unity fractions of the mass), so its normalization changes negligibly even if we vary the properties of the turbulence dramatically; moreover the power-law regime is in the near-universal range so it also does not vary much.  But increasing $\Machdisk$ has a large effect on $R_{\rm sonic}$ and $M_{\rm sonic}$, correspondingly extending the MF to lower masses with increasing $\Machdisk$. For $\Machdisk\ll1$ or $b\ll1$, the MF begins to be suppressed rapidly. In this limit, the variance in the density field is small, so $\nu$ becomes $\gg1$ even near the most unstable scale. As such, we transition from sampling the ``core'' of the density distribution to the tails, and the probability of collapse is exponentially suppressed. Because, within the tail, the probability of a ``crossing event'' is a strong function of $\nu$, the MF also rapidly becomes sharply peaked around the most unstable scale.

The MF also depends on $Q$. For $Q<1$, the fact that the system is unstable even at $\rho=\rho_{0}$ means that essentially all of the mass is in self-gravitating objects on the maximal instability scale (making a much sharper peak). For $Q\gg1$, we again move into the exponentially suppressed regime as with $\Machdisk\ll1$.

Unsurprisingly, over the turbulence-dominated regime, there is essentially no dependence of the MF on $\gamma$. Below the sonic mass, though, $\gamma$ makes a large difference, with more mass in low-mass objects with softer $\gamma$.

\begin{figure}
    \centering
    \plotonesize{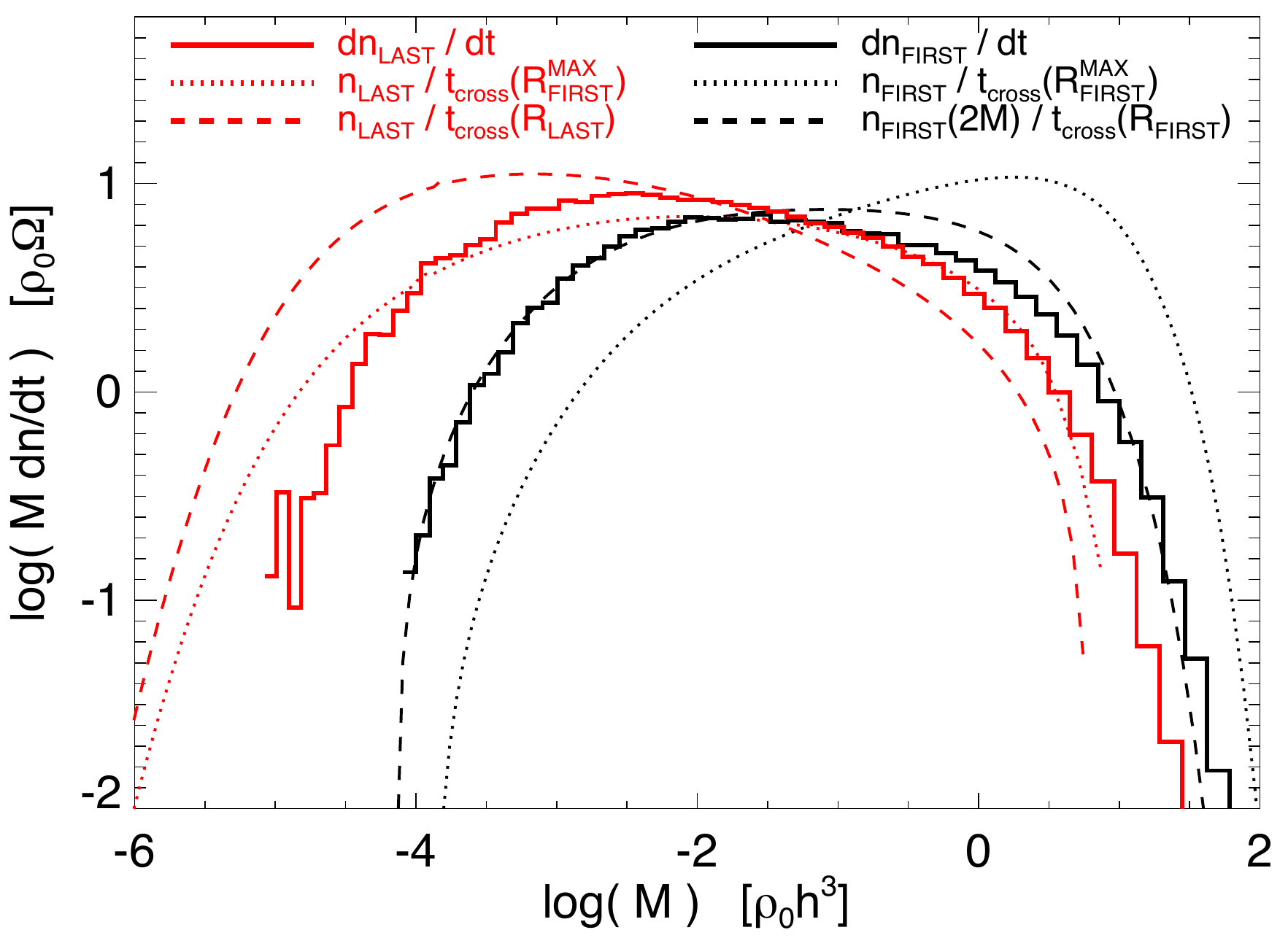}{1.}
    \caption{Global, time-averaged rate of formation of bound structures, as a function of mass, for a model in which all density modes start at $\delta=0$ (median density) and self-gravitating structures are immediately removed (i.e.\ collapse or are destroyed). We show both first (black) and last (red) crossings. 
    Histograms show the full numerical result (for the ``default'' model). 
    Dotted lines show the instantaneous MF predicted for a fully-developed density PDF (Fig.~\ref{fig:mfs}), divided by a constant timescale $t_{\rm cross}(R_{\rm first}^{\rm max})$ (the turbulent crossing time at the scale where the MF peaks in Fig.~\ref{fig:mfs}). 
    Dashed lines show the same ``fully-developed'' MF, divided by the crossing time at each mass/radius scale $t_{\rm cross}(R)$ (for first-crossing we also offset it by $0.3$\,dex to lower masses). 
    This provides a good approximation to first-crossings: they develop approximately independently on each scale on the local crossing time, but with a truncated high-mass MF (because the highest masses depend on merger-driven growth, discussed below). However the constant timescale (dotted) is a better approximation for last-crossing: these are not driven by rapidly-evolving, small-scale fluctuations, but ``pre-seeded'' inside larger objects and so tied to the rate of first-crossing events driven by larger-scale fluctuations. 
    \label{fig:mf.vs.time}}
\end{figure}

\vspace{-0.5cm}
\subsubsection{Last Crossings -- The Smallest Collapsing Objects}

The last-crossing MF is also truncated exponentially at high masses; this always occurs at somewhat smaller masses than the peak of the first-crossing MF. There is also a near-exponential cutoff below the sonic mass. 

In general, the last-crossing MF is ``saturated'' when $\Machdisk\gtrsim1$, and the effect of changing various parameters is largely to shift the sonic mass/lower cutoff of the MF. However as for first-crossings, large $Q\gg1$, $\Machdisk\ll1$ or $b\ll1$ lead to an exponentially suppressed MF with a more narrow mass range around the most unstable mass. This forces convergence between the first and last-crossing MFs. 

There are some generic differences from the first-crossing MF. There, the MF generically had a slope shallower than $\alpha=2$ (i.e.\ had most mass in the most massive objects); here the slope is much closer to $\alpha\approx2$ and even slightly steeper over some range. 

Mathematically, this is related to the exact solutions for a linear barrier in Eq.~\ref{eqn:flast}.\footnote{We can see in Fig.~\ref{fig:mr.scalings} that ${\rm d}B/{\rm d}S\approx$\,constant over much of the dynamic range of interest, so this is not a bad approximation to the full solutions.} The only difference between first and last-crossing MFs in that case is the replacement of the term $B_{0}/S$ in front of the first-crossing MF with ${\rm d}B/{\rm d}S$ for the last-crossing MF. $B_{0}$ and ${\rm d}B/{\rm d}S$ are constant by definition for a linear barrier, but $S=S(R)$ decreases with $R$, so the first-crossing MF should be shallower than the last-crossing MF by a term $\propto S^{-1}$. From Fig.~\ref{fig:mr.scalings} we see this is a weak logarithmic power $\sim M^{0.2-0.4}$. 

Physically, the origin of this difference is that, when there is close to equal power on all scales, the collapsed fraction should be close to uniform on all scales; but the first-crossing MF only ``counts'' objects on the largest scales in which they are self-gravitating. Most of the self-gravitating mass on small scales lies {\em within} larger objects, rather than being isolated. This is closely related to the correlation functions discussed below. 

We see a stronger dependence on $\gamma$ in the last-crossing MF as compared to the first-crossing MF, because more of the mass is concentrated near the sonic mass (where the gas thermal physics is important). For $\gamma < 1$, the low-mass cutoff is significantly more shallow; for $\gamma>1$, it becomes quite sharp. For $\gamma\ge4/3$, collapse is entirely suppressed below $R_{\rm sonic}$, so there is a sharp ``spike'' then cutoff in the MF. 

In the last-crossing MF we see a stronger dependence of the MF slope over the turbulence-dominated range on $p$. The dependence is still weak; but as shown in \papertwo, if we use the linear barrier exact solution and drop the angular momentum and thermal pressure terms, we expect a slope 
\be
\alpha_{\rm last\ cross}\approx -\frac{3}{2}{\Bigl(}1+\frac{1}{p}{\Bigr)} - \frac{(3-p)^{2}\,\ln{(M/M_{\rm maximal})-p\,\ln{2}}}{2\,S(M)\,p^{2}}
\ee
for the values of $M$ and $S$ here, this becomes slightly shallower than $\alpha=2$ at lower $p<2$, and steeper at $p>2$. However over the most likely range $p\approx5/3-2$, the difference in slope is small ($\approx0.15$).

\begin{figure}
    \centering
    \plotonesize{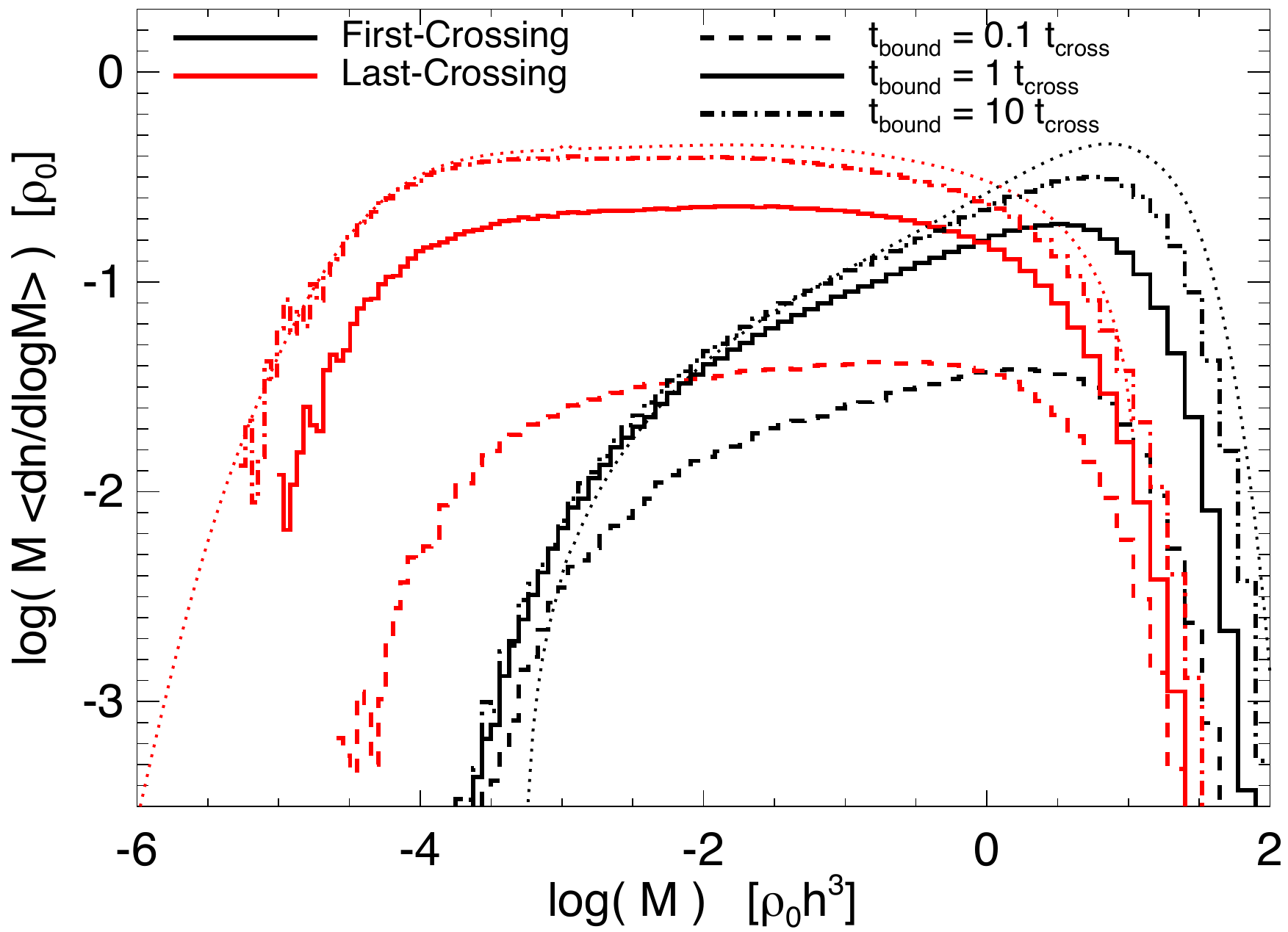}{0.95}
    \caption{The steady-state MF which develops in time from uniform ($\delta=0$) initial conditions. The density field evolves and crossings are recorded as Fig.~\ref{fig:mf.vs.time}, but regions which cross (become self-gravitating) are allowed to survive and evolve for some finite time $t_{\rm bound} \propto t_{\rm cross}(R)$ (with $R$ evaluated at the time of first crossing). For $t_{\rm bound}\ll t_{\rm cross}$, the steady-state MF resembles the instantaneous rate-of-formation (more sharply concentrated near the most unstable mass) multiplied by a lifetime $t_{\rm bound}$. As $t_{\rm bound}$ increases, structure develops ``top-down'' at low masses (building up the low-mass end of the MF) and ``bottom-up'' at high masses. As $t_{\rm bound}\rightarrow \infty$, we recover the analytic prediction for the instantaneous MF in fully-developed turbulence (dotted lines). 
    \label{fig:mf.vs.time.duration}}
\end{figure}

\vspace{-0.5cm}
\subsection{The Dynamic Range of Fragmentation}
\label{sec:frag.dynrange}

From the comparison of the MFs in Fig.~\ref{fig:mfs}, it is tempting to infer the ``dynamic range'' of fragmentation, i.e.\ the range of scales over which fragmentation occurs. This should be crudely related to the ``gap'' between the peak of the first-crossing MF and the sonic scale cutoff in the last-crossing MF. But the MFs alone to not tell us how an individual ``trajectory'' (Eulerian or Lagrangian volume) behaves, in particular the range of scales over which it (individually) is gravitationally unstable and can fragment. 

We calculate this in Fig.~\ref{fig:fragrange}. To do so, we use the Monte Carlo method described in \S~\ref{sec:methods}, and for each parameter set from Fig.~\ref{fig:mfs}, generate $\sim10^{8}$ ``trajectories'' sampling the volume. For each trajectory that is anywhere self-gravitating, we then record the location of first and last crossing, and measure the logarithmic interval in radius $\Delta R \equiv \log{(R_{\rm last}/R_{\rm first})}$ over which that volume element is self-gravitating. Since each trajectory represents a volume element, this corresponds to the distribution of spatial scale ranges over which fragmentation will occur, for a given random {volume} (i.e.\ random point in space) selected in any self-gravitating object. 

We can also compare the distribution of mass scales over which fragmentation will occur: we follow the same procedure, but now define the ``mass range'' $\Delta M\equiv \log{(M_{\rm last}[R_{\rm last}]/M_{\rm first}[R_{\rm first}])}$, and mass-weight so that the distribution reflects the PDF for a random mass element. 

The dynamic range in mass and spatial scale of fragmentation behave broadly as expected: they become more broad with increasing $\Machdisk$, lower $Q$, and softer $\gamma$, all in line with the lower sonic lengths/masses predicted (while the upper mass, at the maximal instability scale, stays about constant). Larger $b$ (more highly compressive turbulence) also increases the dynamic range, as seen in the width of the MFs, by extending the range over which the variance is significant compared to the barrier. 

Interestingly, even though the MF is more broad with lower $p$ (shallower turbulent power spectra), the ``typical'' dynamic range of fragmentation is actually smaller. Shallower power spectra spread the turbulent power over a wider range of scales more uniformly (with $p=1$ corresponding to equal power on all scales), making it more likely that a fluctuation can become self-gravitating (cross above and below the barrier) on small scales, without necessarily being embedded in a more powerful larger-scale fluctuation. Thus even though the MF is more broad, the local dynamic range of fragmentation within each self-gravitating region is not. 

In a mass-weighted sense, the distribution of fragmentation scales is relatively flat, over the range $\Delta M \sim \log{(M_{\rm sonic}/M_{\rm maximal})}$, corresponding to the scale-free MFs. But in the volume-weighted sense, the distributions are significantly more narrow. Most of the volume, selected within a self-gravitating object, is not likely to be self-gravitating on much smaller scales. Fragmentation occurs in dense sub-regions that occupy a small fraction of the ``parent'' volume.

\begin{figure}
    \centering
    \plotonesize{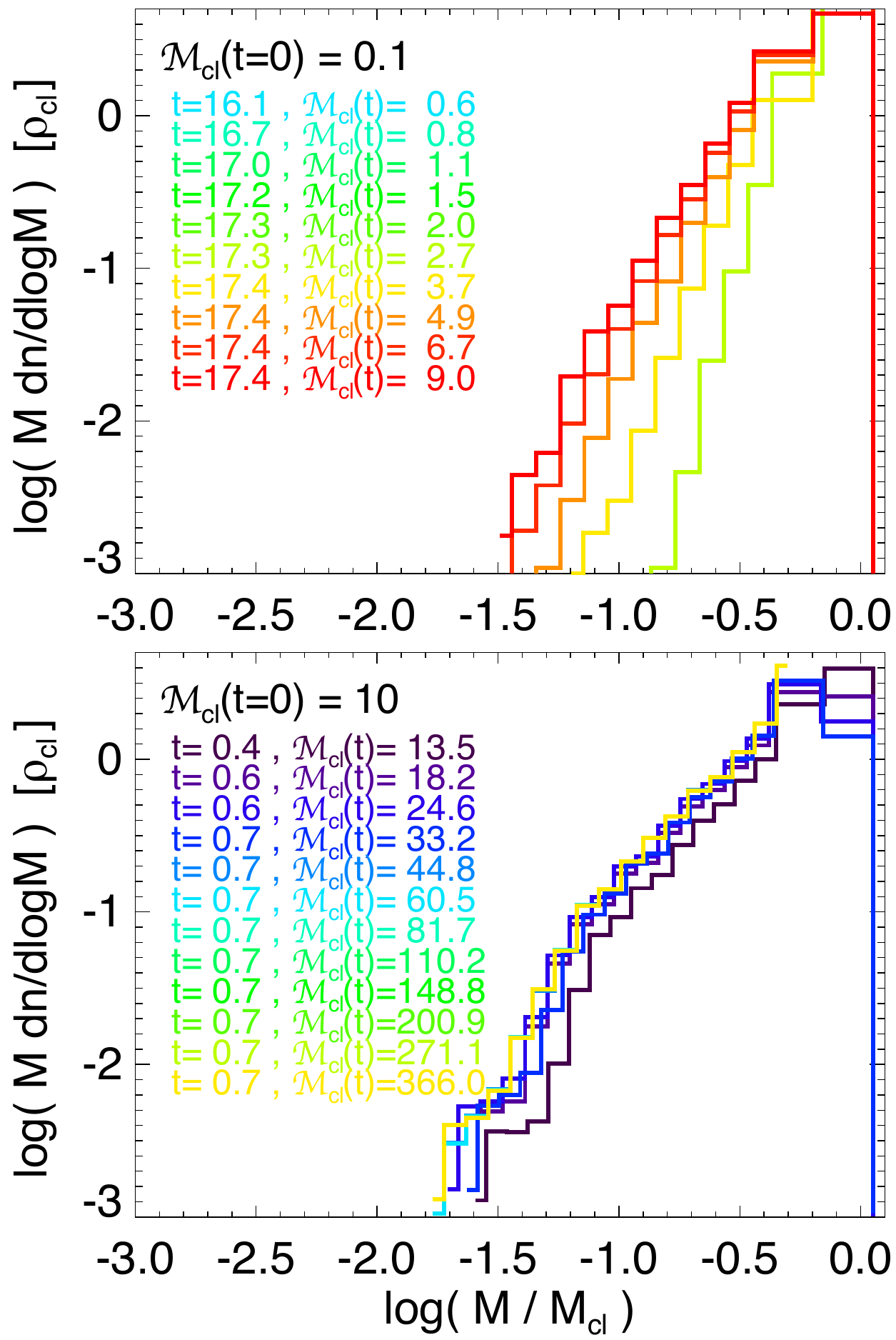}{0.95}
    \caption{Fragmentation mass spectrum (i.e.\ last-crossing mass spectrum) developed as a function of time inside an initially non-fragmenting, self-gravitating cloud (i.e.\ within a previous last-crossing ``region'') as that region collapses (following \S~\ref{sec:collapse:poly} at constant virial parameter). We show the MF at several times (each labeled $t$ in units of $t_{0}$ (Eq.~\ref{eqn:collapse.t.defn}), with the cloud-averaged Mach number $\Machcloud(t)$ increasing as the region collapses), until the entire volume has become self-gravitating on some sub-scale. Mass is in units of the total cloud mass $M\cloudsub$. 
    We compare our standard (isothermal) model, but consider a cloud with initial cloud-scale Mach number $\Machcloud(t=0)=0.1$ ({\em top}) and $10$ ({\em bottom}). The former undergoes essentially no fragmentation until a very large number of dynamical times have passed and the cloud has contracted to where $\Machcloud\gtrsim 1-3$ (collapse by a factor $R\cloudsub\sim 0.1\,R_{0}$). Almost all fragmentations occur near the ``parent'' cloud scale. A cloud which begins highly super-sonic (despite having no initial fragmenting sub-regions) develops fragmentation almost immediately as it collapses.
    \label{fig:core.vs.time}}
\end{figure}

\vspace{-0.5cm}
\subsection{Correlation Functions}
\label{sec:results.corr.fn}

Figs.~\ref{fig:corr.fn.varmass}-\ref{fig:corr.fn.fixmass} show the predicted three-dimensional autocorrelation functions ($\xi(r)$ from Eqs.~\ref{eqn:xi.last.defn}-\ref{eqn:xi.cm.defn}) for first and last-crossing objects, as a function of the parameters varied in Figs.~\ref{fig:mr.scalings}-\ref{fig:mfs} above. 

We first consider $\xi(r)$ for objects with the ``characteristic'' mass at which the mass density $M\,{\rm d}n/{\rm d}\log{M}$ in the corresponding last/first-crossing MF peaks. This is generally close to $M_{\rm sonic}$ and $M_{\rm maximal}$ for last/first crossings, respectively. With this choice, the correlation functions are nearly universal. Their normalization depends only weakly on varied parameters, and the shape is almost constant; $\xi(r)$ is a near power-law, but with some run in slope from a shallower $\xi(r)\propto r^{-1}$ at small separations to a steeper $\xi(r)\propto r^{-2}$ at large separations. 

If we instead consider a constant mass at which to evaluate $\xi(r\,|\,M)$, we see a stronger dependence on the varied parameters; the change in clustering reflects the ``position'' in the MF relative to its peak, then, rather than an intrinsic difference in the clustering structure around the characteristic mass.

We can gain considerable insight from the closed-form solution for a linear barrier, derived in \paperthree: 
\be
\label{eqn:corr.linear}
\xi_{\rm MM}^{\ell}(r,\,M\,|\,B=B_{0}+\mu\,S) = \frac{\exp{{\Bigl(}{B_{M}^{2}}\,{[S_{M}+S_{M}^{2}/S(r)]^{-1}}{\Bigr)}}}{\sqrt{1-[S(r)/S_{M}]^{2}}}-1
\ee
where $S_{M}=S(M)=S(R[M])$ and $B_{M}=B(M)$ for a last-crossing of scale $M[R]$. 
Comparing this to the solutions in Figs.~\ref{fig:corr.fn.varmass}-\ref{fig:corr.fn.fixmass} shows it is a very good approximation.

At large separations $r$ (where $S(r)\ll1$), this becomes $\xi(r\,|\,M)\approx (B_{M}/S_{M})^{2}\,S(r) = \nu_{M}^{2}\,S(r)$. Since we calculate $\xi$ at the characteristic mass of the MF, we are implicitly choosing a point near the minimum in $\nu$, so $\nu_{\rm M}\approx\nu_{\rm min}\sim1$ in each case, hence we obtain a similar normalization for $\xi(r)$. But when we choose different masses, we systematically alter the normalization with $\nu_{M}$; more rare systems (higher-$\nu$) give a correlation function with systematically higher normalization. In this large-separation limit, the shape of $\xi(r)$ simply traces the variance as a function of scale in the density field, with some normalization ``bias'' (that is identical in meaning here to the bias parameter of cosmological correlation functions). At large scales (above the injection scale or characteristic scale $h$ set by the maximum velocity dispersions) this declines as $r^{-2}$ because $S(r)$ is suppressed by the angular momentum barrier in Eq.~\ref{eqn:S.R} (Fig.~\ref{fig:mr.scalings}; a falloff of this nature if required to ensure mass conservation). 

At separations $r\ll h$, Eq.~\ref{eqn:corr.linear} becomes $\xi(r)\approx\exp{(\nu_{M}^{2}/2)}/\sqrt{2\,(1-S(r)/S_{M})}$. The normalization again is just a function of $\nu_{M}$. In \paperthree\ (Eq.~34 therein) we show that expanding this about the sonic scale gives an approximate power-law $\xi(r)\propto r^{\chi}$ over the turbulence-dominated intermediate regime, where $\chi$ smoothly varies from $-2$ at large separations to a smaller value $\sim -(13+p\,(p-6))/(4\,(1-p))$ below $\sim h$. So the slope of $\xi(r)$ at $r\lesssim h$ is set by the turbulent power spectrum controlling the run in the density fluctuations with scale, and is systematically steeper at lower-$p$. Finally $\xi(r)$ steepens as it approaches the sonic length and $S(r)\rightarrow S_{M}$, because there is little range for independent random fluctuations to act. 

In Fig.~\ref{fig:corr.fn.varmass}, we only see a large change in $\xi(r)$ when the global Mach number $\Machdisk$ is subsonic. This is also where we saw the MF in Fig.~\ref{fig:mfs} become exponentially suppressed. This relates to the $\nu_{M}$ dependence discussed above. Recall, for the same linear barrier that predicts Eq.~\ref{eqn:corr.linear}, we have $\flast(M) = (\mu/\sqrt{2\pi\,S_{M}})\,\exp{(-\nu_{M}^{2}/2)}$. So at scales $r \lesssim h$ (where $S(r)\sim S_{M}$), we expect $\xi(r\,|\,M)\propto \exp{(\nu_{M}^{2}/2)}\propto \flast(M)^{-1}$, i.e.\ the clustering amplitude is inversely proportional to the predicted number density! 

Indeed most of the mass dependence and differences in $\xi(r)$ in Fig.~\ref{fig:corr.fn.fixmass} can be explained by a near-universal number density-bias relationship, specifically (for a linear barrier) 
\begin{align}
\nonumber \xi_{\rm MM}^{\ell}(r\,|\,M) &\approx \frac{\mu\,\flast^{-1}(M)}{\sqrt{2\pi\,[S_{M}-S(r)]}} \\ 
&=\frac{{|}{{\rm d}B_{M}}/{{\rm d}\ln{M}}{|}}{\sqrt{2\pi\,[S_{M}-S(r)]}}\,{\Bigl(}V_{M}\,\frac{{\rm d}n(M)}{{\rm d}\ln{M}} {\Bigr)}^{-1}
\label{eqn:numden.bias}
\end{align}
with $V_{M}\equiv M/\rhocrit(M)$.

Physically, fragmentation on small scales is highly clustered because most of the power in the density fluctuations comes from large scales (see \paperthree). Near the sonic length (the characteristic scale of the last-crossing distribution), the probability of an independent fluctuation suddenly arising, localized on small scales, sufficient to cross the barrier, is extremely small. Indeed the fact that it is the characteristic ``last-crossing'' mass {\em implies} that such fluctuations must be vanishingly rare. Thus last-crossings depend on fluctuations on larger scales to ``seed'' most of the necessary density fluctuation, i.e.\ they ``live'' inside larger collapsing objects (as with e.g.\ protostellar cores seeded inside of GMCs). This is true for any plausible bulk properties of the turbulence (any turbulent spectral index $p\ge 1$). The probability of multiple sub-objects forming inside a given region is largely set by the amplitude of the larger-scale fluctuation on the scale of that region.

For this reason, the {\em shape} of the correlation function, as we see in Eq.~\ref{eqn:numden.bias}, is almost entirely set by the run in $S(r)$, which in the proper dimensionless units (rescaled between the maximal instability scale and sonic scale) is nearly universal (with only a weak logarithmic dependence on the shape of the turbulent power spectrum or polytropic index). 

The {\em amplitude} of clustering depends inversely on the average number density, or more specifically from Eq.~\ref{eqn:numden.bias} the volume filling factor, of objects of a given mass, since the higher this is, the ``easier'' it is for totally independent fluctuations localized on a given scale to produce collapsing objects of a given mass (hence it is less dependent on parent-scale fluctuations).

\begin{figure}
    \centering
    \plotonesize{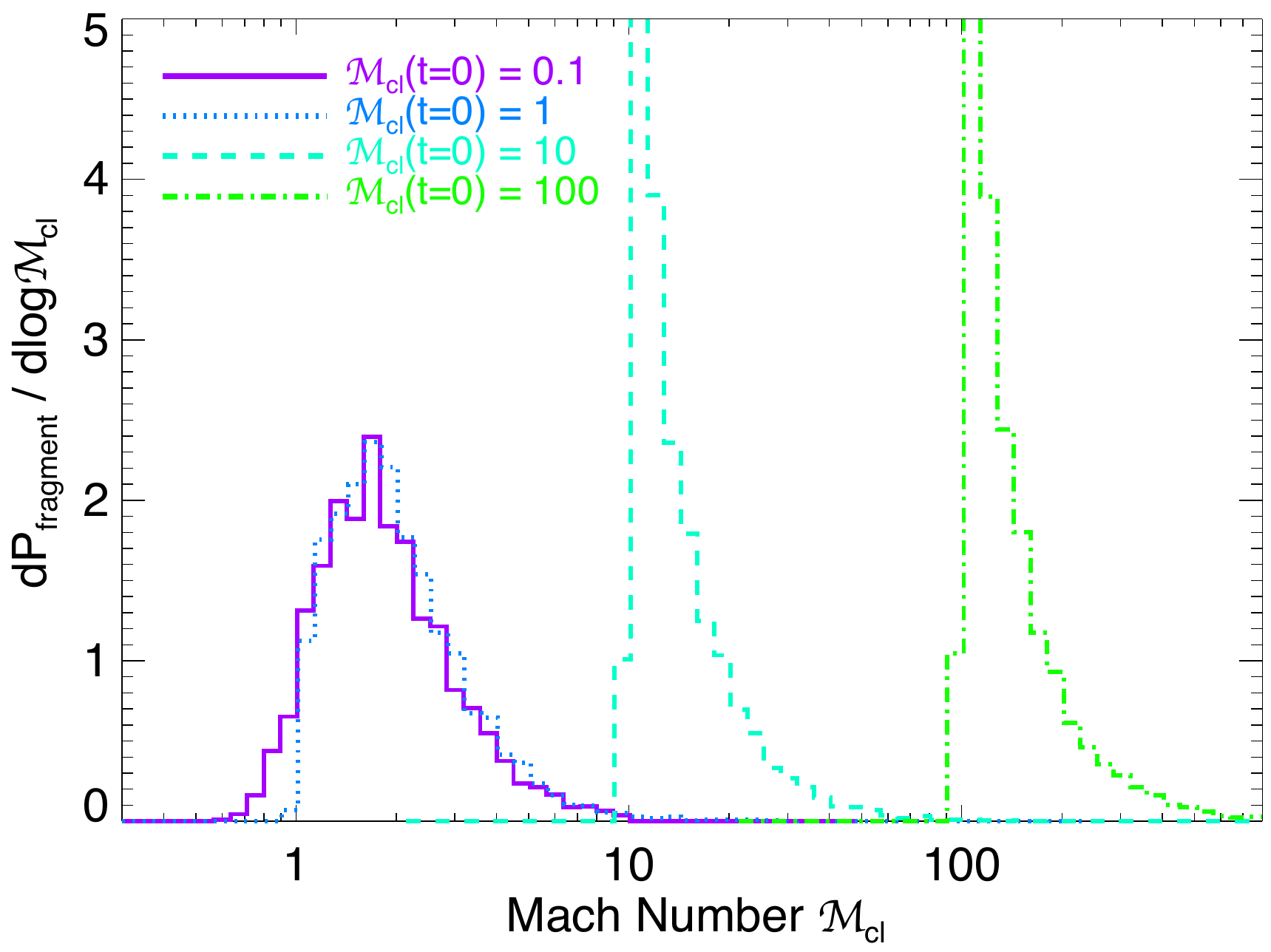}{0.95}
    \caption{Distribution (mass-weighted) of instantaneous Mach numbers $\Machcloud$ at which an initially non-fragmenting, isothermal self-gravitating cloud will fragment as it contracts (as in Fig.~\ref{fig:core.vs.time}). We consider the two cases in Fig.~\ref{fig:core.vs.time} and plot the average fraction of mass which would fragment and collapse per logarithmic interval in $\Machcloud$ (which is directly related to the time since the cloud began to contract and/or total contraction factor), as well as two additional cases. Clouds which begin sub-sonic all contract until $\Machcloud\sim1-3$ (independent of the initial $\Machcloud$) then fragment. Clouds which begin with $\Machcloud\gg1$ begin to fragment very rapidly. 
    \label{fig:core.vs.time.machdist}}
\end{figure}

\vspace{-0.5cm}
\subsection{Non-Gaussianity, Intermittency, and Correlated Turbulent Fluctuations}
\label{sec:nongaussian:text}

In Appendix~\ref{sec:nongaussian}, we consider in detail how different models for non-Gaussian and/or correlated structures in the density field affect the mass and correlation functions predicted here. This is parameterized in our models via either our approximate intermittency treatment (\S~\ref{sec:intermittency}) or the non-Gaussian and correlated fluctuation structure when $\gamma\ne1$ (\S~\ref{sec:frag:poly}).

For realistic intermittency levels and equations-of-state, we find that the effects on the mass functions stemming specifically from non-Gaussian statistics and complicated correlations in the density field are relatively minor (generally equivalent to small changes in the Mach number or $Q$; see Figs.~\ref{fig:mf.intermittency.structfn}-\ref{fig:mf.intermittency.convolved}). We see in Figs.~\ref{fig:corr.fn.varmass}-\ref{fig:corr.fn.fixmass} that the correlation functions predicted are also not very sensitive to the inherent correlation structure of the different scale-modes in the turbulence. 

For a detailed discussion, see \S~\ref{sec:nongaussian}. But this is generally true because the most important parameters in our model (the width of the density distribution, and the run of this width with scale), follow from the lowest-order moments of the density field -- they are essentially set by the power spectrum, which varies only weakly with these higher-order effects. 

We stress that even our simplest models ($\gamma=1$ and $\beta=1$), where we assume uncorrelated Fourier fluctuations, produce large non-zero correlation functions. In other words, the assumption of uncorrelated fluctuation phases in Fourier space is {\em not} equivalent to an assumption of no real-space correlations in the density field. In fact, if we directly calculate the mass or density autocorrelation function from these models, we obtain a prediction in good agreement with that directly measured in full numerical simulations in \citet{vazquez-semadeni:2001.nh.pdf.gmc}, without needing to invoke any additional correlation structure in the field. The same is true for the well-studied cosmological case: models with un-correlated Fourier modes still produce a real-space mass auto-correlation function with significant structure!

\begin{figure}
    \centering
    \plotonesize{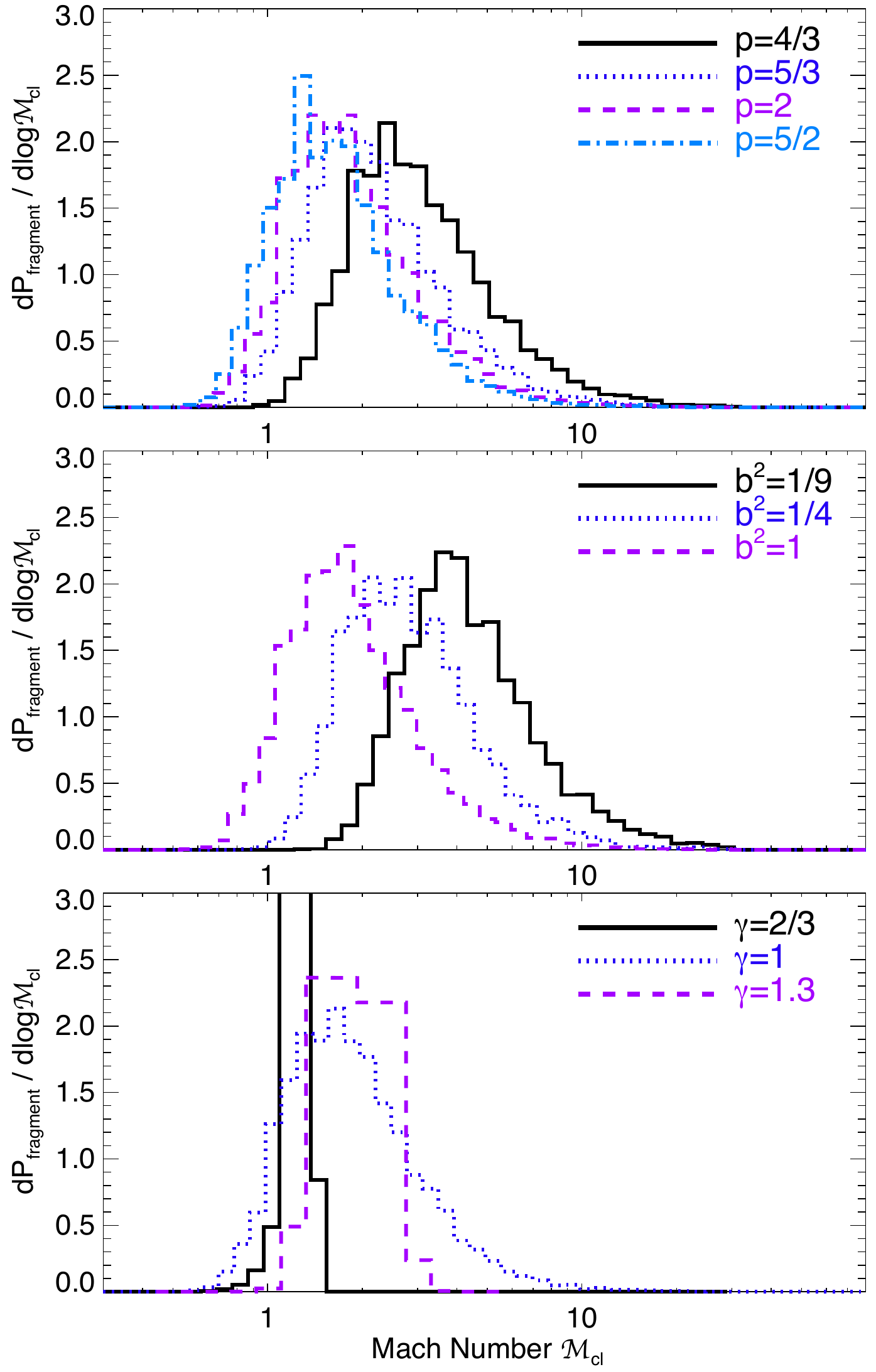}{0.95}
    \caption{Distribution of Mach numbers at which an initially non-fragmenting self-gravitating cloud will fragment, as in Fig.~\ref{fig:core.vs.time.machdist}, for otherwise identical initial conditions but varied model parameters as labeled. All clouds here begin with $\Machcloud=0.3$, but all sub-sonic initial conditions give nearly identical results for each parameter value (and initial $\Machcloud\gg1$ always produces very rapid fragmentation, as in Fig.~\ref{fig:core.vs.time.machdist}). Changing the power spectrum $p$ has weak effects, though very shallow spectra show slightly slower fragmentation. Changing the value $b$ of compressive-to-total fluctuations has the largest effect; collapse requires supersonic {\em compressive} Mach numbers, i.e.\ $b\,\Machcloud\gtrsim1-3$. Changing the polytropic index to either softer or stiffer values leads to a more narrow range of $\Machcloud$ where fragmentation occurs (for different reasons), although the timescale and collapse factor associated with the same $\Machcloud$ is very different for different $\gamma$.
    \label{fig:core.vs.time.mdistvparam}}
\end{figure}

\vspace{-0.5cm}
\section{Time Dependence}
\label{sec:frag:time}

Thus far, all of our calculations have concerned instantaneous properties at a fixed moment in time (assuming fully-developed turbulence). In this section, we extend the models to approximate the time-dependent evolution of turbulent density fields and fragmentation cascades.

\vspace{-0.5cm}
\subsection{In a Stationary Background}
\label{sec:frag:time:general}

First consider a system with constant mean background conditions (average density and steady-state turbulent power spectrum). 

{\em If} the density field is strictly lognormal with un-correlated fluctuations on different (Fourier) scales (as we assume in our simplest model), then we show in \paperone\ that the evolution of the modes of the density field must obey a generalized Fokker-Planck equation. For a (zero mean) Gaussian variable $x$ with variance $S_{x}$ the PDF to find the system with value $x$ at time $t$ given an initial $x_{0}$ at $t_{0} = t-\Delta\,t$ is then\footnote{\label{foot:lagrangian.structfn.constraint} As shown in \paperone, in the limit of small $\Delta\,t$, Eq.~\ref{eqn:timeevol} represents the only form that the evolution of $P(x,\,t)$ can take if we require that $P$ is exactly Gaussian, modes are uncorrelated, $S_{x}$ is conserved in ensemble average (true if the turbulence is steady-state), and the growth in variance between $x$ and $x_{0}$ is independent of our choice of integration step-size.}
\begin{align}
\label{eqn:timeevol}
P(x,\,t)\,{\rm d}x &= \frac{1}{\sqrt{2\pi\,\tilde{S}_{x}}}\,
\exp{{\Bigl(}-\frac{(x-\tilde{x}_{0})^{2}}{2\,\tilde{S}_{x}} {\Bigr)}}\,{\rm d}x \\ 
\nonumber \tilde{S}_{x} &\equiv S_{x}\,[{1-\exp{(-2\,[t-t_{0}]/\tau)}}]\\ 
\nonumber \tilde{x}_{0} &\equiv x_{0}\,\exp{(-[t-t_{0}]/\tau)}
\end{align}

Here, $\tau$ is the correlation time\footnote{By definition, the Lagrangian correlation amplitude between modes decays with $e$-folding time $\tau$.}; equivalently, it is the timescale at which the variance in $x(t)$ with respect to $x_{0}$ grows, or the timescale for turbulent mixing of a non-diffusive passive scalar (see \S~\ref{sec:intermittency:motivation}). This is measured in numerical simulations \citep[see e.g.][]{klessen:2000.pdf.supersonic.turb,ostriker:2001.gmc.column.dist,kitsionas:2009.grid.sph.compare.turbulence}, which show it is approximately the crossing time 
\be
\tau \approx \tilde{\tau} \, t_{\rm cross} = \tilde{\tau}\,R/\langle v_{t}^{2}(R) \rangle^{1/2}
\ee
with $\tilde{\tau}\approx1$ a constant (\citealt{pan:2010.turbulent.mixing.times} find $\tilde{\tau}\approx 0.90-1.05$ over the range $\mathcal{M}\sim1.2-10$; \citealt{zhou:2010.scalar.field.logpoisson.pdfs} obtain similar results from experiment in sub-sonic cases).

Incorporating this approximate form for time dependence into our Monte Carlo approach is straightforward, since we can simply evolve the contribution from each mode, i.e.\ each $\Delta \delta_{j}$ in Eq.~\ref{eqn:trajectory.sum}, according to Eq.~\ref{eqn:timeevol}. For a ``timestep'' $\Delta t$, this is equivalent to taking 
\begin{align}
\label{eqn:delta.timeevol}
\Delta \delta_{j}(t+\Delta t) &=  \Delta \delta_{j}(t)\,\exp{(-\Delta t/\tau)} \\ 
\nonumber & + \mathcal{R} \, \sqrt{\Delta S\,(1-\exp{(-2\,\Delta t/\tau)})}\\ 
\nonumber & \approx 
\nonumber \Delta \delta_{j}(t)\,(1-\Delta t/\tau) + \mathcal{R}\,\sqrt{2\,\Delta S\,\Delta t/\tau}
\end{align}
where $\mathcal{R}$ is a Gaussian random number with unity variance, and the latter equality is the series expansion for small differential timesteps $\Delta t$. 

If we assume that the statistics of density fluctuations follow those of compressive (longitudinal) velocity differences, Eqs.~\ref{eqn:timeevol}-\ref{eqn:delta.timeevol} yield Lagrangian velocity structure functions  (moments of the distribution of velocity differences measured for Lagrangian fluid parcels as a function of time separation) $S_{n}^{L}(\Delta t)\equiv \langle | {\bf v}(t) - {\bf v}(t + \Delta t)|^{n}\rangle \propto \Delta t^{n/2}$, identical to the \citet{kolmogorov:turbulence} model \citep[see][]{yakhot:2008.lagrangian.structfn}. For \citet{kolmogorov:turbulence} this follows from the assumption that the statistics are strictly scale-free in the inertial range (independent of integral scale and viscosity); but this is equivalent to our derivation assuming un-correlated modes.$^{\ref{foot:lagrangian.structfn.constraint}}$ However, as in the time-independent case, experiments indicate that intermittency violates these assumptions and produces distinct structure functions, which we discuss below.

\vspace{-0.5cm}
\subsubsection{Generalization for Polytropic Equations of State}
\label{sec:frag:time:general:poly}

It is straightforward to generalize this for a polytropic equation of state, using the approximation from \S~\ref{sec:frag:poly:walk} that the distribution is locally Gaussian with $\Delta S(R,\,R-\Delta R)\rightarrow \Delta S(R,\,R-\Delta R,\,\rho)$. We still need to solve the nonlinear Langevin equation, but can follow the same procedure as before, simply replacing steps in spatial scale with steps in time.

\begin{figure}
    \centering
    \plotonesize{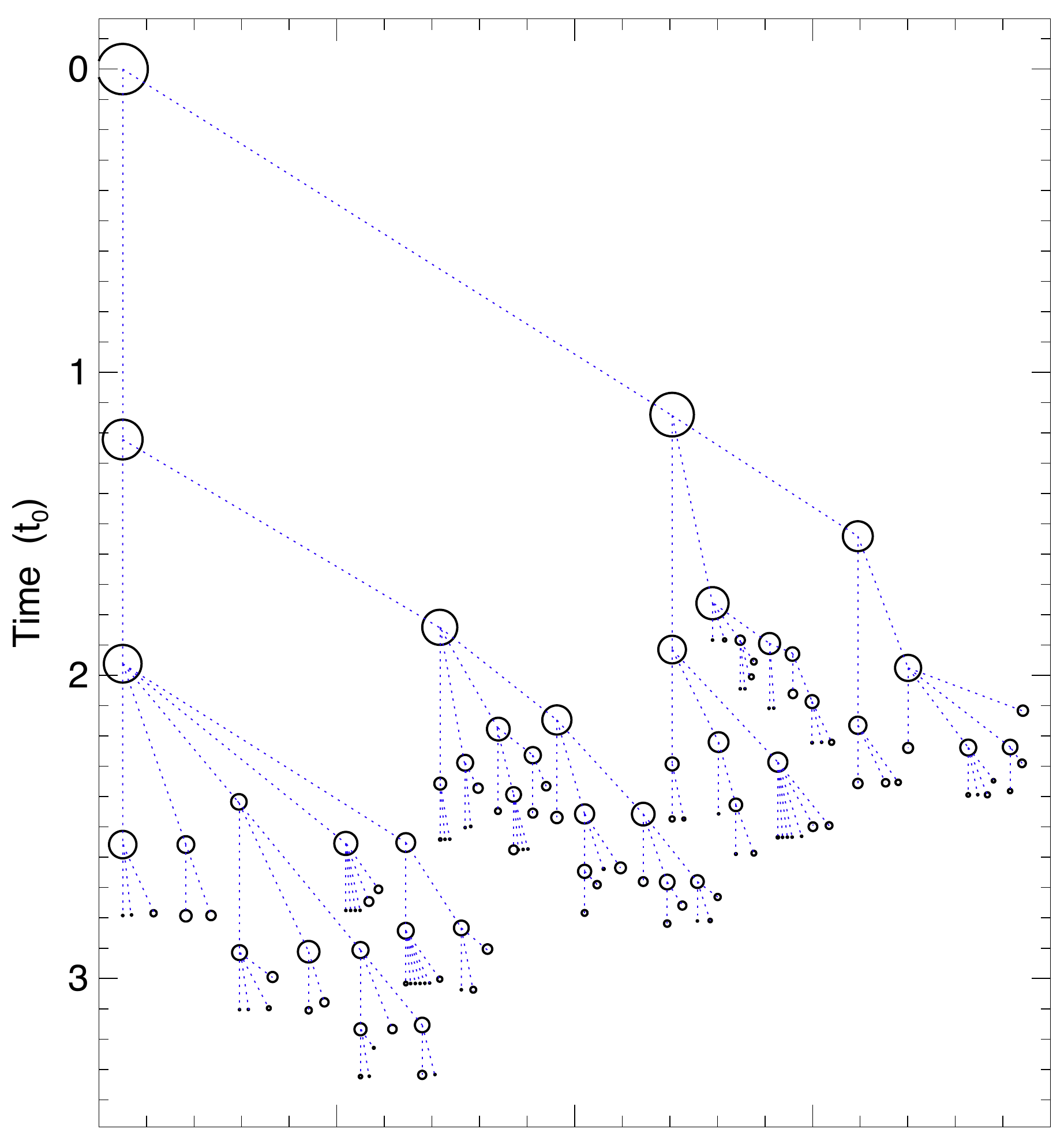}{0.99}
    \caption{Illustration of a ``fragmentation tree,'' constructed as described in \S~\ref{sec:frag.trees}, for a collapsing cloud (as \S~\ref{sec:collapse:poly}) which is initially a last-crossing with $\Machcloudzero=1$ in our standard model. The cloud begins with some initial mass $M\cloudsub$ at $t=0$, with no self-gravitating (fragmenting) sub-regions. Time since $t=0$ (in units of $t_{0}$ from Eq.~\ref{eqn:collapse.t.defn}, approximately the cloud crossing time) is on the vertical axis. Each new circle (``branch'' end) represents the formation of a new self-gravitating sub-region (last-crossing) within the cloud that then internally collapses and fragments, forming at the time where the circle is placed (and connected via the branch to its ``parent'' and descendant sub-clouds). Circle size scales with the log of each sub-region/fragment mass, from $0.01\,M\cloudsub$ (the minimum value plotted) to $M\cloudsub$. 
    \label{fig:frag.tree}}
\end{figure}

\vspace{-0.5cm}
\subsubsection{Generalization for Intermittent Turbulence}
\label{sec:frag:time:general:int}

Generalizing this for intermittent turbulence as we approximate it in \S~\ref{sec:intermittency} is more complicated, because the statistics follow a discrete Poisson distribution rather than a continuous distribution. 

Consider for simplicity the purely quantized log-Poisson case. Recall, when stepping from one scale to another, we have $\rho_{R_{2}}=\beta_{\rho}^{m}\,(R_{1}/R_{2})^{\gamma^{\prime}}\,\rho_{R_{1}}$, where $m$ a Poisson random variable with mean $\lambda$, and $\gamma^{\prime}$ is related directly to the variance ($\lambda$). We can write the step as 
\begin{equation}
\Delta s= \ln{(\rho_{R_{2}}/\rho_{R_{1}})} = \lambda_{\Delta S}\,(1-\beta_{\rho}) + m\,\ln{\beta_{\rho}}
\end{equation}
with $P(m)=(\lambda_{\Delta S}^{m}/m!)\,\exp{(-\lambda_{\Delta S})}$ and $\lambda_{\Delta S}=\Delta S/(\ln{\beta_{\rho}})^{2}$. This suggests that we can think of the change in density across a given scale for a trajectory as the cumulative result of a Poisson rate process. In each timestep $\Delta t$, we can generate a Poisson $\Delta m$ with mean and variance $\lambda(\Delta t,\,\Delta S) = (\Delta S/(\ln{\beta_{\rho}})^{2})\,(2\,\Delta t/\tau)$, and take $m$ to be the sum of $\Delta m$ (i.e.\ result of the time-integral) over the correlation time, i.e.\ from $t=t-\tau/2$ to $t$ (or $N=\tau/(2\,\Delta t)$ steps):\footnote{We include the factor of $2$ here so that $\tau$ has an identical meaning as in the Gaussian case: starting from $\Delta s=0$, the variance for an ensemble of trajectories will grow in the linear regime ($t-t_{0}\ll \tau$) at the same average rate as in either case, and the correlation between fluctuations will decay with the same average correlation time.}
\begin{align}
\nonumber \Delta s_{j}(t+\Delta t) &= \sum_{i=j-\tau/(2\,\Delta t)}^{j}\,{\Bigl[} \frac{2\,\Delta t}{\tau}\,\lambda_{\Delta S}\,(1-\beta_{\rho}) + \Delta m_{i}\,\ln{\beta_{\rho}}{\Bigr]}\\
P(\Delta m_{i}) &= \frac{1}{\Delta m_{i}!}\,{\Bigl(} \frac{2\,\Delta t}{\tau}\,\lambda_{\Delta S} {\Bigr)}^{\Delta m_{i}}\,\exp{{\Bigl(}-\frac{2\,\Delta t}{\tau}\,\lambda_{\Delta S}{\Bigr)}}
\end{align}

In this particular formulation, then, the variation $\Delta S$ is contributed by some ``number of events'' in a Lagrangian volume within a correlation time, with expectation value $\lambda_{\Delta S} = \Delta S/(\ln{\beta_{\rho}})^{2}$. For larger $|\ln{\beta_{\rho}}|$ (i.e.\ more highly intermittent turbulence), the ``number of events'' is smaller, hence larger non-Gaussian corrections appear; while as $\beta_{\rho}\rightarrow1$ the variance is assumed to be distributed over an infinite number of small events, leading (via the central limit theorem) to Gaussian statistics.\footnote{If we allow for continuous exponential damping of fluctuations (on a correlation time) and integrate over sufficiently large times, rather than simply counting fluctuations in a fixed time window, we recover the thermodynamic intermittency models discussed in Appendix~\ref{sec:appendix:alt.intermittency}, which are qualitatively similar but converge to the lognormal statistics more rapidly.}

This form for time evolution is motivated by cascade models (\S~\ref{sec:intermittency}) for the intermittent Lagrangian moments and structure functions of the compressive velocities ($S_{n}^{L}(\Delta t)\equiv \langle | {\bf v}(t) - {\bf v}(t + \Delta t)|^{n}\rangle$), which appear to also follow a log-Poisson hierarchy \citep[even in highly compressible turbulence; see e.g.][]{konstandin:2012.lagrangian.structfn.turb} as do the Eulerian structure functions  \citep{dubrulle:logpoisson,shewaymire:logpoisson}. Similar velocity scalings can be derived from symmetry considerations based on the Navier-Stokes equations \citep{yakhot:2004.phase.corr.enstrophy.decay,yakhot:2008.lagrangian.structfn}. Applied to the velocity moments, the above scaling predicts $S_{n}^{L}$ comparable to numerical simulations \citep[][]{konstandin:2012.lagrangian.structfn.turb}. Moreover, direct observations of density fluctuations in the solar wind exhibit Lagrangian structure functions with consistent log-Poisson scalings \citep[critically, with consistent parameters in density and velocity fluctuations, although in somewhat better agreement with the alternative intermittency model in Appendix~\ref{sec:appendix:alt.intermittency}; see][and references therein]{marsch:1994.multifractal.solar.wind.fluct.scalar.vel,sorriso-valvo:1999.solar.wind.intermittency.vs.time}. This also appears broadly consistent with the evolution of scalar concentration distributions, which exhibit non-Gaussian features in simulations \citep{pan:2010.turbulent.mixing.times}, though further study is needed.\footnote{As noted in \S~\ref{sec:nongaussian:text} \&\ \ref{sec:nongaussian}, since our predictions depend primarily on the second-order and lower moments of the density field, the exact form of the intermittency model we adopt makes little difference provided that the variance (second-order Lagrangian structure function) obeys a similar linear scaling in time ($S_{2}^{L}\sim \Delta t$) to our assumption. This is equivalent to extended self-similarity, so is true in nearly all multi-fractal models.}

\vspace{-0.5cm}
\subsection{Spherical Collapse of a Barotropic Cloud}
\label{sec:collapse:poly}

We now consider the evolution of fields in time in a collapsing (or expanding) background. The specific case of interest for many applications in gravo-turbulent fragmentation is a contracting, self-gravitating cloud or ``clump.''

Consider a spherical, marginally self-gravitating overdensity or cloud, with mass $M\cloudsub$, radius $R\cloudsub$, cloud-scale averaged rms turbulent velocity $\vtcloud(R\cloudsub)$, average sound speed at the mean cloud density $\cscloud(\rhocloud)$, and virial parameter $Q^{\prime}$:
\be
\cscloud(\rhocloud)^{2} + \vtcloud(R\cloudsub)^{2} = \cscloud^{2}\,(1 + \Machcloud^{2}) = Q^{\prime}\,\frac{G\,M\cloudsub}{R\cloudsub}
\ee
where $Q^{\prime}\sim1$ is constant and $\Machcloud\equiv \vtcloud(R\cloudsub)/\cscloud(\rhocloud)$ is the cloud-averaged Mach number. The total cloud energy is then  $-G\,M\cloudsub^{2}/R\cloudsub$. Assume that the gas can be approximated as a polytrope $c_{s}^{2}\propto \rho^{\gamma-1}$ over some limited dynamic range in $\rho$ (it is trivial to allow $\gamma$ to change as we follow collapse). This gives 
\be
\cscloud^{2} = \cscloudzero^{2}\,{\Bigl(}\frac{R\cloudsub}{R_{0}}{\Bigr)}^{-3\,(\gamma-1)}\equiv \cscloudzero^{2}\,\xcloud^{-3\,(\gamma-1)}
\ee
where $R_{0}$ and $\cscloudzero$ are an initial radius and (mean) sound speed of the cloud and $\xcloud\equiv R\cloudsub/R_{0}$. The virial relation above gives 
\be
1+\Machcloud^{2} = Q^{\prime}\,\frac{G\,M\cloudsub}{R\cloudsub\,\cscloud^{2}} = (1 + \Machcloudzero^{2})\,\xcloud^{-1+3\,(\gamma-1)} 
\ee
where $\Machcloudzero\equiv \vtcloud(R_{0})/\cscloudzero$ is the initial cloud-scale Mach number.

Let us assume that there is no source to ``pump'' turbulence other than the gravitational contraction of the cloud itself; then it is well-established that the kinetic energy of the turbulence decays in about a crossing time (dissipating in shocks or via the cascade). So
\begin{align}
\frac{{\rm d}E_{t}}{{\rm d}t} &= - \frac{\eta}{2}\,\frac{M\cloudsub\,\vtcloud^{2}}{R\cloudsub/\vtcloud} \\
\nonumber &= -\frac{\eta\,Q^{\prime\,3/2}}{2}\,\frac{G^{3/2}\,M\cloudsub^{5/2}}{R_{0}^{5/2}}\,\xcloud^{-5/2}\,{\Bigl(}1 - \frac{\xcloud^{1-3\,(\gamma-1)}}{1+\Machcloudzero^{2}} {\Bigr)}
\end{align}
where $\eta\sim1$ is a constant calibrated from numerical simulations \citep[e.g.][]{kitsionas:2009.grid.sph.compare.turbulence,pan:2010.turbulent.mixing.times}. 

Since (by assumption) the cloud is dissipating and not gaining energy from outside, it must contract. If this occurs with approximately constant virial parameter $Q^{\prime}$, then the change in energy from contraction is given by 
\be
\frac{{\rm d}E_{g}}{{\rm d}t} = \frac{G\,M\cloudsub^{2}}{R\cloudsub^{2}}\,\frac{{\rm d}R\cloudsub}{{\rm d}t} = \frac{G\,M\cloudsub^{2}}{R_{0}}\,\frac{1}{{\xcloud}^{2}}\,\frac{{\rm d}{\xcloud}}{{\rm d}t}
\ee

Without external energy input, these must be equal, so (after some simplifying algebra) we obtain the following differential equation for the evolution of the cloud size: 
\be
\label{eqn:size.evol}
\frac{{\rm d}{\xcloud}}{{\rm d}\tau} = - {\xcloud}^{-1/2}\,{\Bigl(}1 - \frac{{\xcloud}^{1-3\,(\gamma-1)}}{1+\Machcloudzero^{2}} {\Bigr)}^{3/2}
\ee
where we define $\tau \equiv t/t_{0}$ and $t_{0}$ is the characteristic timescale 
\be
\label{eqn:collapse.t.defn}
t_{0}\equiv \frac{2}{\eta\,Q^{\prime\,3/2}}\,{\Bigl(}\frac{G\,M\cloudsub}{R_{0}^{3}} {\Bigr)}^{-1/2}
\ee
which is proportional to the initial dynamical time of the cloud.

\vspace{-0.5cm}
\subsubsection{Some General Consequences}
\label{sec:frag:collapse:notes}

Some general behaviors can be discerned from Eq.~\ref{eqn:size.evol}. Since $\xcloud$ and $\Machcloudzero$ are positive and $\xcloud(\tau=0)=1$, ${\rm d}{\xcloud}/{\rm d}t < 0$ and the cloud shrinks. However for $\gamma>4/3$, $1-3\,(\gamma-1)<0$, so the second term will vanish (or, if there is some overshoot, become negative and drive expansion) and the contraction will cease at 
\be
\xcloud_{\rm stable} = \frac{R_{\rm stable}}{R_{0}} = (1+\Machcloudzero^{2})^{\frac{-1}{3\,(\gamma-1)-1}};\ \ \ (\gamma>\frac{4}{3})
\ee
Physically, $c_{s}$ grows faster than $\vtcloud$ in this case, stabilizing against any further contraction (until this energy can dissipate) and fragmentation at this $\gamma$. 

For $\gamma<4/3$, however, the cloud collapses to $\xcloud\rightarrow 0$ in a finite time $t \sim [\Machcloudzero^{2}/(1+\Machcloudzero^{2})]^{1/2}\, t_{0}$. As it does so, $\vtcloud$ grows faster than $\cscloud$ so $\Machcloud$ increases as $\xcloud$ decreases. Thus {\em either} with large initial $\Machcloudzero\gg1$ or as $\xcloud\rightarrow0$, we reach $\Machcloud\gg1$. In this limit the $\xcloud^{1-3\,(\gamma-1)}$ term vanishes and ${\rm d}{\xcloud}/{\rm d}\tau \approx - {\xcloud}^{-1/2}$, i.e.\ collapse becomes scale-free. In this limit, since the cloud dynamical time at any instant is $\sim (G\,M\cloudsub/R\cloudsub^{3})^{-1/2} \propto {\xcloud}^{3/2}$, and $|{\rm d}\tau/{\rm d}\ln{{\xcloud}} |\approx {\xcloud}^{3/2}$, the cloud spends an equal number of local dynamical times at each logarithmic interval in size as it collapses (even though the collapse to ${\xcloud}=0$ proceeds in finite absolute time). And in this limit, because the turbulent term dominates, collapse becomes independent of $\gamma$. 

For isothermal gas, we can analytically solve Eq.~\ref{eqn:size.evol}: 
\begin{align}
\nonumber \tau({\xcloud})_{\gamma=1} &= 2\,{\Bigl [}\sqrt{z\,(1-z)^{-1}}-\sin^{-1}(\sqrt{z}) {\Bigr]}{\Bigr|}_{\xcloud\,(1+\Machcloudzero^{2})^{-1}}^{\,(1+\Machcloudzero^{2})^{-1}}\\
&= 2\,{\Bigl [}\frac{1}{\Mach^{\prime}}-\sin^{-1}{{\Bigl(}\frac{1}{\sqrt{1+\Mach^{\prime\,2}}}{\Bigr)}}\,{\Bigr]}{\Bigr|}_{\Machcloud}^{\Machcloudzero}
\end{align}

We can trivially follow collapse for a cloud with one $\gamma$ up to some $\rhocloud$ and/or $R\cloudsub$, then switch to a different $\gamma$ and continue to follow it \citep[see][]{jappsen:2005.imf.scale.thermalphysics,hennebelle:2009.imf.variation,veltchev:2011.frag}. Thus the derivation above can be used to treat collapse with a complicated barotropic or bivariate equation of state $c_{s}(\rho,\,R)$. 


\vspace{-0.5cm}
\subsubsection{Modification to Density Field Evolution}
\label{sec:frag:time:cloud}

As a parent cloud globally contracts, then, all scales $R$ shrink, and densities correspondingly scale up. However, recall that at fixed $\Machcloud$ the dimensionless quantities that determine the last-crossing distribution depend only on the {\em relative} values $R/R\cloudsub$ and $\rho/\rhocloud$, which are independent of the cloud shrinking. The only effect, then, on the fragmentation process in these dimensionless units in a collapsing or expanding system is through the evolution in $\Machcloud$. 

So we should consider all quantities in terms of the scaled variables $R/R\cloudsub$ rather than the absolute $R$: essentially following ``Lagrangian'' modes $k/k\cloudsub=\,$constant. Up to variation in $\Machcloud$, then, these Lagrangian modes follow the evolution derived above for a stationary system, and we can trivially construct the trajectories $\delta(x\equiv R/R\cloudsub)$ at each time and compare them to the dimensionless $B(x \equiv R/R\cloudsub)=\,$constant. 

However as the cloud contracts $\Machcloud$ does evolve; we therefore must take $B=B(x,\,\Machcloud)$ and $\Delta S=\Delta S(x,\,\Machcloud)$ at each time. This leads to time-dependence in the barrier $B$ and variance $S$ at fixed $x=R/R\cloudsub$: from Eqs.~\ref{eqn:S.R.new} \&\ \ref{eqn:rhocrit.box} it follows that
\begin{align}
\nonumber \frac{{\rm d}B}{{\rm d}\tau}{\Bigr|_{x}} &= \frac{x^{p-1}-\tilde{\rho}_{\rm crit}^{\gamma-1}}{(1+\Machcloud^{2})\,[(2-\gamma)\,\tilde{\rho}_{\rm crit}^{\gamma-1}+\Machcloud^{2}\,x^{p-1}]}\,\frac{{\rm d}\,\Machcloud^{2}}{{\rm d}\tau} + \frac{1}{2}\frac{{\rm d}S}{{\rm d}\tau}{\Bigr|_{x}}\\ 
\frac{{\rm d}S}{{\rm d}\tau}{\Bigr|_{x,\tilde{\rho}}} &= \ln{{\Bigl[}\frac{\tilde{\rho}^{\gamma-1}+b^{2}\Machcloud^{2}}{\tilde{\rho}^{\gamma-1}+b^{2}\Machcloud^{2}\,x^{p-1}} {\Bigr]}}\,\frac{1}{(p-1)\Machcloud^{2}}\,\frac{{\rm d}\,\Machcloud^{2}}{{\rm d}\tau}
\end{align}
where ${\tilde{\rho}}\equiv \rho/\rhocloud$ and $\tilde{\rho}_{\rm crit}\equiv \rhocrit(x,\,\Machcloud)/\rhocloud$.
At fixed $x$, this means it becomes {\em easier} to cross the barrier at all scales as the cloud contracts when $\gamma<4/3$ (and the reverse for $\gamma>4/3$). If $\Machcloud\gg1$, ${\rm d}B/{\rm d}\tau\rightarrow 0$, i.e.\ the barrier becomes self-similar in the turbulence-dominated regime, while the variance $S$ systematically increases with $\Machcloud$ as $R\cloudsub$ decreases.

\vspace{-0.5cm}
\section{Results for Time-Dependent Fields}
\label{sec:timedep.results}

We now consider some basic consequences for the time-dependent fragmentation process in a time-evolving turbulent field. 

\vspace{-0.5cm}
\subsection{Global Mass Function Evolution: The Rate of Formation of Bound Structures and ``Fragments''}
\label{sec:global}

First, we consider the global evolution of the first-crossing and last-crossing distributions. However, care is needed. For the stochastic differential equations above, the {\em instantaneous} rate of change of a quantity like the mass function is always formally divergent (because the discrete ``events'' operate as delta functions). Time evolution can only be defined over some finite interval. But averaging over any finite interval requires some decision regarding both the initial conditions of the turbulence, and what happens to regions that do become self-gravitating. These choices are {\em not} unique and will lead to different conclusions.

\vspace{-0.5cm}
\subsubsection{Instantaneous ``Formation Rate'' from Smooth Initial Conditions}
\label{sec:global:formation.rate}

One example is shown in Fig.~\ref{fig:mf.vs.time}. Here we consider the rate of {\em initially} forming first/last crossings of mass $M$, in a field that develops from initially uniform density. We evolve our standard model assuming all trajectories begin at $\rho=\rho_{0}$, and immediately count the resulting bound objects the moment the trajectory has a crossing. We stress that if we began with more developed turbulence/inhomogeneity, we would obtain different results, and if we allowed the bound objects themselves to evolve, they could become more massive in time (we do not allow this, simply recording their mass when they first become self-gravitating). 

The interesting question is how the resulting rates-of-formation (${\rm d}n/{\rm d}t$) compare to the instantaneous MFs ($n(M)={\rm d}n/{\rm d}\log{M}$) in fully-developed turbulence. For first-crossings, ${\rm d}n_{f}/{\rm d}t$ is flattened relative to $n_{f}(M)$. If we simply multiply $n_{f}(M)$ by a constant (say, the inverse crossing time at the most unstable scale $R[\nu_{\rm min}]$ or, closely related, the scale where $M\,{\rm d}n_{f}/{\rm d}t$ is maximized $R_{\rm first}^{\rm max}$), we predict a ${\rm d}n_{f}/{\rm d}t\propto M^{-\alpha}$ with slope $\alpha<2$, in poor agreement with the full calculation. This is not surprising: we should expect objects to ``develop'' on something like the crossing time $\tau(R)$ {\em on each scale}, since this is the timescale for the turbulence to develop (and, formally, the correlation time over which the density field on that scale ``resets''). If instead we compare ${\rm d}n_{f}/{\rm d}t$ to $n_{f}(M)/\tau(R[M])$, the shapes are in much better agreement. But while $n_{f}(M)/\tau(R[M])$ has the right shape, it is biased towards higher-mass objects than ${\rm d}n_{f}/{\rm d}t$ by a factor $\approx2$. We obtain quite a reasonable fit (to $\approx0.1\,$dex) if we shift the MF by this factor, i.e.\ ${\rm d}n_{f}(M)/{\rm d}t\approx n_{f}(2\,M)/\tau(R[2\,M])$. We discuss this high-mass truncation below.

Based on this, we might also expect the last-crossing rate-of-formation to follow the ``static'' $n_{\ell}(M)$ divided by the local crossing time $\tau(R[M])$; but it does not. Instead, the rate of formation ${\rm d}n_{\ell}(M)/{\rm d}t$ is much closer to the instantaneous MF $n_{\ell}(M)$ multiplied by a constant. This constant is approximately the inverse crossing time at the maximum of the {\em first-crossing} rate-of-formation function (i.e.\ $1/\tau[R_{\rm first}^{\rm max}]$). 

Why does this occur? Recall, as discussed above regarding the predicted correlation functions, isolated last-crossings (i.e.\ last-crossings arising independently on small scales) are very rare, since most of the power in density fluctuations is on large scales. This means that last-crossings are ``seeded'' by first crossings on a larger scale. This will occur whenever ${\rm d}B/{\rm d}S\gg1$. Consider the following (extreme) illustrative example of this. On some large scale $R_{0}$, the variance in density fluctuations contributed by modes with $k\sim 1/R_{0}$ is a large $S_{0}\gg1$. Between this and some much smaller scale $R_{1}\ll R_{0}$, the variance contributed from modes on each scale drops very sharply, so the total contributed is some very small $S_{1}\ll 1\ll S_{0}$. The value of the log density $\delta=\ln{(\rho/\rho_{0})}$ at $R_{1}$ is therefore a gaussian random variable with variance $S=S_{0}+S_{1}$, the convolution of the contribution from large scales $R_{0}$ and smaller scales $R_{1}<R<R_{0}$. But since $S_{0}\gg S_{1}$, $\delta(R_{1})$ is hardly altered by the modes on scales $R<R_{0}$ -- i.e.\ it is almost entirely set by the fluctuations around $R_{0}$. But assume that the barrier $B(R)$ rises smoothly and continuously at $R<R_{0}$, so it continues to increase while $S$ increases very weakly on smaller scales (${\rm d}B/{\rm d}S\gg1$). If $B(R)$ is a monotonic function of $R$, then the ``location'' $R$ where this trajectory $\delta$ will have its last crossing just depends on $\delta$; but the value of $\delta$ is almost entirely determined by the fluctuations at $\sim R_{0}$. In other words, the number density and scale of last-crossings is set not by fluctuations near that scale (which evolve on the crossing time at that scale), but by fluctuations on much larger scales, and so are ``seeded'' on these longer timescales.

\vspace{-0.5cm}
\subsubsection{The Integrated ``Collapse Rate''}
\label{sec:global:collapse.rate}

An important integral quantity is the rate, integrated over the mass function, at which mass is collapsed into bound objects. This is straightforward to calculate for a given assumed initial condition (e.g.\ the constant-density ICs above) by simply integrating over the rate-of-formation in the MF.

We can derive a reasonable approximation to the full numerical solution with the following simple arguments. Recall, the first-crossing mass is concentrated near the most unstable scale $R[\nu_{\rm min}]$. At a given instant near this scale, we can approximate the MF with the solution for a linear barrier (Eq.~\ref{eqn:ffirst.linear.barrier}); we can also note from Fig.~\ref{fig:mr.scalings} that around this scale $\rhocrit$ is approximately constant (${\rm d}B/{\rm d}S$ is small, or in the linear barrier $B_{0}\gtrsim \mu\,S$). Integrating the mass density down to some $S_{\rm max}$ with these approximations gives a total collapsed mass density $F\equiv \rho_{\rm collapsed}/\rho_{0}={\rm erfc}[B_{0}/\sqrt{2\,S_{\rm max}}]$ (just the integral over the ``tail'' of the log-normal). Using $R[\nu_{\rm min}]\approx h$ (Eq.~\ref{eqn:r.maximal}) in the expression for $\rhocrit$ (Eq.~\ref{eqn:rhocrit}) gives $B_{0}\approx \ln{[Q\,\tilde{\kappa}^{-1}(1+\tilde{\kappa}^{2})]}$. In Fig.~\ref{fig:mr.scalings}, we see that the $S_{\rm max}$ of interest (the largest $S$ before $\nu$ begins to rise rapidly) corresponds to the maximum in ${\rm d}S/{\rm d}\ln{R}$ (below this scale, $S$ grows very slowly, so ${\rm d}B/{\rm d}S$ becomes large and MF is suppressed). For $S(R)$ of the form in Eq.~\ref{eqn:S.R}, where we can write ${\rm d}S/{\rm d}\ln{R} = \ln[1 + b^{2}\,v_{t}^{2}/(c_{s}^{2}+\kappa^{2}\,k^{-2})] = \ln[1 + b^{2}\Machdisk^{2}\,|kh|^{1-p}/(1 + \tilde{\kappa}^{2}\,(1+\Machdisk^{2})\,|kh|^{-2})]$, this is approximately $S_{\rm max}\approx \ln[1 + 0.5\,b^{2}\,\Machdisk^{2}/(\tilde{\kappa}^{2}\,(1+\Machdisk^{2}))^{(p-1)/2}]$. 

Finally, since the density field is randomized on a correlation time about equal to the turbulent crossing time, the rate of ``filling'' the high-density tail of the PDF is $\approx F/t_{\rm cross}(R[\nu_{\rm min}])\approx F/t_{\rm cross}(h)$, where $t_{\rm cross}(h) = h/v_{t}(h) = \sigma_{g}[h]\,\Omega^{-1}/v_{t}[h]=({1+\Machdisk^{2}})^{1/2}/(\Machdisk\,\Omega)$. This gives the approximate integrated collapse rate 
\begin{align}
\label{eqn:fcoll.time}
&\frac{1}{M_{\rm tot}\,\Omega}\frac{{\rm d}\,M_{\rm collapsed}}{{\rm d}t} \approx \\
\nonumber &{\hfill}\frac{\Machdisk}{\sqrt{1+\Machdisk^{2}}}\,
{\rm Erfc}{\Bigl[}\frac{\alpha\,\ln{{[}{Q\,(1+\tilde{\kappa}^{2})}/{\tilde{\kappa}}{]}}}
{\sqrt{2\,\ln{(1+0.5\,b^{2}\Machdisk^{2}/[{\tilde{\kappa}^{2}\,(1+\Machdisk^{2})}]^{(p-1)/2})}}} {\Bigr]}
\end{align}
where the $(1+\tilde{\kappa}^{2})/\tilde{\kappa}$ term should be replaced with ``$1$'' if a ``hard'' upper barrier/injection scale is used instead of an angular momentum barrier (i.e.\ if $\tilde{\kappa}=0$), and $\alpha\approx1.0-2.0$ is a fudge factor representing the detailed integration over the shape of $\nu(R)$. 

In super-sonic turbulence ($\Machdisk\gg1$), the collapse fraction is order-unity (tens of percent) per dynamical time, with weak dependence on $Q$ or $\kappa$. And it is a logarithmically {\em increasing} function of $\Machdisk$ at fixed $Q$, while it decreases with $\Machdisk$ if $Q\propto \Machdisk$ and $Q\ge1$ (as in a $Q>1$ disk primarily supported by supersonic turbulence). This is discussed in more detail in \paperone; there this is shown to agree well with the results of a number of idealized ``turbulent box'' simulations (Fig.~11 therein).\footnote{Specifically, Eq.~\ref{eqn:fcoll.time} appears to agree well with the results of idealized forced turbulent box simulations from $\Mach\sim1-100$ with pure hydrodynamics plus gravity \citep{vazquez-semadeni:2003.turb.reg.sfr} as well as hydro, gravity, and MHD \citep{padoan:2011.new.turb.collapse.sims}, as well as galactic simulations with cooling, star formation, and feedback from stellar evolution \citep{hopkins:clumpy.disk.evol,hopkins:rad.pressure.sf.fb,hopkins:stellar.fb.mergers,hopkins:fb.ism.prop}, with the simulations scattering by about a factor $\sim2$ about the predicted relation. The scatter appears to be largely related to the non-constant rates of collapse in the simulations.}

In sub-sonic turbulence, however, the collapse fraction is cut off rapidly (at fixed $Q$), in line with the exponentially suppressed MF we saw previously. Expanding the above gives a mass fraction collapsed per dynamical time of $\approx \Machdisk\,\exp{(-\tilde{\nu}^{2}/2)}/(\tilde{\nu}\,\sqrt{2\pi})$ where $\tilde{\nu}\equiv \Machdisk^{-1}\,\ln{[Q\,(1+\tilde{\kappa}^{2})/\tilde{\kappa}]}$. Note, however, that if there were {\em no} angular momentum support ($(1+\tilde{\kappa}^{2})/\tilde{\kappa}\rightarrow1$) and the disk were marginally stable or unstable ($Q\le1$), then the suppression is far weaker, with collapsed fraction $\sim \Machdisk$ per dynamical time. The exponential suppression in sub-sonic turbulence relies critically on a thermal or angular momentum barrier sufficient to stabilize the disk in the ``traditional'' Toomre sense.

\vspace{-0.5cm}
\subsection{``Top-Down'' vs.\ ``Bottom-Up'' Structure Formation} 
\label{sec:top.vs.bottom}

Another key general question in the time histories of forming objects is whether bound objects evolve in ``top-down'' or ``bottom-up'' fashion. In ``top-down'' formation, objects initially form at large masses and then fragment into smaller objects -- so much so that the ``first-crossing'' mass is typically depleted in time. In ``bottom-up'' formation, small objects assemble hierarchically into larger objects (this is the standard ``hierarchical'' structure formation in cosmology). 

We examine this in Fig.~\ref{fig:mf.vs.time.duration}. In order for bound objects to ``evolve'' in any sense, of course, we must allow them to have a finite lifetime (unlike the case in Fig.~\ref{fig:mf.vs.time}, where bound objects are immediately removed). We therefore follow the same procedure as Fig.~\ref{fig:mf.vs.time}, assuming all trajectories grow from uniform initial $\rho=\rho_{0}$, but in this case when a structure becomes bound, we allow it to ``survive,'' and continue to evolve (freely allowing the density modes to evolve in time) for some timescale $t_{\rm bound}$ equal to a fixed multiple of the crossing time at the first-crossing scale ($t_{\rm cross}(R_{\rm first})$). When this time expires, the object is removed and the trajectory is ``replaced'' with one at the mean density. Since the lifetimes are finite, we plot the time-averaged MF that results in each case. 

Unsurprisingly, when $t_{\rm bound}\ll t_{\rm cross}$, the MF normalization is suppressed (since objects form on finite times but have short lifetimes), and the shape is close to the ``rate of formation'' in Fig.~\ref{fig:mf.vs.time}, because objects have little time to evolve after they initially cross the barrier. As we increase $t_{\rm bound}$, objects are not only more common (being longer-lived), but allowing the turbulence to evolve within such objects for a longer time, we see the first-crossing MF extend to higher ``maximum'' masses, while the last-crossing MF extends to lower ``minimum'' masses. This continues, eventually saturating when $t_{\rm bound} \gg t_{\rm cross}$, in which limit we recover the analytic prediction for the MF in fully-developed (non-time dependent) turbulence. It appears, therefore, that with increasing time for bound objects to evolve, structures ``develop'' at high masses in ``bottom-up'' fashion, and at low masses in ``top-down'' fashion.

Recall for lognormal density distribution, the integrated mass fraction which is self-gravitating on any given scale (independent of first/last-crossings) is ${\rm Erfc}(\nu_{m}/\sqrt{2})$, where we use $\nu_{m}\equiv B_{m}/S$ to denote the variable $\nu$ with respect to the {\em mass-weighted} density PDF (so $B_{m}=\ln{(\rhocrit/\rho_{0})} - S/2$, as opposed to the volume-weighted PDF where $B_{v} =\ln{(\rhocrit/\rho_{0})} + S/2$). If all trajectories begin at $\nu(R)\approx0$, then the dispersion in $\nu$ grows at each radius with the crossing time $\tau(R)$. So if we truncate at some fixed fraction of the mass distribution, the ``upper envelope'' $\nu_{\rm upper}$ grows with ${\rm d}\nu_{\rm upper}/{\rm d}t \approx \tau(R)^{-1}$. In some differential timestep ${\rm d}t$, then, this moves the $\nu_{\rm upper}(R)$ crossed by the upper envelope by some $\Delta \nu$ corresponding to a shift $\Delta R$ or $\Delta M$ in the size/mass scale which is now collapsing at the threshold rate. The ``collapse scale'' therefore migrates at a rate 
\be
\frac{{\rm d}R_{\rm coll}}{{\rm d}t}= \frac{{\rm d}\nu_{\rm upper}}{{\rm d}t}\,{\Bigl(}\frac{{\rm d}\nu_{m}}{{\rm d}R}{\Bigr)}^{-1} \approx \frac{v_{t}(R_{\rm coll})}{R_{\rm coll}}\,{\Bigl(}\frac{{\rm d}\nu_{m}}{{\rm d}R}{\Bigr)}_{R_{\rm coll}}^{-1}
\ee
If ${\rm d}R_{\rm coll}/{\rm d}t<0$, then collapse cascades ``top-down'': large scales become self-gravitating first and, with time, smaller and smaller scales subsequently develop self-gravitating structure within these parent scales. This is the typical Jeans-style fragmentation cascade. 

But if ${\rm d}R_{\rm coll}/{\rm d}t>0$, then collapse flows ``bottom-up,'' with larger and larger ``parent structures'' collapsing subsequently. This corresponds, physically, to two processes: (1) accretion of larger-scale material by the central, already-bound clump when compressions make the ``parent'' region sufficiently dense, and (2) clump-clump mergers. Indeed, in simulations of galaxy structure that resolve only the largest molecular clouds and in which those clouds are predominantly supported by rotation, they grow in time via mergers and accretion as predicted here \citep[see e.g.][]{dobbs:2008.gmc.collapse.bygrav.angmom,tasker:2009.gmc.form.evol.gravalone,hopkins:fb.ism.prop}.

Since ${\rm d}\nu_{\rm upper}/{\rm d}t\sim \tau^{-1}$ is always positive, the sign of ${\rm d}R_{\rm coll}/{\rm d}t$ is the same as that of ${\rm d}\nu_{m}/{\rm d}R$. This gives a time-independent criteria for ``bottom-up'' growth, as a function of $B$ and $S$ alone:
\be
\frac{{\rm d}R_{\rm coll}}{{\rm d}t} > 0 \ \ \ {\rm iff} \ \ \ \frac{{\rm d}B}{{\rm d}S}< \frac{1}{2}\,\frac{B}{S}
\ee
For a linear barrier ($B=B_{0}+\mu\,S$), this is just $S<B_{0}/\mu$. So on sufficiently large scales (small $S$), a system with an approximately linear barrier grows ``bottom-up,'' while on smaller scales, structures grows ``top-down.'' 

This corresponds well with what we see in Fig.~\ref{fig:mf.vs.time.duration}. Below $\sim h$, $S$ increases with decreasing $R$ (because it measures the sum of contributions from fluctuations on different scales), while $B$ also increases (in either the turbulence-dominated or thermal-dominated regimes), hence ${\rm d}B/{\rm d}S$ is large and positive (see Fig.~\ref{fig:mr.scalings}). Since (as we show earlier) the linear-barrier approximation is not bad, it must be the case as $S$ becomes larger that ${\rm d}B/{\rm d}S \gg B/(2\,S)$, so structure grows ``top-down.'' 

Above $\sim h$, we see in Fig.~\ref{fig:mr.scalings} a sharp transition where ${\rm d}B/{\rm d}S\ll0$. The presence of an angular momentum barrier ($\kappa$) makes collapse more difficult on large scales ($B(R)$ increases with $R$, while $S$ must decrease to ensure mass conservation). In this limit, ${\rm d}B/{\rm d}S < B/(2\,S)$ so structure growth must proceed ``bottom-up.''\footnote{In the case of cosmological structure, $B\approx$\,constant, so the condition for ``bottom-up'' growth is always satisfied and hence traditional ``hierarchical'' structure formation emerges, even though the form of the time evolution of perturbations (linear growth vs.\ the stochastic growth here) is quite different.} Since this follows the sign of ${\rm d}\nu/{\rm d}R$, the scale where growth transitions from ``top-down'' to ``bottom-up'' is just the maximal instability (quasi-Toomre) scale defined above. If the system is turbulent and has no angular momentum barrier, then all scales collapse ``top-down'' -- this is the standard Jeans collapse-style form of fragmentation.

\vspace{-0.5cm}
\subsection{Time-Dependent Fragmentation of Collapsing ``Last-Crossing'' Clouds}
\label{sec:cloud.collapse}

Now we consider the behavior of a collapsing polytropic cloud. If the cloud already has ``last-crossings'' contained within it, each such crossing collapses on its own appropriate timescale, while the ``parent'' cloud continues to collapse itself. We can track each such collapsing object separately, so the key problem that needs to be addressed is the evolution of a cloud which is collapsing on its last-crossing scale, i.e.\ we can without any loss of generality focus on the problem of a single cloud which is a ``last-crossing'' within some parent cloud. Shortly below, we will show how multiple such histories can be ``stitched together'' into a full history for a parent cloud with multiple sub-clouds. 

\vspace{-0.5cm}
\subsubsection{Methodology}
\label{sec:cloud.collapse.methods}

To do so, we consider the collapsing cloud background as derived in \S~\ref{sec:frag:subcmf}, i.e.\ a collapsing spherical polytropic cloud. For simplicity, we neglect the angular momentum ($\kappa$) term, since this is usually negligible at the scale of last-crossings in any case. 

First, we construct a cloud with appropriate initial conditions. Since the largest collapsing scale without fragmenting sub-scales in the initial conditions is the last-crossing scale, we have by definition $\rho=\rhocrit$ at $R=R\cloudsub=R_{\rm last}$, i.e.\ $Q^{\prime}=1$ in \S~\ref{sec:frag:subcmf}. We need to assign some initial conditions for the density field on scales $R<R\cloudsub$. One option would be to assume the initial density is uniform on all scales. But this is not consistent, strictly speaking, with clouds that form out of a ``parent'' density field. To match that case, we can simply generate an ensemble of trajectories in the usual Monte Carlo fashion, chose those which have a last crossing on some scale, and (for now) simply discard all information on scales above the last-crossing scale (retaining the trajectory ``below'' the last crossing as the ``initial'' density field $\delta(R)$). We can also generate an ensemble of these trajectories to fully sample the density PDF, by repeatedly drawing trajectories with the same last-crossing scale. In practice, it is numerically less expensive (but mathematically identical) to simply generate trajectories starting at a parent scale of $R\cloudsub$, and iteratively discard and re-generate those which include subsequent crossings on smaller scales. 

This defines the initial density field. For a given $\Machcloudzero$ of the cloud (the mach number at the initial last-crossing/cloud scale), then, the cloud evolves according the equations above (\S~\ref{sec:collapse:poly}). The absolute size shrinks, and the density field undergoes the random walk described above, but with $\Machcloud$ increasing and this introducing corresponding evolution in the variance and barrier with time. 

As the density field evolves while the cloud collapses, eventually, new crossings will appear on scales $R<R\cloudsub$. Each of these should itself collapse on its own appropriate timescale; this is just a new collapsing ``last-crossing'' cloud with its own appropriate initial Mach number. We therefore record the point at which each such crossing appears (along with the mass, size scale, and Mach number of the crossing), and remove that trajectory. The surviving trajectories continue to evolve as before. 

Because, for the simple model considered here (for $\gamma<4/3$), the cloud will collapse to arbitrarily small size ($\Machcloud\rightarrow \infty$), and even though collapse proceeds in finite absolute time it spends an equal number of dynamical times in every logarithmic interval in radius, all trajectories will eventually develop a new crossing. We therefore evolve each last-crossing until all trajectories have crossed. 

Finally, note that in this scenario, for a given (fixed) $\gamma$, $b$, and $\beta$, the only parameter that enters into the equations governing the cloud evolution is the initial Mach number $\Machcloudzero$ of the cloud (which, for a fixed parent disk in which the cloud forms, is equivalent to the initial size and/or mass scale of the cloud). So the ``fragmentation history'' we calculate in this manner is a one-parameter family, in $\Machcloudzero(R\cloudsub[t=0])$. 

\vspace{-0.5cm}
\subsubsection{Results}
\label{sec:cloud.collapse.results}

Figs.~\ref{fig:core.vs.time}-\ref{fig:core.vs.time.machdist} show the resulting fragmentation histories for last-crossing clouds with different initial Mach numbers, in our standard model. 

First, we consider last-crossing clouds with initial Mach numbers $\Machcloudzero=0.1,\,10$, and follow their fragmentation in time. We specifically plot the MF of fragments produced following the cloud collapse, at several times. Most fragmentations initially occur near the ``parent'' cloud scale -- this is the maximal instability scale within the collapsing cloud (in Eq.~\ref{eqn:r.maximal}, the maximal instability scale for $p=2$ and $\tilde{\kappa}=0$ is $\approx0.7\,R\cloudsub$ for $\Machcloud\sim0.1-1$). This is consistent with both the clustering and time evolution seen thus far -- isolated last-crossings on small scales do not typically develop ``independently,'' but are ``seeded'' by larger scales. Thus even allowing last-crossings to collapse evolve, and fragment in time, their characteristic last-crossing scale remains similar at each stage of collapse (``evolving'' only logarithmically as each sub-cloud continues in its own collapse). Last-crossings on a wide range of scales, evident in the global last-crossing MF, are a consequence of modes ``frozen in'' by fluctuations on much larger scales, rather than the evolution of a cloud or clouds with fixed large-scale modes (the isolated collapsing cloud considered here).

Note also that when the initial Mach number is small $\Machcloudzero\lesssim1$, it takes considerable time for fragmentation to develop. This does not occur until the cloud has collapsed so far that $\Machcloud\gtrsim1$, i.e.\ until the turbulence is super-sonic; at this point new fragmentation events begin to develop. For initially low Mach numbers, this means fragmentation does not develop until $R\cloudsub(t)\ll R\cloudsub(t=0)$. 

Since most subsequent crossings develop in a narrow range of scales, we can temporarily neglect the mass dependence and consider the distribution in time (or, equivalently, in cloud Mach number or size relative to the initial value) of crossings, shown in Fig.~\ref{fig:core.vs.time.machdist}. Here we clearly see the effects above: for clouds which are initially sub-sonic or transsonic, fragmentation does not develop until $\Machcloud\sim2-3$; once clouds are super-sonic the process is self-similar, with fragmentations occurring over a factor of $\sim2$ in $\Machcloud$ (before all trajectories have crossed/fragmented). This corresponds to a factor of $\sim5-10$ collapse in $R\cloudsub$, in a time about equal to the cloud free-fall time.

In Fig.~\ref{fig:core.vs.time.mdistvparam}, we examine how this fragmentation history depends on other properties of the turbulence. We show for each case the Mach number dependence, but vary $\gamma$, $b$, and $p$ (because these are defined as last-crossings at the cloud scale, $Q$ factors out completely). In general we see that the histories are quite similar independent of these choices, albeit with some interesting subtle differences. We have also performed this calculation for different values of $\beta$, but we find almost no effect here for any plausible values.

There is no exact closed-form analytic expression for the time-dependence of fragmentation, but we can approximate it via the following. The collapsed fraction is $\int {\rm d}S^{\prime}\ffirst(S^{\prime})$; approximate $\ffirst(S)$ with the solution for a linear barrier, assume that the density distribution can reach equilibrium at each logarithmic stage of collapse, and approximate $S$ by Taylor expanding near the scales $R\sim R\cloudsub$ where fragmentation events occur. This leads to:
\begin{align}
\frac{{\rm d}f_{\rm frag}}{{\rm d}t} &\approx 2\,B_{0}\,e^{-B_{0}\,(1+2\,S_{0}^{-1})}\,{\Bigl(}{S_{0}^{-2}}\,\frac{{\rm d}S_{0}}{{\rm d}t}{\Bigr)} \\ 
\nonumber S_{0} &\equiv {(3-p)^{-1}}\,\ln{(1+b^{2}\,\Machcloud^{2})} 
\end{align}

For small $\Machcloud\ll1$, this simplifies to approximately ${\rm d}f_{\rm frag}/{\rm d}\tau \sim \Machcloud^{-1}\exp{[-2\,b^{-1}\,\Machcloud^{-2}]}\ll1$ (where again $\tau\equiv t/t_{0}$ defined in Eq.~\ref{eqn:collapse.t.defn}). Fragmentation is exponentially suppressed, because there is simply not enough variation in the density field on scales below the last-crossing ($\Machcloud\ll1$, so $S\ll 1$ within the cloud); in other words, the last-crossing scale is ``frozen'' and though it collapses, it cannot develop new self-gravitating structure on small scales. 

For large $\Machcloud\gtrsim1$, ${\rm d}f_{\rm frag}/{\rm d}\tau \sim \psi^{-2}\,\exp{(-\psi^{-1})}$ where $\psi = \tau\,[1-3\,(\gamma-1)]/[2\,B_{0}\,(3-p)]$; the collapse rate grows exponentially in time then declines when most trajectories have already crossed. Integrating, the fragmented fraction $f_{\rm frag}\approx \exp{(-B_{0}\,[1-\tau_{0}/\tau])}$ where $\tau_{0} \equiv 2\,(3-p)\,[1-3\,(\gamma-1)]^{-1}$. So the entire cloud fragments within a time $t = t_{0}\,\tau_{0}$ that is a couple free-fall times. Within that time, the Mach number grows by a logarithmic factor $\Delta \ln{\Machcloud} \approx 3-p$, and the cloud collapses in size by a factor $\Delta\ln{R\cloudsub}\approx \tau_{0}$. So shallower power spectra and stiffer equations of state produce fragmentation over a wider range in time and Mach numbers, by suppressing the power in density fluctuations on the maximal instability scale, but the dependence is only manifest in a weakly dependent pre-factor (because turbulence dominates in this regime, and so collapse is approximately scale-free).

\vspace{-0.5cm}
\section{Constructing ``Fragmentation Trees''}
\label{sec:frag.trees}

In \paperone, we outline how time-evolution models of the sort described above can be used to construct ``merger and fragmentation trees'' analogous to dark matter halo merger trees in the extended Press-Schechter formalism. This is just the procedure we used above in \S~\ref{sec:global} to follow the development of first-crossings with finite lifetimes. As in \S~\ref{sec:global}, we considered in \paperone\ only the global statistics of first-crossings, and did not allow for self-gravitating clouds to undergo any time-dependent evolution (contraction or successive fragmentations on the last-crossing scale). Having further developed these models to allow for the full fragmentation cascade, we now outline the methodology for construction of a complete ``fragmentation tree.''

(1) Select (or construct) the initial conditions within the ``cloud.'' This may vary depending on the problem of interest. But they can be generated as described in \S~\ref{sec:cloud.collapse.methods}, for either smooth initial conditions or ``fully developed'' turbulence within the cloud.

(2) Evolve the system forward by one time step $\Delta t$ (evolving the density PDFs per \S~\ref{sec:timedep.results}), and calculate the new MF ${\rm d}N/{\rm d}M=V_{\rm cl}\,{\rm d}n/{\rm d}M$ for last-crossings below the parent cloud scale at the updated time (as \S~\ref{sec:cloud.collapse.results}). This can be done either via Monte Carlo methods or (if the barrier is sufficiently simple) analytic solution. Additional physics in the model can be applied at this stage.\footnote{Many possibilities for this are presented in \paperone; for example one could make additional assumptions about time-dependent variations in the equation of state, damping of the turbulent velocities, continuous accretion or expulsion of gas from the clouds.}

(3) Draw a population from this MF. There are several ways to do this, outlined in \citet{somerville:merger.trees}. For example, one can bin the entire MF in narrow mass bins and draw from a Poisson distribution with $\langle N(M,\,M+\Delta M) \rangle = \Delta\,M\,{\rm d}N/{\rm d}M$, but one must be careful in doing so to construct the ``draw'' so that mass is conserved. A simpler (although slightly less accurate) approximation is to assume all fragmentations are binary, for sufficiently small timesteps $\Delta t$. Draw an $M_{1}$ from the MF by matching a uniform random variable to the cumulative $N(>M)$ distribution, and divide the total mass $M$ into $M_{1}$ (which is the ``fragmented'' sub-unit) and $M-M_{1}$ (the remainder of the ``parent''). The latter continues on the same ``collapse trajectory'' as before the fragmentation event. However, if $M_{1}>M/2$ (assuming the draw was from a random volume element, as in this case), then $M-M_{1}<M/2$ and the ``residual'' is not self-gravitating independent of the unit $M_{1}$, so the cloud should be treated as if no fragmentation has occurred (i.e.\ $M-M_{1}$ is still bound directly to $M_{1}$, so the total collapsing mass is still $M$). 

If desired, however, the smaller region $M_{1}$ now collapsing can be followed, speeding up the collapse rate, but tracking the residual mass $M-M_{1}$ still bound but now above the last-crossing scale. This will then represent a subsequent, time-dependent accretion onto the unit $M_{1}$ after it collapses. Repeating this process for all ``levels'' in a collapsing first-crossing unit, every bound parcel of gas can be assigned a last-crossing sub-unit to which it is most tightly bound, and a timescale on which it will accrete onto this sub-unit. Thus accretion information is implicitly included in the fragmentation tree.

(4) Remove each ``fragmented'' sub-unit. These are now independent collapsing sub-clouds, each of which can be treated with steps (1)-(4) in the same manner as the original parent cloud. The mass remaining in the ``original'' cloud can then be used to repeat steps (1)-(4) as well, until some desired threshold is reached (e.g.\ mass or size falling below some threshold, or some physical criteria such as a density threshold where collapse is completed or shut down). In sending the new clouds to (1), either new initial conditions can be generated, or the updated density field (with fragmenting/last-crossing sub-regions removed) can be used as the initial condition for the next timestep. Note, though, that the collapse rate for the ``parent'' follows the same ``track'' as for the mass with which the cloud originally formed (despite the removal of the sub-unit), since the fragmented sub-units still exist within the parent region and contribute to its total mass.

This can be followed for an ensemble of clouds (corresponding to e.g.\ the ``initial'' last-crossing population instantaneously predicted for fully developed turbulence within a disk and/or more massive initial cloud), sampling different Monte-Carlo histories at a given set of initial conditions as well as the dependence on those conditions. 

Fortunately, for a polytropic gas with constant index $\gamma$, in a spherical cloud collapsing as described in \S~\ref{sec:collapse:poly}, the dimensional parameters of the cloud (size, mass, density, etc.) factor out and the statistics of the fragmentation histories form a one-dimensional family in the initial cloud mach number $\Machcloudzero$. So the tree construction can be simplified as follows: generate a set of trees by iterating steps (1)-(4) for an ensemble of clouds on a grid of $\Machcloudzero$ values; within each tree, when a fragment forms, record its own $\Machcloudzero^{i}=\Mach(R_{\rm fragment},\,t_{\rm fragment})$ (and absolute dimensional properties) but do not follow it further. Each such fragment can then be assigned the history of one of the grid ensemble with the same $\Machcloudzero$, and so on until some final collapse threshold is reached for all fragments. 

Fig.~\ref{fig:frag.tree} shows one illustrative example fragmentation tree, constructed as described above, for an initially $\Machcloudzero=1$ last-crossing cloud, with initial conditions corresponding to fully developed turbulence in our standard model.


\vspace{-0.5cm}
\section{Discussion}
\label{sec:discussion}

We have developed a flexible, analytic framework to understand fragmentation and development of self-gravitating structures in turbulent media. In \paperone-\paperthree, we used the fact that the density distribution in isothermal, turbulent gas is approximately lognormal to show how the mathematical excursion-set formalism could be applied to calculate certain properties of turbulent density fluctuations. In those papers, we specifically considered the relatively simple case of a galactic disk with supersonic, isothermal hydrodynamic turbulence; nevertheless, we showed that this could explain a remarkable range of properties observed in giant molecular clouds, protostellar cores, and the ISM. 

In this paper, we generalize this in a number of important aspects. We develop a framework that allows for different levels of rotation (and disk mass profiles), complicated gas equations of state (not just barotropes but multivariate equations of state as well), magnetic fields, turbulent power spectra, turbulent driving mechanisms (e.g.\ ratio of compressive vs.\ solenoidal forcing), intermittency, non-Gaussian statistics (i.e.\ arbitrarily non-lognormal density PDFs), correlated density fluctuations on different scales, and collapsing (or expanding) backgrounds. We also generalize the theory to follow time-dependent development of fragmentation, and show how to construct ``fragmentation trees,'' analogous to cosmological ``merger trees'' for the growth of dark matter halos. 

This should enable the application of this theory to a wide variety of astrophysical contexts. As discussed in \paperone\ \&\ \papertwo, there are a number of applications to study the structure, formation, and evolution of GMCs, protostellar cores, voids, and other features in the large-scale ISM. The formalism developed here allows for a first approximation of the effects of stellar feedback changing both the thermodynamic properties and turbulence within, say, collapsing GMCs and/or cores. This is critical for a detailed calculation of the formation of stellar clusters or the development of the stellar initial mass function. Following time-dependent collapse and development of subsequent fragmentation within cores is necessary to study the formation of binary and multiple stellar systems. Allowing for non-isothermal gas, magnetic fields, and Keplerian rotation is key to generalize to the case of proto-planetary disks and questions of the role of turbulence in planet formation (either by direct collapse or by acceleration of planetesimal growth). Highly intermittent fluctuations in magnetized, barotropic media are particularly interesting for studying shocks, mixing, transport, and possibly even triggered detonations in convective stars at the late stages of their evolution. Many of these systems will be studied in more detail in future work. Our intention here is not to specify to any single astrophysical system but to develop a general mathematical framework that can be applied to a wide range of problems where turbulence is important.

Here, we focus on two key scales relevant for self-gravity: the ``first-crossing'' scale, i.e.\ the largest scale on which objects are self-gravitating, and the ``last-crossing'' scale, the smallest scale on which they are self-gravitating (below which they will not fragment), hence the smallest objects into which ``parent'' regions will fragment as they collapse. We show how a wide variety of properties of first and last-crossing ``objects'' can be analytically estimated, including: mass functions, size-mass relations, linewidth (dispersion)-mass relations, correlation functions/clustering amplitudes, ``dynamic range'' of fragmentation (the dynamic range over which typical sub-regions will undergo subsequent fragmentations as they collapse), formation rates with time, typical growth/fragmentation histories, collapse and fragmentation rates, and nature of the ``hierarchy'' of structure formation (i.e.\ whether or not these objects will tend to develop ``top-down'' via a traditional fragmentation cascade or ``bottom-up'' via mergers with other self-gravitating objects). 

To first approximation, many of these properties depend surpisingly weakly on the properties of the medium. This is because both gravity and turbulence (over the inertial range) are fundamentally scale-free processes. As a consequence, the mass functions of first and last-crossings tend towards near-universal shapes, close to Schechter functions (power-laws with exponential cutoffs). Over the power-law range the slope is close to ${\rm d}n/{\rm d}M\propto M^{-2}$, corresponding to equal mass per logarithmic interval in mass, exactly what we would expect for a completely scale-free process. First-crossing mass functions are slightly more shallow (slopes closer to $\propto M^{-1.8}$), because for any physically reasonable turbulent power spectrum there is more ``power'' in turbulent fluctuations on large scales (so small-scale high-density regions are more likely to live inside of larger-scale ``parent'' regions, rather than be isolated). This is closely related to a near-universal behavior of the correlation functions: both first and last-crossings are strongly correlated on small scales, and the correlation function has a nearly universal power-law like shape (running as $\xi(r)\propto r^{-1}$ at small scales and $\propto r^{-2}$ on larger scales). The normalization of $\xi(r)$ is a function of mass, but scales in nearly-universal fashion inversely with the abundance of a given population (Eq.~\ref{eqn:numden.bias}). Last-crossings, in particular, are always strongly clustered, because most of the power in turbulent velocity fluctuations (hence density fluctuations) is at large scales -- so collapsing, small-scale objects are preferentially formed inside larger-scale density fluctuations. Power-law correlations between mass and radius (Eq.~\ref{eqn:mass.radius.approx} and Fig.~\ref{fig:mr.scalings}; corresponding to approximately constant surface density in the turbulence-dominated regime), and between velocity dispersion and radius, naturally emerge.

There are two characteristic scales in the problem which set most of the important physics. The first is the ``maximal instability scale'' (Eq.~\ref{eqn:r.maximal}). This is the characteristic scale of first-crossings, and the scale on which the medium is most unstable to fragmentation. This has some similarities to, but is {\em not} the same as, the traditional Toomre length/mass scale (in fact it scales in opposite fashion with respect to some parameters like the Toomre $Q$). In a turbulent disk, this is near the disk scale height, or in a driven turbulent ``box'' near the driving scale. Above this scale, density fluctuations are suppressed by mass conservation and so the number density of collapsing objects is exponentially suppressed. If the presence of this scale owes to an angular momentum barrier (as in a rotating disk), then above this barrier the sense of structure formation is also reversed from ``standard'' fragmentation cascades: more massive objects are more likely to form via the hierarchical merger of incompletely collapsed (angular momentum-supported) clouds, rather than form or fragment out of spontaneous density fluctuations or larger objects. For example, many simulations of GMC formation in galaxies, which can often only resolve the most massive objects, see mergers dominating the formation of such systems, while simulations with smaller mass or force resolution see far fewer events \citep[compare e.g.][]{dobbs:2008.gmc.collapse.bygrav.angmom,hopkins:fb.ism.prop}. 

The second key scale is the sonic scale (Eq.~\ref{eqn:r.sonic}), the spatial scale below which the rms {\em compressive} Mach number becomes sub-sonic (and corresponding mass scale defined by the minimum self-gravitating mass on that spatial scale). This has some dimensional terms in common with the traditional Jeans mass, but it is not the same. This defines the characteristic scale of the last-crossing distribution. Below this scale, density fluctuations are suppressed by thermal and magnetic pressure, so once again the number density of collapsing objects is exponentially suppressed. We stress that this suppression is {\em not} because thermal and magnetic pressure suddenly become much ``better'' at resisting gravity on these scales. Rather, it is the suppression of density fluctuations, with sub-sonic Mach numbers, that is critical. 

Together, these two scales define the ``dynamic range'' of fragmentation (Fig.~\ref{fig:fragrange}). While most of the volume, under any conditions, tends to be non self-gravitating, most of the mass in self-gravitating objects is embedded in ``parent'' objects on the maximal instability scale and undergoes a fragmentation cascade down to the sonic scale. 

The most important effects of varying the global turbulent power spectrum (spectral slope, Mach number, compressive versus solenoidal forcing), global disk (or ``parent'' cloud) stability (Toomre $Q$ or virial parameter), and gas equation of state, are captured in their effects on shifting the maximal instability scale and sonic scale. In fact, most of the differences in the predicted mass functions (Fig.~\ref{fig:mfs}) and correlation functions (Fig.~\ref{fig:corr.fn.varmass}) with different model parameters can be eliminated if we simply ``renormalize'' the spatial or mass scales (``stretching'' or ``compressing'' each in scale to match the maximal instability scale and sonic scale in each case). 

This gives a simple intuition for the effects of most parameter variations. Higher Mach numbers produce fragmentation over a wider range of scales because the turbulence is super-sonic (able to produce large density fluctuations) down to smaller scales; if the global Mach numbers are sub-sonic, the total mass fraction involved in fragmentation at all becomes exponentially suppressed. A shallower turbulent power spectrum, anchored to the same large-scale Mach number, implies more power at small scales (a smaller sonic scale) and so promotes fragmentation to smaller scales as well. Increasing the global stability of the system (i.e.\ raising the ``barrier'' required for a fluctuation to collapse) leads to a narrower range around the maximally unstable scale on which collapse is likely, and suppresses the overall mass fraction involved in collapse. Since it is the compressive component of the Mach number which drives density fluctuations, changing the balance of compressive (longitudinal) vs.\ solenoidal modes is nearly degenerate with changes to the Mach number. Changing the gas equation of state, while it introduces non-Gaussian density PDFs and complicated correlations between modes on different scales, has the most important effect of shifting the sonic scale (i.e.\ fragmentation is promoted over a wider dynamic range with a softer equation of state, simply because the cascade can proceed further before hitting the sonic scale). In the super-sonic scale regime, the thermal physics of the gas has almost no effect. 

For these reasons, phenomena such as intermittency, non-Gaussian statistics, correlations between turbulent fluctuations on different scales, and anisotropic collapse are predicted to have suprisingly little impact on the statistics of fragmentation (Figs.~\ref{fig:mf.intermittency.structfn}-\ref{fig:mf.intermittency.convolved}). To the extent that there are some effects, they are largely degenerate with smaller variations in the Mach number, equation of state, or global stability of the system. 

We show that fragmentation develops globally (in an initially smooth system) starting near the maximal instability scale. First-crossings tend to form on the turbulent crossing time at each scale (Fig.~\ref{fig:mf.vs.time}). {\em If} objects have finite subsequent lifetimes (Fig.~\ref{fig:mf.vs.time.duration}), then a fragmentation cascade proceeds from the largest to the smallest scales, with the dynamic range of self-gravitating object sizes/masses increasing with that lifetime. 

Last-crossings, on the other hand, tend to be born ``fully formed'' when their ``parent'' region undergoes a density fluctuation that pushes it above first crossing. As noted above, because the variance on the characteristic scale of last-crossings (the sonic scale) is small (by definition), they only very rarely arise via an entirely local (small-scale) fluctuation, but rather ``ride on'' larger-scale, larger-amplitude density fluctuations (i.e.\ first crossings). As such, they are not formed on the {\em local} crossing or dynamical time, but rather ``seeded'' by first crossings, and so form at a rate regulated by larger-scale fluctuations in the turbulent medium (the characteristic timescale of the maximal instability scale, rather than the sonic scale). In other words, the mass spectrum of fragmentation in objects at or below the sonic scale is ``frozen in'' by fluctuations on larger scales. 

We show how to follow the development of subsequent fluctuations in these objects as they collapse. Recall, fragmentation which is ``pre-seeded'' on all scales is captured in the statistics described above; but it is possible that a last-crossing region (which at a given instant contains no fragmenting sub-regions) could develop such sub-regions later in time as it collapses. Formally speaking, for ``soft'' equations of state ($\gamma<4/3$), if collapse proceeds infinitely (to zero size) at constant virial parameter, the system must eventually fragment. However, even under these conditions, collapsing last-crossing objects which are initially at or below the sonic scale (i.e.\ have cloud-scale compressive Mach numbers $\lesssim1$) develop such fragmentation very slowly (Figs.~\ref{fig:core.vs.time}-\ref{fig:core.vs.time.mdistvparam}). In fact, significant sub-fragmentation {\em only} occurs when -- as a consequence of these assumptions (in particular collapse at constant virial parameter) -- the cloud Mach number becomes super-sonic ($\Machcloud\sim2-3$). {If} some other physics eventually enters to prevent this from occurring -- for example, if the equation of state becomes stiffer as the system collapses, or the collapse proceeds directly rather than converting all energy to turbulent motions (i.e.\ the virial parameter is not constant during collapse), then collapse will proceed to $r\rightarrow0$ within such an ``initial last-crossing'' without ever developing significant sub-fragmentation. In this case, the last-crossing distribution predicted at a given instant will indeed represent the final collapsed-fragment distribution. Thus the statistics of fragmentation, and quantities like the last-crossing distribution, are ``frozen in'' at these scales, and only weakly modified by the collapse process itself. In contrast, a cloud which is initially highly supersonic (but somehow contrived to have no self-gravitating sub-regions) will develop sub-fragmentation very rapidly (in one crossing time) as it began to collapse, independent of the turbulent power spectrum or gas equation of state. 

Our intention here is to develop and present a robust, general framework to understand ``gravo-turbulent'' fragmentation and structure formation. In future work, we will apply the results here to many specific observed systems of astrophysical interest. It is also important to test many of the assumptions and analytic calculations in the model here by comparing to fully non-linear numerical simulations and calculations, especially idealized cases (for example, an isothermal driven turbulent box with self-gravity), in which the relevant input parameters of the model (the turbulent power spectrum, gas equation of state, etc.) can be completely specified. Some particularly important questions include how the density power spectrum relates to the velocity power spectrum, how this is altered in the presence of rotation and magnetic fields, and how intermittency is manifest in the log-density field (as opposed to just the velocity field). Many of these are quite interesting questions in their own right, which should inform our general understanding of compressible turbulence. In a companion paper (in preparation), we will present a preliminary series of such comparisons useful both for developing a deeper understanding of many behaviors seen in simulations, as well as testing and calibrating the theoretical assumptions necessary here. 

\vspace{-0.7cm}
\acknowledgments 
We thank Ralf Klessen, Chris McKee, and Eliot Quataert for many helpful discussions during the development of this work. We also thank the anonymous referee for a number of insightful suggestions. And we thank Claude-Andr{\'e} Faucher-Gigu{\`e}re and Robert Feldmann for many stimulating discussions that have raised topics of future research. Support for PFH was provided by NASA through Einstein Postdoctoral Fellowship Award Number PF1-120083 issued by the Chandra X-ray Observatory Center, which is operated by the Smithsonian Astrophysical Observatory for and on behalf of the NASA under contract NAS8-03060.\\

\bibliography{/Users/phopkins/Documents/work/papers/ms}

\clearpage
\begin{appendix}

\vspace{-0.5cm}
\section{Detailed Effects of Non-Gaussianity, Intermittency, and Correlated Turbulent Fluctuations}
\label{sec:nongaussian}

\begin{figure}
    \centering
    \plotonesize{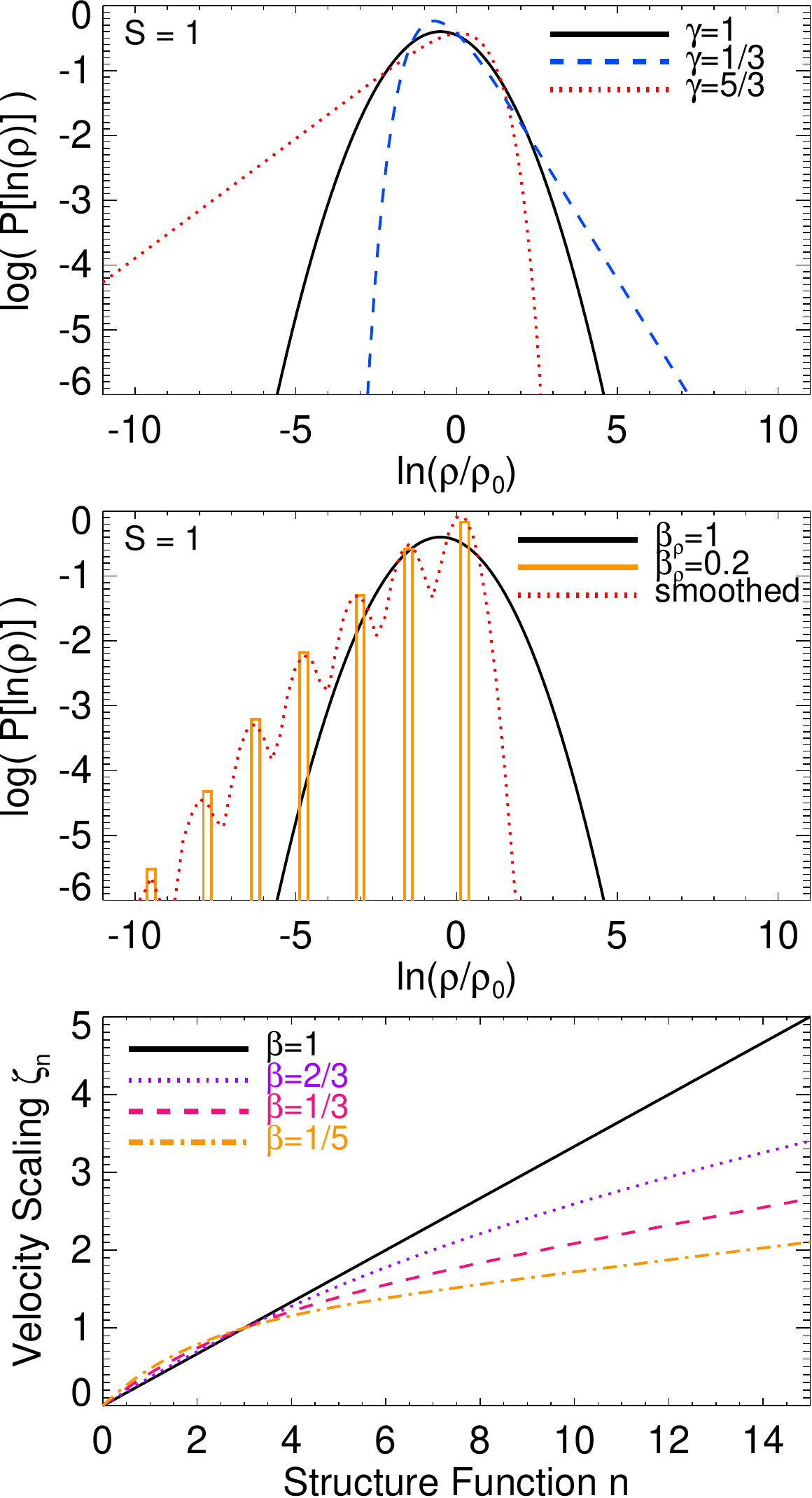}{0.95}
    \caption{Density PDFs and structure functions for the polytropic ($\gamma\ne1$) and intermittent ($\beta\ne1$) models in Figs.~\ref{fig:mfs}-\ref{fig:mf.intermittency.convolved}. 
    {\em Top:} Density PDFs for different polytropic index $\gamma$ (\S~\ref{sec:frag:poly}), normalized to the same Mach number at $\rho_{0}$ (such that the variance $S(R,\,\rho_{0})=1$). Non-isothermal gas leads to power law-like tails at low ($\gamma>1$) or high ($\gamma<1$) densities and steeper suppression of opposite fluctuations. 
    {\em Middle:} Density PDFs at fixed $S=1$ for isothermal ($\gamma=1$) gas with no intermittency ($\beta=\beta_{\rho}^{3}=1$) and (probably unphysically) strong intermittency according to the quantized log-Poisson \citet{sheleveque:structure.functions} model ($\beta_{\rho}=0.2$). The latter is skewed to low densities, discretized, and vanishes entirely for densities above a maximum near $\sim \rho_{0}$. We compare the same $\beta_{\rho}=0.2$ model convolved with small Gaussian scatter such that $10\%$ of $S$ is Gaussian. This is still highly non-Gaussian but removes the most severe discreteness effects. 
    {\em Bottom:} Velocity structure functions $S_{n}\equiv\langle |\delta v^{n}|\rangle\propto r^{\zeta_{n}}$, for different choices of $\beta=\beta_{\rho}^{3}$. 
    Kolmogorov turbulence with lognormal statistics is $\beta=1$.
    \label{fig:mf.intermittency.structfn}}
\end{figure}

\begin{figure}
    \centering
    \plotonesize{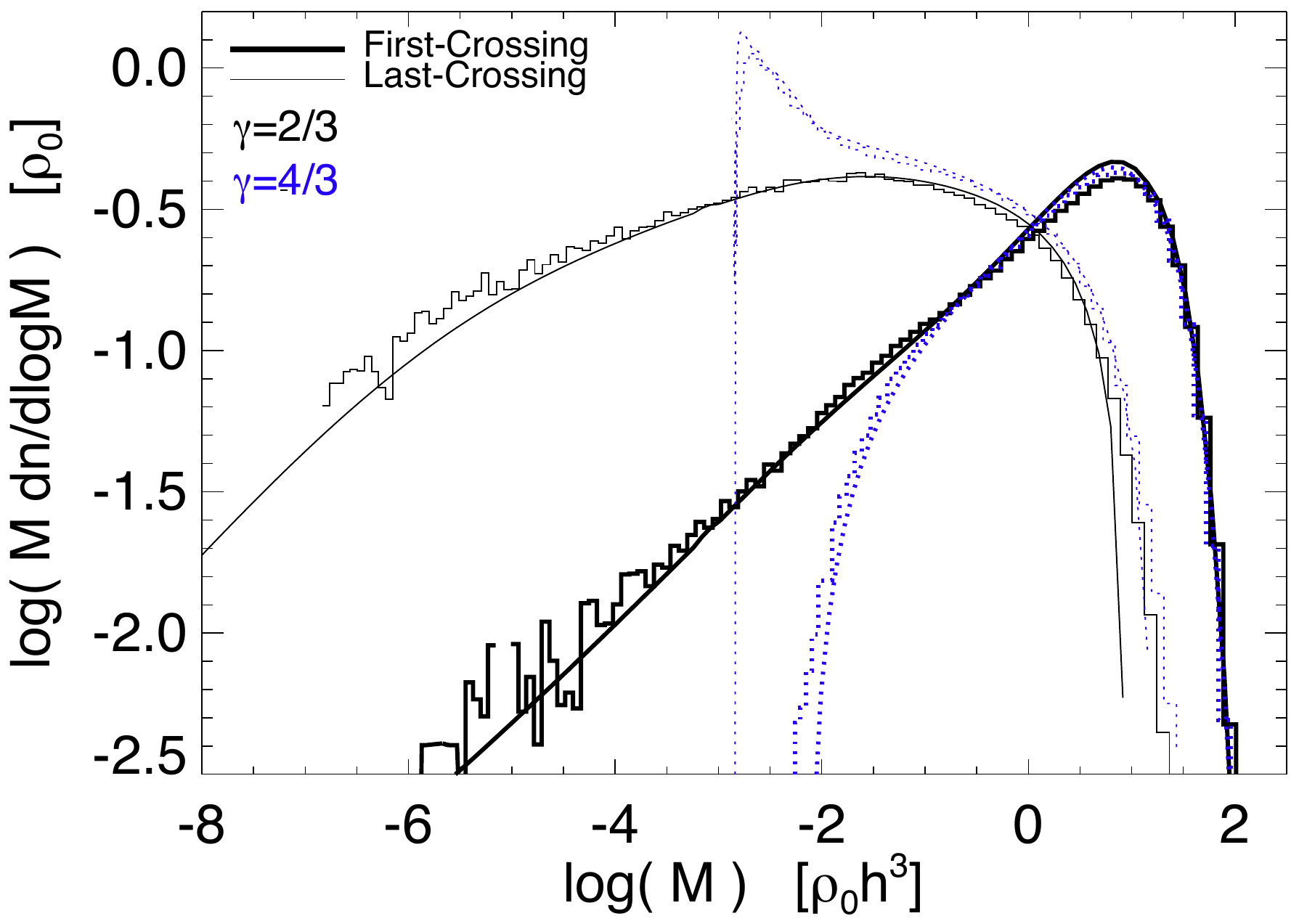}{0.95}
    \caption{First \&\ last-crossing MFs in non-isothermal gas ($\gamma=2/3$ and $\gamma=4/3$). We show the exact Monte Carlo numerical solution (\S~\ref{sec:methods.general}; histograms), accounting fully for a non-Gaussian density distribution and (necessarily) correlated modes in the density field between different scales. We compare the analytic result using Eqs.~\ref{eqn:flast}-\ref{eqn:ffirst}, derived (for $\gamma=1$) assuming {\em locally} Gaussian statistics and fully un-correlated mode structure between scales, but using the appropriate $\rhocrit(R)$ for $\gamma\ne1$, and replacing the variance $S(R)$ (for $\gamma=1$) with $S(R,\,\rhocrit)$ (see Fig.~\ref{fig:mr.scalings}). The two agree well. Although changing $\gamma$ can significantly change the MF, the dominant effect is {\em not} the correlation structure of the density field nor the local non-Gaussianity in the density statistics.
    \label{fig:mf.eos.exact.vs.num}}
\end{figure}

\begin{figure}
    \centering
    \plotonesize{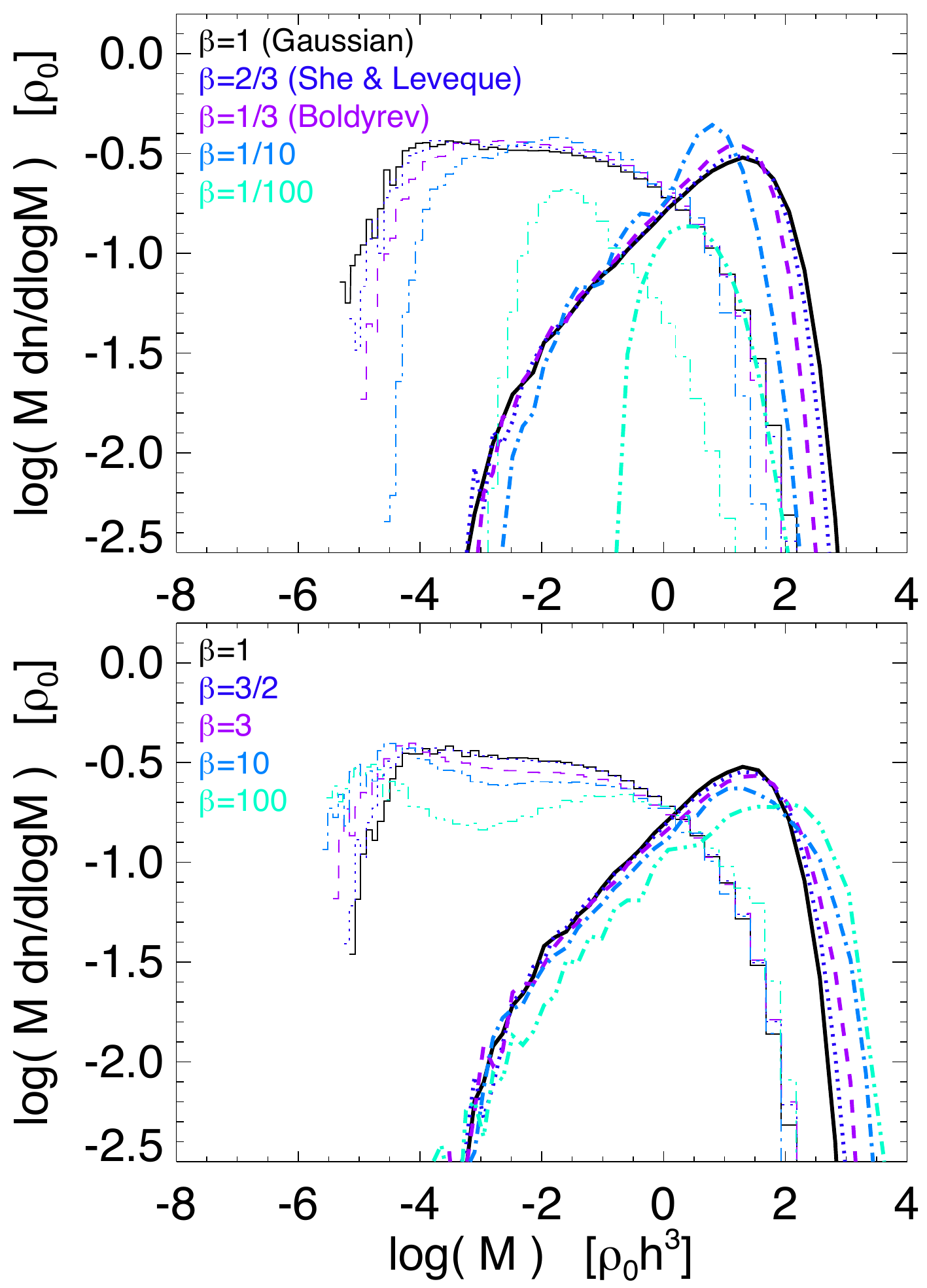}{0.95}
    \caption{First \&\ last-crossing MFs in isothermal turbulence with different levels of intermittency (\S~\ref{sec:nongaussian}). {\em Top:} We show the exact Monte Carlo distribution accounting for the non-Gaussian density distribution and correlated modes between different scales corresponding to our ``standard model'' but different $\beta=\beta_{\rho}^{3}$ and $\gamma^{\prime}$ parameterizing the intermittency (according to the quantized log-Poisson \citet{sheleveque:structure.functions} model for intermittency). $\beta=0-1$ parameterizes the strength of intermittency and correlation between fluctuations: $\beta=1$ is non-intermittent, uncorrelated modes with Gaussian statistics; $\beta\rightarrow0$ gives fully intermittent, perfectly correlated fluctuations on different scales with infinitely skew statistics. $\beta=2/3$ is the ``standard'' intermittency seen in experiments and simulations of isotropic hydrodynamic turbulence. $\beta=1/3$ corresponds to more singular turbulence with infinitely thin sheets and shocks or strongly magnetically-dominated collisionless media. Smaller $\beta$ are generally not seen but are shown for comparison. Strong intermittency -- as parameterized in the quantized log-Poisson model -- creates a sharp (but possibly artificial) discrete cutoff in the density distribution above $\rho\sim\rho_{0}$, truncating the MF at low/high masses. The effect is similar to increasing $Q$. 
      {\em Bottom:} Same, but using ``inverse $\beta$'' models which represent identical levels of intermittency, but reverse the sign of the skewness; this biases the PDF towards high densities and the MFs to a broader range, but the effect is weaker.
    \label{fig:mf.intermittency}}
\end{figure}

\begin{figure}
    \centering
    \plotonesize{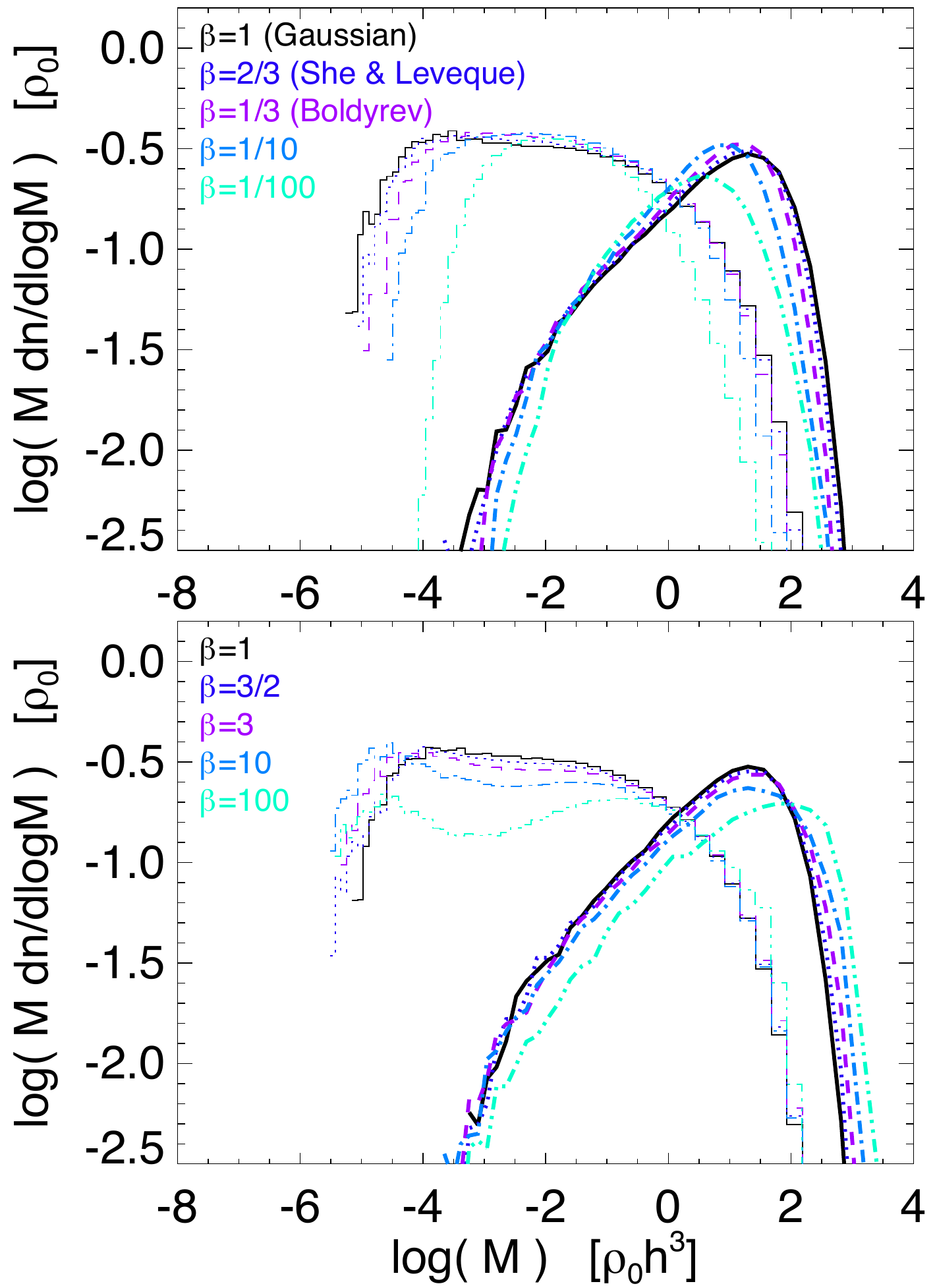}{0.95}
    \caption{As Fig.~\ref{fig:mf.intermittency}, but assuming $10\%$ of the variance in the density is contributed by normal fluctuations, while the remaining $90\%$ comes from quantized log-Poisson fluctuations with the labeled $\beta$. While the qualitative effect is the same (narrower MFs with lower $\beta$, as the density PDF is skewed towards progressively lower densities), the dependence on $\beta$ is substantially weaker. The most severe effects in Fig.~\ref{fig:mf.intermittency} arise from the strictly discrete density PDF in the simple log-Poisson model; these are eliminated by even a small amount of variation which allows both positive and negative fluctuations. The skewness is almost identical (within $\sim2\%$) in each case to the ``pure quantized'' model, so non-Gaussianity alone is not the dominant effect. Here, the effect is almost degenerate with small changes in $\Machdisk$ (from $\sim6-10$ for $\beta<1$) or $Q$ (from $1-2$). 
    \label{fig:mf.intermittency.convolved}}
\end{figure}

\subsection{Overview}

In two places in the text we have considered non-Gaussian, inherently correlated structures in turbulence: first, for non-isothermal $\gamma\ne1$ equations of state (\S~\ref{sec:frag:poly}), and second, the case with intermittency, $\beta_{\rho}\ne1$ (\S~\ref{sec:intermittency}). Both of these produce non-Gaussian density PDFs and non-normal statistics. In fact, for certain choices the density PDFs resemble one another -- however we stress that the intrinsic correlation structures of the turbulence implied are quite different. 

Recall, in taking a ``random walk'' with $\gamma\ne1$ to sample the density field, we must allow for a density-dependent local dispersion $S=S(R,\,\rho)$ and solve each trajectory as a correlated random walk -- in other words, the ``steps'' depend on the absolute position at each step and the position at every other step. With intermittency, in the context of the \citet{sheleveque:structure.functions} model, it is still true (indeed it is an assumption of the model) that the ``stepping'' is independent of position (in log space); however the statistics are highly non-Gaussian and spatially correlated by the skew in the parameter $\beta_{\rho}$ and the discreteness of the Poisson distribution. 

Some effects of different assumptions about intermittency are illustrated in Fig.~\ref{fig:mf.intermittency.structfn}. First, we compare the density PDF, at (for convenience) a fixed total variance $S=1$. Non-intermittent isothermal turbulence produces a lognormal distribution with median $-S/2$. We compare this to the density PDF resulting from strictly quantized log-Poisson statistics -- the simplest mathematically allowed PDF that matches the \citet{sheleveque:structure.functions} structure functions -- as assumed in \citet{shewaymire:logpoisson} and \citet{dubrulle:logpoisson} (but applied to the density statistics as described in \S~\ref{sec:intermittency}). In this case the density PDF is discretized (rather than continuous), with ``jumps'' quantized in units of $\ln{\beta_{\rho}}$. It is also highly skewed, with a large power law-like tail to low densities and a sharp, absolute cutoff at modest $\rho>\rho_{0}$. Clearly, the discreteness effect is not realistic (it is an artifact of our simplifying assumptions); recall from \S~\ref{sec:intermittency} that more generally the statistics in intermittent turbulence should be log-Poisson $P(m)$ convolved with some function $G_{R}(\ln{(\rho/\rho_{0})}/\ln{\beta_{\rho}}-m)$ reflecting the variation in strength of dissipative events. We therefore also consider the same model, but assuming this convolution function $G_{R}$ is Gaussian, with a dispersion equal to $10\%$ of the total variance (the other $90\%$ coming from the log-Poisson distribution). Physically, this corresponds to $10\%$ the variance coming from variation in the ``level of dissipation'' associated with a given structure (e.g.\ the size and/or strength of a shock or vortex), while the remaining $90\%$ comes from variance in the ``number of structures'' in some Lagrangian volume. This convolution is sufficient to make the density PDF continuous, though it preserves the multiple peaks, skewness, and sharp high-density cutoff. 

Fig.~\ref{fig:mf.intermittency.structfn} also shows the results of intermittency on the velocity structure functions, which are well-studied. Recall, $\beta=1$ corresponds to our standard (Gaussian/non-intermittent) assumption. In \citet{sheleveque:structure.functions}, the authors argue that the symmetries in the Navier-Stokes equations lead to an analytically predicted $\beta=2/3$, for isotropic, hydrodynamic turbulence with filamentary shocks; this predicts structure functions in good agreement with a wide range of experiments (including pipe flows with/without boundaries, longitudinal and transverse shear flows, buoyancy-driven turbulence, Taylor-Couette flows, and both two and three-dimensional rotating turbulence) as well as numerical simulations \citep[for a review, see e.g.][and references therein]{shezhang:2009.sheleveque.structfn.review}. \citet{boldyrev:2002.structfn.model} used the same model to derive a scaling with $\beta=1/3$, for highly super-sonic (Burgers) turbulence, with singular sheet-like shocks as the primary dissipative structures; this appears to agree with measurements of turbulence in molecular clouds \citep{boldyrev:structfn.tests} and numerical simulations of other interstellar medium processes (\citealt{pan:boldyrev.structfn.tests}; again for a review see \citealt{shezhang:2009.sheleveque.structfn.review}). This also happens to be the result for highly-magnetized MHD turbulence \citep{muller.biskamp:mhd.turb.structfn} and agrees with the structure functions observed in the solar wind and corona. For the sake of comparison, we consider the more extreme choices of $\beta=1/10$ and $1/100$. The former may correspond to some situations where large, non-hydrodynamic motions create density voids that fill the large majority of the volume \citep[e.g.\ the inter-galactic medium; see][]{liufang:2008.logpoisson.cosmic.baryons}, though this is clearly not a ``turbulent'' medium in the traditional sense. The latter is purely illustrative, as (to our knowledge) no turbulent self-gravitating fluid systems exhibit such extreme behavior. In each case, we note that while formally our model with $G_{R}$ containing $\sim10\%$ of the variance produces small corrections to the structure functions, they are generally negligible.

In either case ($\beta\ne1$ or $\gamma\ne1$), it is important to examine how the non-Gaussianity and different inherent correlation structures change our results.

\vspace{-0.5cm}
\subsection{Effects on the Mass Functions}

\subsubsection{Non-Isothermal Equations of State}

We saw in \S~\ref{sec:mf} (Fig.~\ref{fig:mfs}) that changing $\gamma\ne1$ has a significant effect on the MF. However, this could entirely owe to the effects on the barrier (changing the density threshold for collapse), or the broadening/shrinking of the density distribution (i.e.\ the non-Gaussian PDF simply ``looking like'' a Gaussian PDF with a different variance $S$ near the densities of interest), rather than the effects of statistics being locally non-Gaussian and/or the effect of the modes on different scales being inherently correlated. 

To test this, we examine in Fig.~\ref{fig:mf.eos.exact.vs.num} the first and last-crossing MFs for two choices $\gamma<1$ and $\gamma>1$. We first show the full numerical result (from Fig.~\ref{fig:mfs}), using our Monte Carlo method (\S~\ref{sec:frag:poly:walk}) that includes the inherent correlation structure of the modes on different scales, their dependence on density, and the non-Gaussianity in the statistics. 

We compare this to the prediction if we just apply a slightly modified form of our analytic solution for first and last-crossing (Eq.~\ref{eqn:flast} \&\ \ref{eqn:ffirst}), derived {\em assuming} locally Gaussian statistics and completely un-correlated fluctuations on different scales. In the analytic solution, we simply modify the barrier $B$ to account for the appropriate critical density for collapse ($\rhocrit(R)\rightarrow\rhocrit(R\,|\,\gamma)$) and replace the variance $S$ appearing in the equation is with the ``effective'' $S(R)\rightarrow S(R,\,\rhocrit)$ near the critical density (as defined in \S~\ref{sec:frag:poly}).\footnote{We do have to be careful to modify the integration to avoid where $\Delta S(R,\,\rhocrit)$ between two steps becomes negative; in which case no objects should appear on this scale.}

We see that this captures essentially all of the key information, and reproduces the full numerical solution quite well. In other words, the intrinsic correlation structure of modes has a quite small effect on the MF. Non-Gaussianity, to the extent that it matters, is only important insofar as it makes the density distribution more or less broad at the densities of interest -- re-mapping to a Gaussian with some different dispersion gives similar results.

\vspace{-0.5cm}
\subsubsection{Intermittency}

In Figs.~\ref{fig:mf.intermittency}-\ref{fig:mf.intermittency.convolved}, we examine the effects of intermittency (as approximated in \S~\ref{sec:intermittency}) on the predicted MF, for various values of $\beta=\beta_{\rho}^{3}$, and both the ``standard'' \citet{shewaymire:logpoisson,dubrulle:logpoisson} formulation (which produces a distribution skew towards excess low-density gas) and a mirrored ``inverse beta'' formulation described below (reversing the skew). In Fig.~\ref{fig:mf.intermittency}, we consider strictly quantized log-Poisson statistics (so the density ``jumps'' are discrete, as described in \S~\ref{sec:intermittency:steps}). In Fig.~\ref{fig:mf.intermittency.convolved} we compare the same models but making the density PDF continuous by assuming $10\%$ of the total variance is associated with a Gaussian convolution function $G_{R}$ (as in Fig.~\ref{fig:mf.intermittency.structfn}). 

Figs.~\ref{fig:mf.intermittency}-\ref{fig:mf.intermittency.convolved} show that ``realistic'' levels of intermittency have weak effects on our results. At $\beta=1/10$ ($\beta_{\rho}=0.46$), we begin to see larger effects, truncating the MF at the highest-mass scales ($R>h$) and shifting the last-crossing by a factor $\sim2$, and by $\beta=1/100$ ($\beta_{\rho}=0.21$) we see the MF is highly compressed. In these models, lowering $\beta$ skews the density distribution towards lower values and so ``tightens'' the MF range; this is very similar to decreasing the Mach number of increasing Toomre $Q$ (see Fig.~\ref{fig:mfs}). 

Even at extreme $\beta$ values, much of the effect seen in Fig.~\ref{fig:mf.intermittency}  depends on the strict discreteness of the distribution; when we consider in Fig.~\ref{fig:mf.intermittency.convolved} the same models with just a small fraction of the variance in a smooth Gaussian component, the differences are significantly smaller. Thus, for ``typical'' intermittency models, the most important quantity determining the MFs appears to be the total variance in the density distribution, not the detailed form of the multi-fractal cascade model.

\vspace{-0.5cm}
\subsubsection{``Inverse Beta'' Models}
\label{sec:intermittency:negbeta}

As shown in Fig.~\ref{fig:mf.intermittency.structfn}, the model from \S~\ref{sec:intermittency:steps}, for $0<\beta_{\rho}<1$, produces a density PDF skewed towards low densities (because the distribution of $m$ is not symmetric). However, there are physical situations where the skew might be reversed.\footnote{Although this may not be related to actual intermittency: for example when $\gamma<1$, or when self-gravitating regions are allowed to collapse but still included in the density PDF \citep{vazquez-semadeni:2001.nh.pdf.gmc,bournaud:2010.grav.turbulence.lmc,ballesteros-paredes:2011.dens.pdf.vs.selfgrav}.} 

It is trivial to show that the skewness of the density PDF model will be reversed, while the degree of non-Gaussianity in log-space is preserved, if we simply take $\beta_{\rho} \rightarrow 1/\beta_{\rho}$ (or $\ln{\beta_{\rho}}\rightarrow-\ln{\beta_{\rho}}$), i.e.\ $\beta_{\rho}>1$. For fixed variance $S(R)$, this reverses the sign of $\gamma^{\prime}$, but otherwise the equations in \S~\ref{sec:intermittency:steps} are well-behaved. We denote these ``inverse beta'' models. In this case, a larger value of $\beta_{\rho}$ (since this model corresponds to its ``mirror'' with $1/\beta_{\rho}$) represents larger deviations from Gaussianity. We stress that we are not arguing for a particular physical interpretation of these models, only noting that they are provide a potentially useful illustration of how certain results depend on the skew of the density PDF.

We show the predicted MF from these models in Figs.~\ref{fig:mf.intermittency}-\ref{fig:mf.intermittency.convolved}. As expected, they produce deviations from the strictly log-normal case with the opposite sense of the ``standard'' intermittency models ($\beta_{\rho}<1$); but the magnitude of the effects are somewhat smaller. 

\vspace{-0.5cm}
\subsection{Effects on Correlation Functions}

We might expect to see a larger effect from non-Gaussian statistics in the correlation functions. But in Figs.~\ref{fig:corr.fn.varmass}-\ref{fig:corr.fn.fixmass}, we actually see very little dependence of $\xi(r)$ on either the equation of state $\gamma\ne1$ or intermittency parameter $\beta\ne1$. And to the extent that dependencies appear in Fig.~\ref{fig:corr.fn.fixmass} at intermediate scales and some masses, they appear consistent with the near-universal dependence of clustering amplitude on number density from Eq.~\ref{eqn:numden.bias}.  

This is rather surprising at first: recall the density fluctuations on all scales are {\em intrinsically correlated} when $\gamma\ne1$, and for $\beta\ne1$, the structure functions (higher-order moments of the velocity/density field) deviate substantially from that predicted by self-similar, Gaussian statistics. In Fig.~\ref{fig:mf.intermittency.structfn} we show these, in fact, to highlight how large the deviation actually is.

Why then do the correlation functions not change? Recall, as discussed in \S~\ref{sec:nongaussian:text}, most of the correlation function structure in Fig.~\ref{fig:corr.fn.varmass} is {\em not} be a consequence of inherent correlations in the turbulence, because in non-intermittent, isothermal turbulence we {\em assume} fluctuations on all scales are uncorrelated. But the correlation functions reflect the run of variance with scale, simply as a basic consequence of statistics: if the run of variance with scale is weak on small scales, then the probability of ``isolated'' strong fluctuations (large positive overdensities) arising on these small scales is low, and most such fluctuations will be embedded in larger-scale fluctuations. For a given run in $S(r)$, the correlation function is, to first order, fixed. Inherent correlation structures in the turbulence produce only second-order corrections to this in any of the models considered here. 

\vspace{-0.5cm}
\subsection{Summary}

In short, the predicted quantities in this work -- the width of the density PDF as a function of scale, and two-point clustering -- depend primarily on the {\em lowest-order} structure functions of turbulence (second-order and below). These are relatively insensitive to the observed correlation structure and non-Gaussian statistics measured in intermittency (in either experiment or simulations; see Fig.~\ref{fig:mf.intermittency.structfn}). At realistic levels of intermittency and non-Gaussianity, it therefore seems unlikely that these effects will dramatically change our conclusions. 

We should expect that the differences between models with different $\beta$ and/or other multi-scale correlated structures will be more pronounced if we considered higher-order correlation functions (the three-point function, etc.). As shown in Fig.~\ref{fig:mf.intermittency.structfn}, the structure functions at $n=2$ (the order of the two-point correlation function) differ relatively little even for large differences in $\beta$, but the differences grow with increasing order $n$. It requires going to much higher order $n\gg5$ before large differences become apparent.

\begin{figure}
    \centering
    \plotonesize{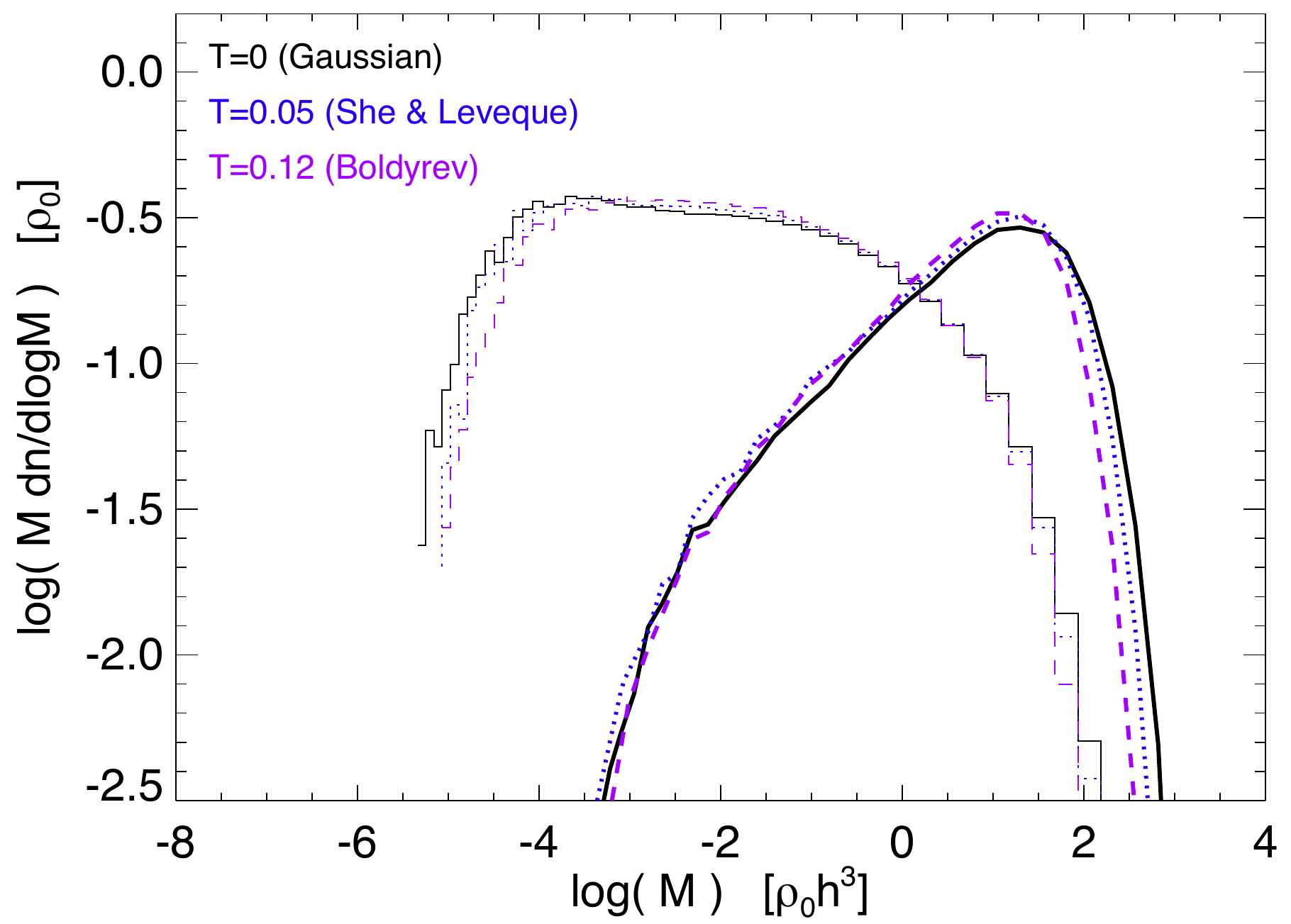}{0.95}
    \caption{Mass functions in our standard model with different degrees of intermittency (as Fig.~\ref{fig:mf.intermittency}), but using the alternative, continuous ``thermodynamic'' intermittency model from \citet{castaing:1996.thermodynamic.turb.cascade} described in Appendix~\ref{sec:appendix:alt.intermittency}. This provides a better fit to the density PDFs in simulations. We consider three values of the parameter $T$ described in the text, $T=0$ (no intermittency, our standard Gaussian statistics case), $T=0.05$ (which gives identical structure functions to the $\beta=2/3$ \citet{sheleveque:structure.functions} model), and $T=0.12$ (identical structure functions to the $\beta=1/3$ \citet{boldyrev:2002.structfn.model} model). The results are very similar to those in Fig.~\ref{fig:mf.intermittency}, with slightly smaller deviations from the Gaussian case for the same ``degree of intermittency.'' 
    \label{fig:mf.intermittency.alt}}
\end{figure}

\vspace{-0.5cm}
\section{Alternative Models for Intermittency}
\label{sec:appendix:alt.intermittency}

As discussed in \S~\ref{sec:intermittency}, a quantized intermittency model for density ``steps,'' while capturing intermittency in the structure functions, gives an unphysically discrete density distribution. 

A different model of intermittency, with continuous variations, is the ``thermodynamic'' model proposed in \citet{castaing:1996.thermodynamic.turb.cascade}. In that work, the assumptions are quite different from \citet{sheleveque:structure.functions}. However, as shown by \citet{hedubrelle:1998.thermo.equiv.logpoisson}, this is just a more general consistent formulation of the same hierarchy with a different choice of $G_{R}$. In this model, the predicted longitudinal (compressive) velocity increments scale as 
\begin{align}
\label{eqn:thermo.var}
P(u)\,{\rm d}u &= \sum_{m=0}^{\infty}\,\frac{\lambda^{m}\,e^{-\lambda}}{m!}\,\frac{u^{m-1}\,e^{-u}}{(m-1)!}\,{\rm d}u \\ 
\nonumber &= I_{1}(2\,\sqrt{\lambda\,u})\,\exp{[-(\lambda+u)]}\,\sqrt{\frac{\lambda}{u}}\,{\rm d}u  
\end{align}
where $u\equiv T^{-1}\,|\ln{(\sigma_{r}/\sigma_{L})}|$ with $T$ an ``intermittency'' constant and $\sigma_{r}$ the characteristic velocity amplitude (such that $\langle |\delta v|^{n} \rangle \propto \sigma_{r}^{n}$); $\sigma_{L}$ is this amplitude at the driving scale. $I_{1}$ is the modified Bessel function of the first kind. 

The variance in $u$ is just $2\,\lambda$, with the corresponding variance in $|\ln{(\sigma_{r}/\sigma_{L})}|$ equal to $2\,\lambda\,T^{2}$. For large-$\lambda$ and/or large total variance, this approaches a Gaussian in $u$ with mean $\langle u \rangle = \lambda$ and variance $2\,\lambda$ (i.e.\ a lognormal in $(\sigma_{r}/\sigma_{L})$ with variance $S$ and mean $-S/2$). 

But it is straightforward to see that this is just the general form of the log-Poisson statistics with 
\begin{align}
G_{R}(Y\,|\,m)\,{\rm d}Y &\rightarrow \frac{u^{m-1}}{(m-1)!}\,e^{-u}\,{\rm d}u \\ 
&=
\nonumber {\rm d}u \int {\rm d}^{m}u_{i}\,\delta{\Bigl(}u-\sum_{i}u_{i}{\Bigr)}\,\prod_{i=1}^{m}\,\exp{(-u_{i})}
\end{align}
where $T\rightarrow (\gamma^{\prime}/6)\,|\ln{\beta}| \approx -\ln{(\beta)}/9$. The predicted structure functions scale as
\be
\label{eqn:structfn.castaing}
\zeta_{n} = \frac{n}{3}\,\frac{1+3\,T}{1+n\,T}
\ee
which are nearly identical to the \citet{sheleveque:structure.functions} structure functions for $T\approx -(\gamma^{\prime}/6)\,\ln{\beta}$, at least to $n\sim 20$.\footnote{In \citet{hopkins:2012.intermittent.turb.density.pdfs} we show that $T=0.05$ gives nearly identical structure functions to the standard $\gamma^{\prime}=2/3$, $\beta=2/3$ quantized log-Poisson model in \citet{sheleveque:structure.functions}; and $T=0.12$ gives structure functions identical to the $\gamma^{\prime}=2/3$, $\beta=1/3$ \citet{boldyrev:2002.structfn.model} model.}

If we follow the assumptions in \S~\ref{sec:intermittency}, that the statistics of log-density fluctuations follow the statistics of longitudinal velocity fluctuations, this leads to the volumetric density PDF: 
\begin{align}
\label{eqn:PV} 
P(\ln{\rho})\,{\rm d}\ln{\rho} 
&= I_{1}(2\,\sqrt{\lambda\,\omega})\,\exp{[-(\lambda+\omega)]}\,\sqrt{\frac{\lambda}{\omega}}\,{\rm d}\omega \\ 
\nonumber &= \sum_{m=0}^{\infty}\,\frac{\lambda^{m}\,e^{-\lambda}}{m!}\,\frac{\omega^{m-1}\,e^{-\omega}}{(m-1)!}\,{\rm d}u \\ 
\nonumber \omega &\equiv \frac{\lambda}{1+T} -  \frac{\ln{\rho}}{T}\ \ \ \ \ \ \ \ \ \ (\omega\ge0) 
\end{align}
with $\lambda = \lambda(R) \equiv S(R)/(2\,T^{2})$.

Because this distribution is infinitely divisible, incorporating it into our Monte-Carlo trajectory calculations is a straightforward extension of the model in \S~\ref{sec:frag:poly:walk} \&\ \ref{sec:intermittency:steps}. In each ``step'' $R\rightarrow R-\Delta R$ with change in variance $\Delta S$, we simply draw the differential change in density $\Delta\ln{\rho}$ from the distribution in Eq.~\ref{eqn:PV} with $\lambda = \Delta \lambda = \Delta S/(2\,T^{2})$ (instead of a Gaussian). Physically, this has a simple interpretation in terms of the density and velocity fields as continuous-time multiplicative random relaxation processes. The variable $\lambda$ represents some ``number of events'' while $G_{R}$ represents a convolution over a Poisson waiting time distribution for each number of events. In other words, Eq.~\ref{eqn:PV} is a stationary result for exponentially damped perturbations driven by discrete random events with a constant average rate; where $T^{-1}$ is a dimensionless ``rate parameter'' (higher-$T$ corresponding to rarer, more highly intermittent ``events'').

In \citet{hopkins:2012.intermittent.turb.density.pdfs}, we specifically show that this form for the density PDF provides an excellent fit to numerical simulations spanning $\Mach\sim0.1-15$, with a variety of forcing schemes, numerical methods, and magnetic field strengths. Moreover, the same $T$ fit from \ref{eqn:PV} to the density PDF appears consistent with the values fit from Eq.~\ref{eqn:structfn.castaing} to the longitudinal velocity fluctuations, suggesting that our simple assumption of mapping between the compressive velocity and density statistics may be plausible (at least for the lower-order statistics that matter here). And this form for density fluctuations is also favored by direct observations of the solar wind \citep{forman:2003.castaing.fits.solar.wind.data,leubner:2005.solar.wind.pdf.castaing,leubner:2005.solar.wind.castaing.fits.density}.

For now we simply consider whether this form for the density PDF changes our conclusions. Fig.~\ref{fig:mf.intermittency.alt} plots the MFs for our standard model, allowing for different levels of intermittency (as in Fig.~\ref{fig:mf.intermittency}), but with the density PDF model above. For $T=0.05$ and more highly intermittent $T=0.12$, the resulting MFs show very weak deviations from the no-intermittency ($T\rightarrow0$) case. The sense of the deviations is identical to the corresponding quantized log-Poisson model, albeit smaller. So we conclude that the detailed form of the multi-fractal model for intermittency makes little difference to our conclusions here.


\vspace{-0.5cm}
\section{General First/Last Crossing Solutions in Continuous Turbulence}
\label{sec:appendix:analytic.continuous}

We now present the general solutions for first and last crossing for any infinitely divisible and continuous PDF. 

First, consider first-crossing; the derivation follows \citet{zhang:2006.general.moving.barrier.solution}, but without simplifying by assuming Gaussian statistics. Again consider trajectories integrated from $S=0$ ($R\rightarrow\infty$) and let $\delta = \ln{(\rho/\rho_{0})}$, 
and as before consider $\ffirst$ and $\Pi$, the probability of a given $\delta<B(S)$ {\em without} having a previous crossing:
\begin{align}
1 &= \int_{0}^{S}\ffirst(\Sprime)\,{\rm d}\Sprime + \int_{-\infty}^{B(S)}\,\Pi(\delta|S)\,{\rm d}\delta \\ 
\Pi(\delta|S) &= P_{0}(\delta|S) - \int_{0}^{S}{\rm d}\Sprime\,\ffirst(\Sprime)\,P_{10}(\delta[S]\,|\,B^{\prime}[\Sprime])
\end{align}
where $P_{10}(\delta[S]\,|\,B^{\prime}[\Sprime])$ is the probability of having $\delta$ at $S$ given an earlier value $B^{\prime}[\Sprime]$; infinite divisibility gives $P_{10}(\delta[S]\,|\,B[\Sprime]) = P_{0}(\delta-B^{\prime}|S-\Sprime)$. Taking the derivative of both sides with respect to $S$ and performing some simplifying algebra gives:
\begin{align}
\label{eqn:ffirst.full.continuous}
\int_{0}^{S}&{\rm d}\Sprime\,\ffirst(\Sprime)\,g_{2}(S,\Sprime) = g_{1}(S) + \alpha\,\ffirst(S) \\ 
\nonumber g_{1}&(S) \equiv \frac{{\rm d}B}{{\rm d}S}\,P_{0}(B|S) + g_{3}(B|S) \\ 
\nonumber g_{2}&(S,\Sprime) \equiv \frac{{\rm d}B}{{\rm d}S}\,P_{0}(B-B^{\prime}|S-\Sprime) + g_{3}(B-B^{\prime}|S-\Sprime) \\ 
\nonumber g_{3}&(x|y) \equiv \int_{-\infty}^{x}{\rm d}\delta\,\frac{\partial}{\partial y}\,P_{0}(\delta\,|\,y) \\ 
\nonumber \alpha &\equiv 1 - {\rm Limit}{\Bigl[}\int_{-\infty}^{B(S)}{\rm d}\delta\,P_{0}(\delta-B^{\prime}[\Sprime]\,|\,S-\Sprime){\Bigr]}_{\Sprime\rightarrow S}
\end{align}
For the distributions here $\alpha$ depends simply on the symmetry properties of $P_{0}$; for $P_{0}$ symmetric in $\delta$ (e.g.\ Gaussian statistics), $\alpha=1/2$; for $P_{0}$ which is one-sided, $\alpha=0$ (the $\beta<1$ or ``normal'' intermittency models) or $\alpha=1$ (the $\beta>1$ or ``inverse'' intermittency models). 

For any continuous function $P_{0}(\delta\,|\,S)$ and $B(S)$, the functions $g_{1}$, $g_{2}$, $g_{3}$, and $\alpha$ are also continuous, and can be tabulated so that Eq.~\ref{eqn:ffirst.full.continuous} (a Volterra integral equation of the second kind) can be solved by standard numerical methods. 

Now, consider last-crossing. We begin the walk at some sufficiently large $S_{i}$ as $R\rightarrow0$ such that the probability of crossing vanishes, and proceed in ``reverse'' direction. The integral constraint is
\begin{align}
1 &= \int_{S}^{S_{i}}\flast(\Sprime)\,{\rm d}\Sprime + \int_{-\infty}^{B(S)}\,\Pi(\delta|S)\,{\rm d}\delta \\ 
\Pi(\delta|S) &= P_{0}(\delta|S) - \int_{S}^{S_{i}}{\rm d}\Sprime\,\flast(\Sprime)\,P_{01}(\delta[S]\,|\,B^{\prime}[\Sprime])
\end{align}
where $P_{01}(\delta[S]\,|\,B^{\prime}[\Sprime])$ represents the probability of having $\delta$ at $S$ given an earlier value $B^{\prime}[\Sprime]$, but for a step in the opposite direction from the first-crossing case. This is related to $P_{10}$ by Bayes's theorem: 
\begin{align}
P_{01}(\delta[S]\,|\,B^{\prime}[\Sprime]) &= 
P_{10}(B^{\prime}[\Sprime]\,|\delta[S])\frac{P_{0}(\delta|S)}{P_{0}(B^{\prime}|\Sprime)} \\ 
&= 
\nonumber P_{0}(B^{\prime}-\delta|\Sprime-S)\,\frac{P_{0}(\delta|S)}{P_{0}(B^{\prime}|\Sprime)}
\end{align}
Again taking the derivative of both sides and simplifying, we obtain
\begin{align}
\label{eqn:flast.full.continuous}
\int_{S}^{S_{i}}&{\rm d}\Sprime\,\flast(\Sprime)\,\tilde{g}_{2}(S,\Sprime) = g_{1}(S) - \tilde{\alpha}\,\flast(S) \\ 
\nonumber \tilde{g}_{2}&(S,\Sprime) \equiv \frac{{\rm d}B}{{\rm d}S}\,P_{01}(B[S]\,|B^{\prime}[\Sprime]) + \int_{-\infty}^{B[S]}{\rm d}\delta\,\frac{\partial}{\partial S}\,P_{01}(\delta[S]\,|\,B^{\prime}[\Sprime]) \\ 
\nonumber \tilde{\alpha} &\equiv 1 - {\rm Limit}{\Bigl[}\int_{-\infty}^{B(S)}{\rm d}\delta\,P_{01}(\delta[S]\,|\,B^{\prime}[\Sprime]){\Bigr]}_{\Sprime\rightarrow S}
\end{align}
with $g_{1}$ the same as in the first-crossing case and $P_{01}$ defined above.
From this form, again, for any continuous $P_{0}(\delta|S)$ and $B(S)$, the input functions are straightforward to tabulate and $\flast(S)$ can be determined by standard numerical methods.

\vspace{-0.5cm}
\section{Analytic First/Last Crossing Solutions in Quantized, Intermittent Turbulence}
\label{sec:appendix:analytic.beta}

Here we derive the exact analytic solutions for the first \&\ last-crossing distributions in intermittent turbulence as approximated in \S~\ref{sec:intermittency:steps} in the text. Consider intermittency of the form characterized by the \citet{shewaymire:logpoisson,dubrulle:logpoisson} model, i.e.\ some characteristic $\beta_{\rho}$ and $\gamma^{\prime}$, with $G_{R}(x)\rightarrow \delta(x)$ (i.e.\ pure quantized log-Poisson processes). Our derivations will closely follow those for the first and last-crossing distributions for the Gaussian (non-intermittent) cases given in \citet{zhang:2006.general.moving.barrier.solution} and \papertwo, respectively. We refer to those papers for more details.

\vspace{-0.5cm}
\subsection{First Crossing}

\subsubsection{Exact Discrete Solution} 

Consider first the case with $\beta_{\rho}<1$. Rather than $\gamma^{\prime}$, it is more convenient to work in the directly related variable $S$ (the variance as a function of scale) or $\lambda$ (the variance of the related Poisson distribution). We begin at some sufficiently large scale where $S=0$, and consider steps in scale (successively smaller radial averaging) that correspond to some increasing $S\rightarrow S + \Delta S$. At each scale, we can evaluate the density distribution and determine the fraction $\ffirst$ of trajectories which are experiencing a first crossing in that step. Obviously when $S=0$, $\delta=0$ (i.e.\ all densities are at the mean on this scale). From Eq.~\ref{eqn:logpoisson.step}, $\rho_{r_{2}} = \beta_{\rho}^{m}\,(r_{1}/r_{2})^{\gamma^{\prime}}\,\rho_{r_{1}}$, and from the definition of $\lambda$ (Eq.~\ref{eqn:logpoisson:lambda.def}), we have at each scale -- equivalently for any trajectory -- that 
\begin{align}
\delta(R) &\equiv \ln{{\Bigl(}\frac{\rho(R)}{\rho_{0}}{\Bigr)}} \\
&= m(R)\,\ln{\beta_{\rho}} + \lambda(R)\,(1-\beta_{\rho}) \\
&= m(R)\,\ln{\beta_{\rho}} + S(R)\,\frac{1-\beta_{\rho}}{(\ln{\beta_{\rho}})^{2}}
\end{align}
where $m$ is a Poisson random variable with mean and variance $\lambda=\lambda(R) = S/(\ln{\beta_{\rho}})^{2}$ ($S=S(R)$ defined as throughout this paper as the variance in the logarithmic density field at each scale). Because $\ln{\beta_{\rho}}<0$, this value being above some critical density $\delta_{\rm crit}$ (i.e.\ crossing the barrier $B(S)$ requires 
\be
m < m_{c}(\lambda) \equiv \frac{1}{\ln{\beta_{\rho}}}\,{\Bigl(}B(S) -  S\,\frac{1-\beta_{\rho}}{(\ln{\beta_{\rho}})^{2}}{\Bigr)}
\ee
Note here $B\equiv \ln{(\rhocrit/\rho_{0})}$ strictly (there is no $\pm S/2$ term, because of how we define the PDF in what follows). 

It must be true that at the ``initial'' ($S=0$) scale $m_{c}<0$ (true for any $B>0$), so that not all trajectories are crossed. A critical difference from the Gaussian case is that while $\delta$ can increase or decrease in a step, $m$ is a Poisson variable and so is a positive integer which cannot decrease. So there are no crossings until some scale $\lambda$ where $m_{c}\rightarrow0$; at this point the fraction of all the volume with $m=0$ (the fraction $=\exp{(-\lambda[m_{c}=0])}$) instantly ``crosses.'' All trajectories with $m>0$ have $\delta < B$ at this scale. Going further in scale, $\lambda$ continues to increase. If $m_{c}(\lambda)$ at that scale decreases, it is irrelevant, because there can be {\em no} decrease in $m$ values for any ``trajectory,'' hence anything which now falls above $m>m_{c}$ already had a first crossing. If $m_{c}(\lambda)$ eventually increases, it is also irrelevant as long as $0<m_{c}<1$, because all trajectories which have not already crossed (i.e.\ had $m=0$) still lie above $m_{c}$. However at $m_{c}=1$ there will suddenly be additional crossings, and so on. Crossings can {\em only} occur at discrete $\lambda_{m_{c}}$ where $m_{c} = 0,\,1,\,2,\,....$. This means $\ffirst$ is {\em nowhere} differentiable. We must define it strictly in integral/sum formulation. 

Define the discrete fraction of crossings at each subsequently increasing positive integer value of $m_{c}$, 
\be
\Delta {F_{f}}_{m_{c}} \equiv {\Bigl |}\int_{\lambda[m_{c}-1]}^{\lambda[m_{c}]}\,{\rm d}S\,\ffirst(S[\lambda]) {\Bigr|}
\ee

Since crossings occur at the discrete $m_{c}$, the fraction crossing at a given $m_{c}$ must be equal to the expected fraction having $m=m_{c}$ at that $\lambda$, 
\be
P(m\,|\,\lambda) = P_{0}(m\,|\,\lambda) \equiv \frac{\lambda^{m}}{m!}\,\exp{(-\lambda)}
\ee
{\em after} subtracting the fraction which previously crossed at $m_{c}^{\prime}< m_{c}$ and now have the value $m$. For Poisson statistics, the probability to have a value $m$ at $\lambda$, given an earlier value $m^{\prime}$ at $\lambda^{\prime}< \lambda$ is just 
\be
P(m[\lambda]\,|\,m^{\prime}[\lambda^{\prime}<\lambda]) = P_{0}(m-m^{\prime}\,|\,\lambda-\lambda^{\prime})
\ee

So we obtain the integral equation
\begin{align}
\label{eqn:ffirst.beta.exact}
\Delta {F_{f}}_{m_{c}} &= P_{0}(m_{c}\,|\,\lambda[m_{c}]) - \sum_{k=0}^{m_{c}-1}\,
\Delta {F_{f}}_{k}\,P_{0}(m_{c} - k\,|\,\lambda[m_{c}]-\lambda[k])
\end{align}
defined for $m_{c} = 1,\,2,\,3,\,...$, 
with 
\be
\Delta {F_{f}}_{0} = P_{0}(0\,|\,\lambda[m_{c}=0]) = \exp{(-\lambda[m_{c}=0])} \\ 
\ee
This is defined {\em at} the $\lambda=\lambda[m_{c}]$ with integer values of $m_{c}$ only; $\ffirst(S)=0$ everywhere else.  

Note that if $m_{c}(\lambda)$ equals a specific integer value at multiple $S$ or $\lambda$, $\lambda[m_{c}]$ is defined as the {\em smallest} such value of $\lambda$ (i.e.\ the largest physical scale, since we are ``counting'' in the direction of increasing $\lambda$). 

Once again we note that this function is everywhere non-differentiable, therefore $\ffirst(S)$ is nowhere defined. However we can define the discretely {\em averaged} $\ffirst(S)$ by assigning a ``width'' $\Delta S$ corresponding to the $\Delta \lambda$ between $m_{c}$ values, then defining $\langle \ffirst(S) \rangle \equiv \Delta {F_{f}}_{m_{c}} / \Delta S$ at each $m_{c}$. For any $B(S)$, it is now straightforward to evaluate $\langle \ffirst(S) \rangle$.

\vspace{-0.5cm}
\subsubsection{Continuum Limit \&\ Linear Barrier Solution}

To compare to our derivation for the Gaussian statistics case, we can take the continuum limit of Eq.~\ref{eqn:ffirst.beta.exact}, i.e.\ $m\gg1$ so that the discreteness effects between $m$ can be ignored. First note that since $P_{0}(0\,|\,\lambda\rightarrow0)\rightarrow 1$, we can write Eq.~\ref{eqn:ffirst.beta.exact} as 
\be
\sum_{k=0}^{m_{c}}\Delta {F_{f}}_{k}\,P_{0}(m_{c}-k\,|\,\lambda[m_{c}]-\lambda[k]) = P_{0}(m_{c}\,|\,\lambda[m_{c}]) 
\ee
From the definition of $\Delta F_{f}$ it is trivial to see that this becomes, in the continuum case 
\be
\int_{0}^{S}\,{\rm d}S^{\prime}\ffirst(S^{\prime})\,P_{c}(m_{c}[S]-m_{c}[\Sprime]\,|\,\lambda-\lambda^{\prime})
= P_{c}(m_{c}[S]\,|\,\lambda)
\ee
where again $\lambda\equiv S/(\ln{\beta_{\rho}})^{2}$ and we now allow $m_{c}$ to assume the continuum of values for all $S$, and define 
\be
P_{c}(m\,|\,\lambda) = \frac{\lambda^{m}}{\Gamma[m+1]}\,e^{-\lambda} \rightarrow \frac{1}{\sqrt{2\pi\lambda}}\,\exp{{\Bigl(}-\frac{(m-\lambda)^{2}}{2\,\lambda} {\Bigr)}}
\ee
where the latter is the limit of the Poisson distribution for large-$\lambda$. Note that in this limit 
\begin{align}
P_{c}(m_{c}\,|\,\lambda) &\rightarrow \frac{|\ln{\beta_{\rho}}|}{\sqrt{2\pi S}}\,\exp{{\Bigl(}-\frac{(B - \tilde{\mu}\,S)^{2}}{2\,S}{\Bigr)}} \\ 
\tilde{\mu} &\equiv \frac{1-\beta_{\rho} + \ln{\beta_{\rho}}}{(\ln{\beta_{\rho}})^{2}}
\end{align}
i.e.\ this is just the log-normal probability of $B$, for a mean $\langle \delta \rangle = \tilde{\mu}\,S$, and for $\beta_{\rho}\rightarrow1$ (no intermittency), $\tilde{\mu}\rightarrow -1/2$, exactly as in the case of pure lognormal statistics.

For either form of $P_{c}$, this is a Volterra integral equation of the first kind, and is straightforward to solve via standard numerical methods for {\em any} arbitrary barrier $B(S)$. This is explained in \citet{zhang:2006.general.moving.barrier.solution} in detail. 

Now consider a linear barrier of the form $B=\ln{(\rhocrit/\rho_{0})} = B_{0}+\mu_{0}\,S$. It is straightforward to verify that the continuum limit equation is solved by 
\be
\ffirst(B=\ln{(\rhocrit/\rho_{0})} = B_{0}+\mu_{0}\,S) = \frac{B_{0}}{S}\,P_{c}(m_{c}\,|\,\lambda)
\ee
As $\beta_{\rho}\rightarrow1$, this becomes identical to the exact solution for the log-normal (non-intermittent) case, as it should.

\vspace{-0.5cm}
\subsection{Last Crossing}

\subsubsection{Exact Integral-Discrete Solution} 

Now consider last-crossings. The derivation is very similar, but now we ``begin'' our evaluation at $R\rightarrow 0$, where $S$ takes some finite value and $B$ is sufficiently large that all trajectories are below the barrier. This means again that $m_{c}<0$ must be true (since $B\gg S$ on these scales). Going in the direction of {\em decreasing} $S$, now, all trajectories are un-crossed until $m_{c}=0$, when the fraction $P=P_{0}(0\,|\,\lambda[m_{c}=0])$ cross from being ``below' the barrier (in terms of $\delta$) on all smaller scales to ``above'' the barrier on a larger scale (i.e.\ have a last-crossing). Note that this is {\em not} necessarily the same $\lambda[m_{c}=0]$ as was defined for the first-crossing distribution -- that was the {\em smallest} value of $\lambda$ or $S$ (largest spatial scale) where $m_{c}=0$, and this (because we are evaluating in the opposite direction) is the {\em largest} such value of $\lambda$ or $S$ (smallest spatial scale). 

Unlike in the first-crossing case, because we are evaluating in the opposite direction, trajectories can only {\em decrease} in $m$ in the direction we evaluate. But we still are evaluating when trajectories go from $m>m_{c}$ to $m\le m_{c}$. This, now, can occur {\em between} integer values of $m_{c}$ (e.g.\ a trajectory could suddenly decrease in $m$, crossing from above to below a non-integer value of $m_{c}$. 

At some $\lambda$, it must be true that the sum of all trajectories which have ``last crossed'' from below to above the barrier, plus the sum of all trajectories which are currently below $m_{c}$ but have not at any point crossed, is unity. So 
\begin{align}
1 = &\int_{\lambda}^{\lambda_{R=0}}\,\flast(\Sprime){d\Sprime} + \sum_{m>m_{c}}\,\Pi(m\,|\,\lambda)  
\end{align}
where $\Pi$ represents the probability of being at some $m>m_{c}$, i.e.\ below the barrier, without having previously crossed, so is the standard probability $P_{0}$ after subtracting the probability of crossing at a previous scale $\lambda^{\prime}$ and then arriving at the value $m$
\begin{align}
\Pi(m\,|\,\lambda) &= P_{0}(m\,|\,\lambda) - \int_{\lambda}^{\lambda_{R=0}}\,{{\rm d}\Sprime}\,\flast(\Sprime)\,P_{01}(m[\lambda]\,|\,{\rm FIX}(m_{c}^{\prime}[\lambda^{\prime}]))
\end{align}
Here ${\rm FIX}(m_{c})$ is the integer floor of $m_{c}$ (since an integer crossing from above to below $m_{c}$ ``lands'' at this value). 
The probability $P_{01}$ here represents the probability of a transition from $m^{\prime}$ to $m$ in a step in  the direction of {\em decreasing} $\lambda^{\prime}\rightarrow \lambda$. This is related to the probability $P_{10}$ of a transition from $m$ to $m^{\prime}$ in the direction of increasing $\lambda\rightarrow \lambda^{\prime}$ by Bayes's theorem
\be
P_{01}(m[\lambda]\,|\,m^{\prime}[\lambda^{\prime}>\lambda]) = 
P_{10}(m^{\prime}[\lambda^{\prime}]\,|\,m[\lambda])\,\frac{P_{0}(m\,|\,\lambda)}{P_{0}(m^{\prime}\,|\,\lambda^{\prime})}
\ee
Recall $P_{10}(m^{\prime}[\lambda^{\prime}>\lambda]\,|\,m[\lambda]) = P_{0}(m^{\prime}-m\,|\,\lambda^{\prime}-\lambda)$ for Poisson statistics. Therefore we obtain $P_{01}$ directly. 

The previous equations can be considerably simplified if we use the following relations:
\begin{align}
\sum_{m>m_{c}} P_{0}(m\,|\,\lambda) &= 1 - \sum_{m=0}^{{\rm FIX}(m_{c})}\,P_{0}(m\,|\,\lambda) \\ 
\sum_{m>m_{c}} P_{01}(m\,|\,{\rm FIX}[m_{c}^{\prime}] ) &= 1 - \sum_{m=0}^{{\rm FIX}(m_{c})}\,P_{01}(m\,|\,{\rm FIX}[m_{c}^{\prime}] ) \\ 
\sum_{m>m_{c}}\int{\rm d}\Sprime\,\flast(\Sprime) &P_{01}(m\,|\,{\rm FIX}[m_{c}^{\prime}]) \\ = 
\nonumber \int{\rm d}\Sprime\,&\flast(\Sprime) \sum_{m>m_{c}}\,P_{01}(m\,|\,{\rm FIX}[m_{c}^{\prime}] )
\end{align}

Expanding $\Pi(m\,|\,\lambda)$ and using the relations above, we obtain the following integral equation for $\flast$:
\begin{align}
\label{eqn:flast.beta.exact}
\int_{\lambda}^{\lambda_{R=0}}\,{\rm d}\Sprime\,\flast(\Sprime)\,
\sum_{m=0}^{{\rm FIX}(m_{c})}P_{01}(m[\lambda]\,|\,{\rm FIX}[m_{c}^{\prime}[\lambda^{\prime}]]) &= 
\sum_{m=0}^{{\rm FIX}(m_{c})}P_{0}(m\,|\,\lambda)
\end{align}
with 
\be
P_{01}(m[\lambda]\,|\,m^{\prime}[\lambda^{\prime}]) = \frac{m^{\prime}!}{m!\,(m^{\prime}-m)!}\,
\frac{\lambda^{m}\,(\lambda^{\prime}-\lambda)^{m^{\prime}-m}}{{\lambda^{\prime}}^{m^{\prime}}}
\ee

Unlike in the first-crossing case, this is differentiable at most $S$, since the sum terms vary continuously in $\lambda$ for fixed ${\rm FIX}(m_{c})$. However, when $m_{c}$ crosses a new maximum integer value, ${\rm FIX}(m_{c})$ changes discretely by unity, so both the sum terms change discretely. At these values, $\flast$ is non-differentiable, and must be defined only in terms of the integral, i.e.\ it undergoes a discrete jump in the integral $\Delta F_{\ell}$ as in the first-crossing case. Thus we must again only define $\flast$ completely over some discrete interval average $\langle \flast \rangle$.

\vspace{-0.5cm}
\subsubsection{Continuum Limit \&\ Linear Barrier Solution} 

In the continuum limit, the sum terms in Eq.~\ref{eqn:flast.beta.exact} become 
\begin{align}
\sum_{m=0}^{{\rm FIX}(m_{c})}P_{0}(m\,|\,\lambda)\approx\frac{\Gamma_{\lambda}(m_{c}+1)}{\Gamma(m_{c}+1)} \rightarrow \frac{1}{2}{\Bigl[}1+{\rm Erf}{\Bigl(}\frac{m_{c}-\lambda}{\sqrt{2\lambda}}{\Bigr)} {\Bigr]}
\end{align}
where again the latter is for the $\lambda\gg1$ normal limit of the Poisson distribution. 
and 
\begin{align}
\sum_{m=0}^{{\rm FIX}(m_{c})}P_{01}&(m[\lambda]\,|\,{\rm FIX}[m_{c}^{\prime}[\lambda^{\prime}]]) 
\rightarrow \frac{1}{2}{\Bigl[}1+{\rm Erf}{\Bigl(}\frac{m_{c}-(\lambda/\lambda^{\prime})\,m_{c}^{\prime}}{\sqrt{2\,(\lambda/\lambda^{\prime})\,(\lambda^{\prime}-\lambda)}}{\Bigr)} {\Bigr]}
\end{align}

Again in this limit we obtain a differentiable Volterra integral equation of the first kind, which can be solved by standard numerical methods for any $B(S)$. 

For a linear barrier again (of the same form as we considered for first-crossing), it is straightforward, albeit tedious, to verify (by taking the derivative of both sides of this equation and solving the resulting Volterra integral equation of the second kind) that the continuum solution is given by 
\be
\flast(B=\ln{(\rhocrit/\rho_{0})} = B_{0}+\mu_{0}\,S) \approx {|\mu_{0}+\tilde{\mu}|}\,P_{c}(m_{c}\,|\,\lambda)
\ee
Here note that since the true ``run'' in the numerator of $\nu$ is given by $\nu = (B_{0}+(\mu_{0}+\tilde{\mu})\,S)/S^{1/2}$, this is the same as the form of the solution for lognormal (non-intermittent) statistics, but with an appropriate modification $\tilde{\mu}$ for the $\beta_{\rho}$-dependent run in the mean. In the limit $\beta_{\rho}\rightarrow 1$, this becomes identical to the solution for the non-intermittent case. 

\vspace{-0.5cm}
\subsection{Inverse $\beta$ Models}

For $\beta_{\rho}>1$, the relation between $\delta$ and $m$ (in terms of $\beta_{\rho}$ and $\lambda$) is identical, and we can define the same $m_{c}$. The difference is, since $\ln{\beta_{\rho}}>0$, $\delta$ now {\em increases} with $m$. So $\delta > B$ means $m>m_{c}$. This qualitatively changes our previous solutions. 

For first-crossing, we again begin at $m=0$ and $S=0$; now $B>0$ is required so the mass is not already crossed, but this simply means $m_{c}>0$. As we go down in scale, $m$ can again only monotonically increase; but since $m>m_{c}$ gives $\delta>B$, crossings can occur ``between'' integer values of $m_{c}$. This is symmetric to the last-crossing distribution with $\beta_{\rho}<1$, though still ``stepping'' in the direction of increasing $\lambda$ (using $P_{10}$ instead of $P_{01}$), and switching the sum over $\Pi$ from $m>m_{c}$ to $m<m_{c}$. This gives:
\begin{align}
\label{eqn:flast.betagt1.exact}
\int_{0}^{S}\,{\rm d}\Sprime\,\flast(\Sprime)\,
\sum_{m\ge m_{c}}P_{0}(m-{\rm FIX}^{+}[m_{c}^{\prime}]\,|\,\lambda-\lambda^{\prime}) &= 
\sum_{m\ge m_{c}}P_{0}(m\,|\,\lambda)
\end{align}
where we use ${\rm FIX}^{+}(m_{c}^{\prime})$ to denote the ``rounded up'' value of $m_{c}$, i.e.\ the minimum integer value $n\ge 0$ such that $n\ge m_{c}$. 

For last-crossing, this reversal makes the situation symmetric to the $\beta_{\rho}<1$ first-crossing solution, and is nowhere differentiable. This gives
\be
\Delta {F_{\ell}}_{m_{c}} = P_{0}(m_{c}\,|\,\lambda) - 
\sum_{k=m_{c}+1}^{k_{\rm max}\rightarrow\infty}\,\Delta {F_{\ell}}_{k}\,
P_{01}(m_{c}[\lambda]\,|\,k[\lambda_{k}])
\ee

\vspace{-0.5cm}
\section{The Log-Density Variance as a Function of Scale}
\label{sec:appendix:S.of.R}

In Eq.~\ref{eqn:S.R}, we approximate the run of density variance $S(R)$ with scale by 
\be
S(R,\,\rho) = \int_{0}^{\infty}\,|\tilde{W}(k,\,R)|^{2}\,
\ln{{\Bigl[} 1 + \frac{b^{2}\,v_{t}^{2}(k)}{c_{s}(\rho,\,k)^{2}+\kappa^{2}\,k^{-2}} {\Bigr]}}\,{\rm d}\ln{k}
\ee
Testing this approximation in numerical simulations (specifically, how to map between turbulent velocity power spectrum and density variance as a function of scale) is an important question for future study. Subsequent work should study this approximation not just in simple isothermal cases, but also its extensions to vertically-stratified, rotating shear flows with non-isothermal gas. 

That said, we can make some very preliminary comparisons with published simulations. Most such studies focus on idealized driven turbulent boxes, with isothermal gas and no rotation. In this case, Eq.~\ref{eqn:S.R} becomes
\be
\label{eqn:S.R.appendix}
S(R) = \int_{k_{\rm max}}^{\infty}\,|\tilde{W}(k,\,R)|^{2}\,
\ln{{\Bigl[} 1 + \frac{b^{2}\,v_{t}^{2}(k)}{c_{s}^{2}} {\Bigr]}}\,{\rm d}\ln{k}
\ee
where $k_{\rm max}$ corresponds to the box scale. Without loss of generality, we can use units where $k_{\rm max}=1$. Note, by definition, $b\,v_{t}/c_{s} = \Machcompressive$, i.e.\ only the compressive component of the velocity enters. So we can simplify by stating all quantities in terms of $\Machcompressive$.\footnote{In the text, we have only approximated the relation between $\Machcompressive$ and $\Mach$ with $\Machcompressive \approx b\,\Mach$, as discussed in \S~\ref{sec:frag:driving}. Determining the most accurate mapping between the two is another important question for future study, but for clarity here, we separate it from the distinct question of the mapping between $\Machcompressive$ and $S(R)$.} 

By definition, $S(R)$ is just the convolution over the log-density power spectrum with the window function ($S(R) = \int |\tilde{W}(k)|^{2}\,\tilde{S}_{\ln{\rho}}(k)\,{\rm d}k$). So Eq.~\ref{eqn:S.R.appendix} implies a power spectrum 
\begin{align}
\label{eqn:S.R.box}
\tilde{S}_{\ln{\rho}}(k) &= k^{-1}\ln{[1 + \Machcompressive(k)^{2}]} \\ 
\nonumber &= k^{-1}\ln{[1 + \Mach_{c,\,0}^{2}\,k^{1-p}]}
\end{align}
where the latter assumes a power-law velocity spectrum and $\Mach_{c,\,0}\equiv \Machcompressive(k=k_{\rm max})$. 

Eq.~\ref{eqn:S.R.box} follows from the assumption that a Mach number-dispersion relation $S\approx\ln{[1+\Machcompressive^{2}]}$ \citep[e.g.][]{konstantin:mach.compressive.relation}, applies scale-by-scale. This follows if we assume the density field is strictly self-similar (see \paperone\ for details). Although reasonable, this is not unique. 

For example, consider instead a model where $S(R\rightarrow0)\approx \ln{[1 + \Mach_{c,\,0}^{2}]}$ is true exactly, but only as a box-integrated statement about the variance on small scales. Using $S(R\rightarrow0) = \int_{1}^{\infty} \tilde{S}_{\ln{\rho}}(k)\,{\rm d}k$, and $\Mach_{c,\,0}^{2} \equiv \int_{1}^{\infty} E_{v,\,c}(k)/c_{s}^{2}\,{\rm d}k$ (here $E_{v,\,c}$ is the power spectrum of compressive velocity fluctuations), we estimate
\be
\label{eqn:S.R.box.alt}
\tilde{S}_{\ln{\rho}}^{\rm alt}(k) \approx  \frac{(p-1)\,k\,E_{v\,c}\,c_{s}^{-2}}{k\,(1 + k\,E_{v,\,c}\,c_{s}^{-2})} = \frac{\Mach_{c,\,0}^{2}\,k^{-p}}{1 + \Mach_{c,\,0}^{2}\,k^{1-p}(p-1)^{-1}}
\ee

In the sub-sonic regime ($\Mach_{c,\,0}\lesssim1$ or $k\gg1$), all of these scalings become identical. In either case, $\tilde{S}_{\ln{\rho}}(k)\rightarrow \Mach_{c,\,0}^{2}\,k^{-p} \propto E_{v,\,c}(k)$, the velocity power spectrum. Since the amplitude of fluctuations is small, the density and log-density power spectra are equivalent: $\tilde{S}_{\rho}(k)\propto \tilde{S}_{\ln{\rho}}(k)\propto E_{v,\,c}(k) \propto k^{-p}$. This is a generic and well-known prediction in the weakly-compressible regime \citep[see][]{montgomery:1987.density.fluct.powerspectra.trace.vel}, which has been verified in a range of simulations \citep[see e.g.][]{kowal:2007.log.density.turb.spectra,schmidt:2009.isothermal.turb}. The same conclusion should generalize as predicted here for polytropic equations of state in the weakly compressible limit \citep{biskamp:2003.mhd.turb.book}. And the box-integrated variance $S(R\rightarrow0)\rightarrow\ln{[1 + \Mach_{c,\,0}^{2}]} \approx  \Mach_{c,\,0}^{2}$, in agreement with the standard Mach number-variance relation.

The most appropriate scaling when $\Mach(k)\gg1$ is more ambiguous. \citet{federrath:2010.obs.vs.sim.turb.compare} directly measure $E_{v,\,c}(k)$ and $\tilde{S}_{\ln{\rho}}(k)$ in simulations of compressively and solenoidally-driven turbulence with $\Mach\sim5.5$. We can therefore directly compare Eqs.~\ref{eqn:S.R.box}-\ref{eqn:S.R.box.alt} to the measured $\tilde{S}_{\ln{\rho}}(k)$, using the measured $E_{v,\,c}(k)$ (and $\Machcompressive^{2}(k) = c_{s}^{-2}\int_{k}^{\infty}E_{v,\,c}(k)\,{\rm d}k$). Both Eq.~\ref{eqn:S.R.box} \&\ \ref{eqn:S.R.box.alt} agree reasonably well with both compressive and solenoidal simulations (within $\sim20\%$ at $k$ outside a factor $\approx2$ of the driving and resolution scale). Interestingly, Eq.~\ref{eqn:S.R.box} agrees slightly better with the compressive case, while Eq.~\ref{eqn:S.R.box.alt} agrees slightly better with the solenoidal case. We can perform a similar exercise with the simulations in \citet{kowal:2007.log.density.turb.spectra}, spanning $\Mach\sim0.2-7$ and Alfv{\'e}n Mach number $\Mach_{\rm A}\sim 0.7-7$; here, however, $E_{v,\,c}(k)$ is not measured (only $E_{v}(k)$, the total velocity power spectrum), so we approximate $E_{v,\,c}(k)\sim b^{2}\,E_{v}(k)$. In this case, Eqs.~\ref{eqn:S.R.box}-\ref{eqn:S.R.box.alt} are consistent with the simulated $\tilde{S}_{\ln{\rho}}$ within a factor $\sim2$ (so much of this deviation may owe to $b$ not being constant). 

Note that, by construction, Eq.~\ref{eqn:S.R.box.alt} reproduces the ``usual'' box-integrated variance-Mach number relation $S(R\rightarrow0)\approx \ln{[1 + \Mach_{c,\,0}^{2}]}$, in both sub and super-sonic limits. However Eq.~\ref{eqn:S.R.box} leads to a slightly different box-integrated relation when $\Mach_{c,\,0}\gg1$; for a power-law $E_{v,\,c}(k)$, this is the dilogarithm. This is identical at sub-sonic $\Mach_{c,\,0}$ but rises more steeply (albeit still logarithmically) at large $\Mach_{c,\,0}$. In \citet{hopkins:2012.intermittent.turb.density.pdfs}, we find (from a large compilation of simulations) that the mass-weighted statistics appear to more accurately follow the logarithmic scaling, while the volume-weighted statistics (which we use in this paper) more closely follow the dilogarithm scaling. This motivates our particular choice in this paper. However, the simulation results clearly remain ambiguous, and their disagreement may be related to more fundamental aspects of the density PDF, such as intermittency. 

In either case, these scalings are broadly consistent with observations of the projected surface density power spectrum in galactic gas \citep{stanimirovic:1999.smc.hi.pwrspectrum,padoan:2006.perseus.turb.spectrum,block:2010.lmc.vel.powerspectrum}. However, with the present uncertainties, the ability to distinguish between the two is limited. 

Given this uncertainty, we have re-run our key calculations in this paper with Eq.~\ref{eqn:S.R.box.alt} (re-adding the appropriate $\kappa$ terms), instead of our usual Eq.~\ref{eqn:S.R.appendix}. We find no change in any of our qualitative conclusions. Quantitatively, the correlation functions and last-crossing distributions are largely unchanged; first-crossing distributions tend to ``shift'' slightly in mass, but at the factor $\lesssim2$ level.

\vspace{-0.5cm}
\section{On the Convolution of Lognormals}
\label{sec:appendix:lognormal.convolve}

In this paper, we have approximated the real-space density distribution as a lognormal averaged on all radial scales. But this cannot be exact if the real-space fluctuations on each scale are uncorrelated. In real-space, mass conservation means that the average density in the summed volume of two sub-regions must be $(V_{1}+V_{2})\,{\rho}_{1+2} = V_{1}\,\rho_{1} + V_{2}\,\rho_{2}$ (with $V_{1},\,V_{2}$ the volume of respective sub-regions); i.e.\ the summed real-space variable is the linear $\rho$, not the logarithm $\ln{\rho}$. 

But it is well-known from numerical calculations that the lognormal approximation is quite accurate (especially for the high-density tail of the density distribution, our focus in this paper), over a wide range of spatial and resolution scales \citep[including re-averaging the same simulations over smoothing scales; see e.g.][]{vazquez-semadeni:1994.turb.density.pdf,padoan:1997.density.pdf,scalo:1998.turb.density.pdf,nordlund:1999.density.pdf.supersonic,ostriker:1999.density.pdf,kowal:2007.log.density.turb.spectra,lemaster:2009.density.pdf.turb.review,schmidt:2009.isothermal.turb}. If it were not, re-calculating the density PDF at different resolution would yield fundamentally different shapes. 

The key is the well-known mathematical result that the convolution of two lognormal distributions (i.e.\ the sum of linear variables drawn from lognormals) can be very well-approximated by a lognormal (related to their similar characteristic functions). In particular, the Fenton-Wilkinson approximation \citep{fenton:1960.lognormal.sum.approx}, in which the convolution of lognormals is approximated as a lognormal, captures the high-density tail (positive fluctuations) very accurately (to within $\approx 2\%$ at $\rho\gg\langle\rho\rangle$; see e.g.\ \citealt{mehta:2007.lognormal.sum.approximation.methods}) given the simple constraint that the first two linear moments of the approximate lognormal are matched to the exact convolution result. If we satisfy these moment conditions, then, the residual deviations from log-normal (owing to the convolution) are much smaller than those owing to realistic levels of intermittency or small deviations from $\gamma=1$, let alone dynamic effects of collapsing regions, self-gravity, and large-scale perturbations (global modes) not captured in our analysis. 

For the first moment, our {\em enforcement} of mass conservation in the ``steps'' of the lognormal (by setting the median value in the PDF to $-S/2$ for a lognormal) means that this always automatically satisfied in real-space. The second moment condition implies (in Fourier space) 
\be
\nonumber \langle (\ln{(\rho/\rho_{0})} - \langle\ln{(\rho/\rho_{0})}\rangle)^{2} \rangle =\ln{(1 + \langle ({\rho/\rho_{0}} - \langle{\rho/\rho_{0}}\rangle)^{2} \rangle)}
\ee
i.e.\ $S[\ln{\rho}]=\ln{(1 + S[\rho])}$ (where $S[\ln{\rho}]$, $S[\rho]$ are the variance in $\ln{\rho}$ and the linear $\rho$, respectively). But this is used to derive $S[\ln{\rho}]$ itself (see e.g.\ \paperone, \citealt{padoan:2002.density.pdf}, \citealt{konstantin:mach.compressive.relation}, and references therein), so is also automatically satisfied to the accuracy that our Eq.~\ref{eqn:S.R} represents the log-density power spectrum. 

Moreover, the non-lognormal intermittency models in the text exactly obey the correct linear convolution condition, for specific choices of $\beta$ and $\gamma^{\prime}$ or $T$ in Appendix~\ref{sec:appendix:alt.intermittency} (see Appendix~A in \citealt{hopkins:2012.intermittent.turb.density.pdfs}).

If, however, the behavior of the {\em low-density} tail of the density distribution is desired (for example, calculating the mass function of underdense ``voids'' or ISM ``holes,'' as in \paperone), this approximation becomes much less accurate. This is confirmed in turbulent box simulations \citep{federrath:2010.obs.vs.sim.turb.compare}. In this limit, one should instead adopt the Schwartz-Yeh \citep{shwartz.yeh:1982.lognormal.sum.approx.lowvalues} approximation (matching the first two moments in log-space), which becomes similarly accurate for $\rho\ll \langle \rho\rangle$.

If the volumes summed are discrete real-space spheres (as opposed to our default choice of a the Fourier-space top-hat), some additional care is needed, discussed below (Appendix~\ref{sec:appendix:window}). 

\begin{figure}
    \centering
    \plotonesize{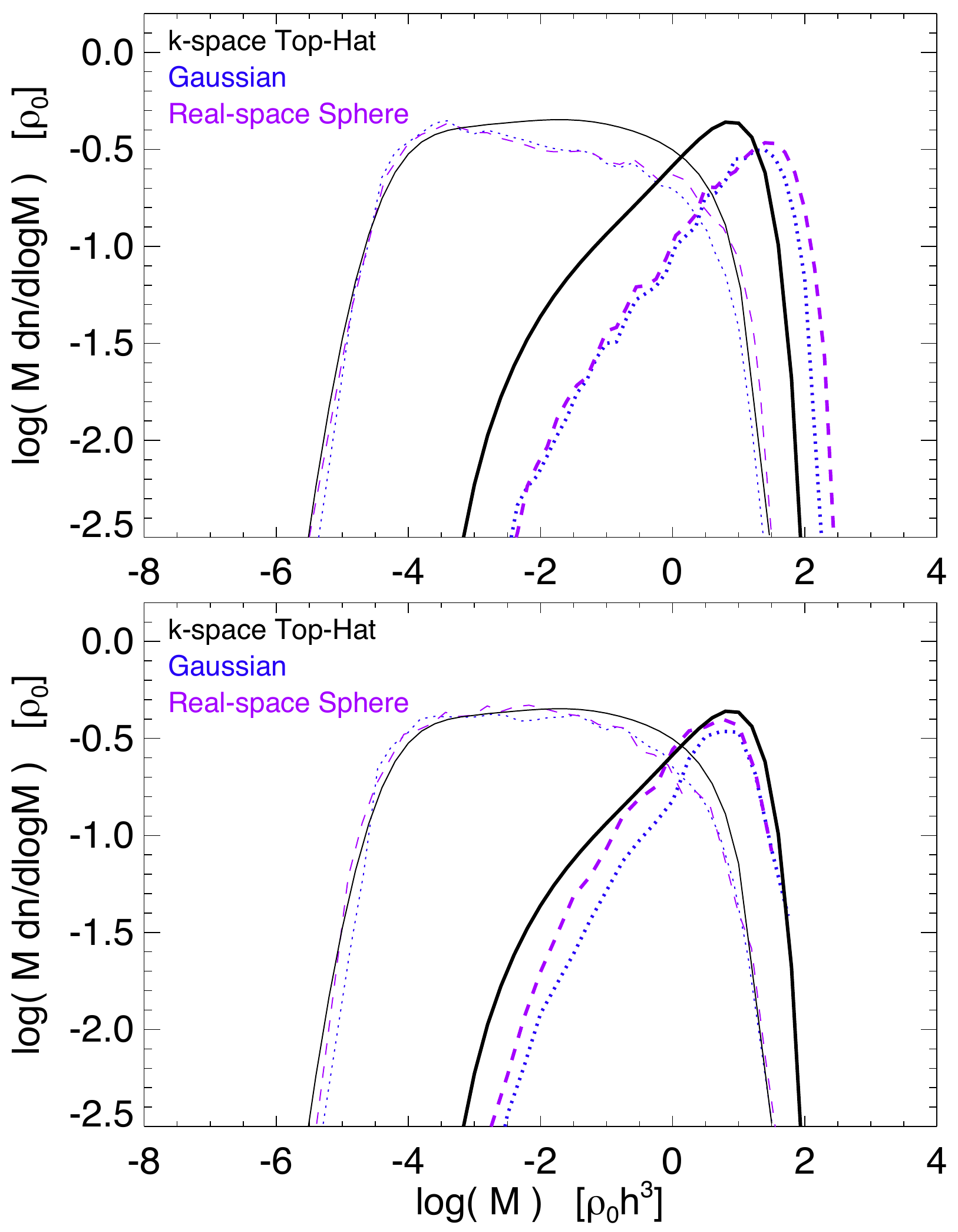}{0.95}
    \caption{Mass functions, as Fig.~\ref{fig:mfs}, with three different window functions $W(x,\,R)$ used to compute the real-space density fluctuations on a scale $R$ (see Appendix~\ref{sec:appendix:window})). We compare a sharp Fourier ($k$)-space top-hat (the default choice in this paper, because it allows exact analytic solutions for the MF), a Gaussian filter (Gaussian weighting in real-space $R$ and Fourier-space $k$), and a real-space top-hat (sharp-edged sphere in real-space). {\em Top:} MF for our standard model with these window functions, with no corrections applied to any quantity. The MFs are nearly identical in shape, but offset in normalization/mass scale at high-mass -- this is because the definition of mass \&\ volume for an ``object'' are different. {\em Bottom:} Same, but enforcing the same definition of mass (\&\ mass-variance relation) for each window function. The larger normalization offsets go away. The Gaussian and real-space sphere windows (themselves giving nearly identical results) produce smoother trajectories in Fig.~\ref{fig:demo} (feature less small-scale noise/structure), so produce fewer ``back-and-forth'' crossings at intermediate scales. This suppresses the low-mass end of the first-crossing MF and high-mass end of the last-crossing MF. 
    \label{fig:mf.vwindow}}
\end{figure}

\vspace{-0.5cm}
\section{The Window Function \&\ Real-Space Averaging}
\label{sec:appendix:window}

In \S~\ref{sec:methods} and throughout this paper, we simplify by assuming that densities are averaged over a window function which is a top-hat in Fourier ($k$) space ($\tilde{W}(k,\,R)$ in Eq.~\ref{eqn:S.R} $=1$ for $k\le R^{-1}$ and $=0$ for $k> R^{-1}$). As discussed in \citet{bond:1991.eps}, this is what allows us to treat fluctuations between scales as an uncorrelated random walk (at least in the isothermal case). This is necessary to obtain closed-form analytic solutions for many of the quantities we consider.

However, in certain situations it might be more appropriate to use a different filter function, for example a simple real-space sphere (a real-space top-hat of radius $R$). But this necessarily means fluctuations between scales are correlated, since ``incrementing'' the averaging scale integrates over contributions from all $k$-modes.\footnote{Consider the sum of a specific pair of discrete real-space volumes $V_{1}$ and $V_{2}$, which have approximately lognormal PDFs for each individual density within the volume, into a larger volume $V_{1+2}=V_{1}+V_{2}$ with mean density $\rho_{1+2}$, and consider the PDF of density that results as an approximate lognormal (with the Fenton-Wilkinson approximation in \S~\ref{sec:appendix:lognormal.convolve}). The first moment condition is just a restatement of mass conservation: $V_{1+2}\,\langle{\rho}_{1+2}\rangle = V_{1}\,\langle\rho_{1}\rangle + V_{2}\,\langle\rho_{2}\rangle$. By our definition of the median $-S/2$ of each lognormal ``step'', $\langle \rho_{1}\rangle = \langle \rho_{2}\rangle = \langle \rho_{1+2} \rangle$, so this is always satisfied. Matching the second moments in the mass conservation-restricted sum, and simplifying, we obtain $S[\ln{\rho_{1+2}}] = \ln{(1+\tilde{S})}$ where $\tilde{S} = f_{1}^{2}\,(\exp{[S_{1}]}-1) + f_{2}^{2}\,(\exp{[S_{2}]}-1) - 2\,f_{1}\,f_{2}\,(\langle\rho_{1}\rho_{2}/\rho_{0}^{2} \rangle -1)$ with $f_{i}\equiv V_{i}/V_{1+2}$ and $S_{i}\equiv S[\ln{\rho_{i}}]$. It is then straightforward to show that, if the variables $\rho_{1}$ and $\rho_{2}$ are uncorrelated, there is only one possible power spectrum shape for $S$, namely that of integrated Poisson fluctuations (with fluctuation ``number'' proportional to volume). The existence of any non-trivial structure in the power spectrum therefore necessarily implies correlated steps in real-space.} Mathematically, this is related to the Fourier transform of the window function. The overdensity in real-space is defined by:
\be
 \langle \ln{\rho({\bf x},\,R_{w})} \rangle = \int {\rm d}{\bf x}^{\prime}\,W(|{\bf x}^{\prime}-{\bf x}|,\,R_{w})\,\ln{\rho({\bf x}^{\prime})}
\ee
so in $k$-space ${\ln{\tilde{\rho}}}({\bf k},\,R_{w}) = \tilde{W}({\bf k},\,R_{w})\,{\ln{\tilde{\rho}}}({\bf k})$. 
For the spherical volume in real-space, $W(x\equiv |{\bf x}^{\prime}-{\bf x}|,\,R_{w}) = (4\pi\,R_{w}^{3}/3)^{-1}$ for $x\le R_{w}$ and $=0$ for $x>R_{w}$, which has Fourier transform
\be
\tilde{W}(k=|{\bf k}|,\,R_{w}) = \frac{3\,[\sin{(k\,R_{w})} - k\,R_{w}\,\cos{(k\,R_{w})}]}{(k\,R_{w})^{3}}
\ee
Thus calculating the increment in $\ln{\rho}$ in real-space involves integrating over contributions from all modes in $k$-space, and ``steps'' are correlated.

The introduction of a non-trivial window function changes our calculation in two ways. First, we must modify the calculation of $S(R)$ in Eq.~\ref{eqn:S.R}, inserting the appropriate $\tilde{W}(k,\,R)$. Second, we must modify our calculation of the density field. Derivations of the appropriate walk are given in \citet{bond:1991.eps,zentner:eps.methodology.review}, but the result is simple. For the Monte-Carlo method of evaluating ``trajectories'' in the (zero-mean) variable $\delta(R)$, each trajectory obeys the integrated Langevin equation:
\be
\delta(R) = \int_{0}^{\infty}\,\mathcal{R}(\ln{k})\,\tilde{W}(k,\,R)\,{\rm d}\ln{k}
\ee
where $\mathcal{R}$ is the ``stochastic force,'' a Gaussian random variable (independently drawn at each ${\rm d}\ln{k}$ interval) with zero mean and variance $\langle \mathcal{R}^{2}(\ln{k}) \rangle = ({\rm d}S/{\rm d}\ln{k})/|\tilde{W}(k,\,R)|^{2}$ (i.e.\ the integrand in Eq.~\ref{eqn:S.R}, without the window function). For a Monte carlo ensemble, each trajectory $\delta(R)$ should be calculated with its own independent set of $\mathcal{R}(\ln{k})$ for all $R$ (but that set is preserved for each $R$ used in the integration).\footnote{For a more detailed discussion of how to treat even more complicated window functions via the path-integral formulation, see \citet{maggiore:2010.path.integral.non.tophat.eps.filter}.}

In Fig.~\ref{fig:mf.vwindow}, we re-calculate the mass functions for our ``standard'' model, using three different window functions: the $k$-space top-hat (standard in the text), the sphere (real-space top-hat) above, and an intermediate (Gaussian) window function.\footnote{This is often used in calculating ``smoothed'' density fields, and is also numerically convenient. 
It is defined by $W(x,\,R_{w}) = \exp{(-x^{2}/2\,R_{w}^{2})/([2\pi]^{3/2}\,R_{w}^{3})}$ and $\tilde{W}(k,\,R_{w}) = \exp{(-k^{2}/2\,R_{w}^{-2})}$.}
First, we simply insert these forms of $W$ into the appropriate equations for $S$ and $\delta$, but keep the rest of the model fixed. As shown, this leads to predicted MFs with very similar shape. Interestingly, the Gaussian and real-space sphere filters predict nearly identical MFs to one another, but with some differences from the sharp $k$-space top-hat. The last-crossing MF predicted in the former cases has a slightly steeper slope (but recall the very large dynamic range plotted and note the difference is quite small, from a logarithmic slope of $\approx -2.0$ to $\approx-2.1$). And the first-crossing MF, while nearly identical in shape, is offset in the former cases to higher masses (by an apparently substantial factor of $\approx3$). However, as cautioned by \citet{bond:1991.eps} and \citet{zentner:eps.methodology.review}, care is needed in this simple comparison, because for otherwise fixed properties, the definition of ``effective mass'' and volume associated with each of these filters is in fact different, as is (as a consequence) the relation between mass or volume scale and the variance associated with that scale. Therefore it may be more appropriate to compare the different filters, enforcing a choice of filter radius such that the mass-volume, and volume-variance (or mass-variance) relations are the same. If we do this, also shown in Fig.~\ref{fig:mf.vwindow}, the large normalization difference, and some of the difference in slope, disappears. 

There is still a non-trivial remaining difference, in the sense that the last-crossing MF has a slightly steeper slope with a Gaussian or real-space sphere filter (less mass at the highest masses) and the first-crossing MF has a slope modified in the opposite sense (less mass at the lowest masses), but this is the regime which does not contain most of the mass in either MF. What happens here is that these filters, by smoothing the contribution of Fourier modes from all scales, exhibit less small-scale ``structure'' -- the trajectories $\delta(R)$ are considerably more smooth than those in Fig.~\ref{fig:demo}, while exhibiting identical variance. As a result, there is less probability of intermediate scale, large random fluctuations which are concentrated over a narrow range in spatial scale $R$ (i.e.\ have a first-crossing just slightly larger in scale than last-crossing). In other words, these different filters further enhance the tendency discussed in the text for the dynamic range of fragmentation to be concentrated in first-crossings at the maximal instability scale and in last-crossings at the sonic scale. 

One potentially important caveat to this comparison is that we have not modified the collapse threshold/barrier (i.e.\ Eq.~\ref{eqn:rhocrit}) for the different window functions. Ideally, one would re-derive this for the appropriate window function. The form we use is based on the dispersion relation derived for single Fourier ($k$) modes, so it is most appropriate to link this to a $k$-space top-hat window function (as we have done in the text). It would be interesting to see whether re-deriving this for Gaussian or real-space sphere overdensities would increase or minimize some of the differences seen here. Moreover, although windows such as a real-space sphere seem physically sensible, in an ideal case the window function (and barrier derivation) would be matched to some typical ``shape'' of density fluctuations, which is certainly not obvious and may not be universal (since we are considering the fully non-linear density field).

\end{appendix}

\end{document}